\input harvmac
\input rotate
\input epsf
\input xyv2
\input temp.defs

\font\teneurm=eurm10 \font\seveneurm=eurm7 \font\fiveeurm=eurm5
\newfam\eurmfam
\textfont\eurmfam=\teneurm \scriptfont\eurmfam=\seveneurm
\scriptscriptfont\eurmfam=\fiveeurm

 \font\teneusm=eusm10 \font\seveneusm=eusm7 \font\fiveeusm=eusm5
\newfam\eusmfam
\textfont\eusmfam=\teneusm \scriptfont\eusmfam=\seveneusm
\scriptscriptfont\eusmfam=\fiveeusm
\def\eusm#1{{\fam\eusmfam\relax#1}}
\font\tencmmib=cmmib10 \skewchar\tencmmib='177
\font\sevencmmib=cmmib7 \skewchar\sevencmmib='177
\font\fivecmmib=cmmib5 \skewchar\fivecmmib='177
\newfam\cmmibfam
\textfont\cmmibfam=\tencmmib \scriptfont\cmmibfam=\sevencmmib
\scriptscriptfont\cmmibfam=\fivecmmib

\writedefs

\noblackbox\input rotate
\let\includefigures=\iftrue
\includefigures
\message{If you do not have epsf.tex (to include figures),}
\message{change the option at the top of the tex file.}
\def\figin{\epsfcheck\figin}\def\figins{\epsfcheck\figins}
\def\epsfcheck{\ifx\epsfbox\UnDeFiNeD
\message{(NO epsf.tex, FIGURES WILL BE IGNORED)}
\gdef\figin##1{\vskip2in}\gdef\figins##1{\hskip.5in}
\else\message{(FIGURES WILL BE INCLUDED)}%
\gdef\figin##1{##1}\gdef\figins##1{##1}\fi}
\def\DefWarn#1{}

\def\underarrow#1{\vbox{\ialign{##\crcr$\hfil\displaystyle
{#1}\hfil$\crcr\noalign{\kern1pt\nointerlineskip}\rightarrowfill\crcr}}}

\def\EUN{\eusm N}


\def\figinsert{\goodbreak\midinsert}
\def\ifig#1#2#3{\DefWarn#1\xdef#1{fig.~\the\figno}
\writedef{#1\leftbracket fig.\noexpand~\the\figno}%
\figinsert\figin{\centerline{#3}}\medskip\centerline{\vbox{\baselineskip12pt
\advance\hsize by -1truein\noindent\footnotefont{\bf
Fig.~\the\figno:} #2}}
\bigskip\endinsert\global\advance\figno by1}
\else
\def\ifig#1#2#3{\xdef#1{fig.~\the\figno}
\writedef{#1\leftbracket fig.\noexpand~\the\figno}%
\global\advance\figno by1} \fi \noblackbox
\input amssym.tex
\def\hat{\widehat}
%
\overfullrule=0pt

\def\M{{\EUM}}

\def\tilde{\widetilde}
\def\bar{\overline}

%

\def\tilde{\widetilde}
\def\bar{\overline}

\def\Tr{{\rm Tr}}

\def\CC{{\Bbb{C}}}

\def\neg{\negthinspace}

\def\M{{\EUM}}

%
\def\tilde{\widetilde}
\def\bar{\overline}
\def\Z{{\Bbb{Z}}}
\def\R{{\Bbb{R}}}
\def\C{{\Bbb{C}}}

\font\zfont = cmss10 
\font\litfont = cmr6 
\def\bigone{\hbox{1\kern -.23em {\rm l}}}
\def\ZZ{\hbox{\zfont Z\kern-.4emZ}}
\def\half{{\litfont {1 \over 2}}}


\def\CC{{\Bbb{C}}}
\def\ZZ{{\Bbb{Z}}}

\font\zfont = cmss10 
\font\litfont = cmr6 
\def\bigone{\hbox{1\kern -.23em {\rm l}}}
\def\ZZ{\hbox{\zfont Z\kern-.4emZ}}
\def\half{{\litfont {1 \over 2}}}

\font\zfont = cmss10 
\font\litfont = cmr6 
\def\bigone{\hbox{1\kern -.23em {\rm l}}}
\def\ZZ{\hbox{\zfont Z\kern-.4emZ}}
\def\half{{\litfont {1 \over 2}}}

\def\N{{\EUN}}

\def\M{{\EUM}}

%
\def\tilde{\widetilde}
\def\bar{\overline}

\font\zfont = cmss10 
\font\litfont = cmr6 
\def\bigone{\hbox{1\kern -.23em {\rm l}}}
\def\ZZ{\hbox{\zfont Z\kern-.4emZ}}
\def\half{{\litfont {1 \over 2}}}


\let\includefigures=\iftrue


\def\Tr{{\rm Tr}}

\def\CC{{\Bbb{C}}}

\newfam\eusmfam
\textfont\eusmfam=\teneusm \scriptfont\eusmfam=\seveneusm
\scriptscriptfont\eusmfam=\fiveeusm
\def\eusm#1{{\fam\eusmfam\relax#1}}

 \noindent
 \Title{hep-th/yymmnnn} {\vbox{ \centerline{
Three-Dimensional Gravity Reconsidered}}} \medskip
\centerline{Edward Witten}
\smallskip
\centerline{\it{School of Natural Sciences, Institute for Advanced
Study}} \centerline{\it{Princeton, New Jersey 08540}}
\bigskip\bigskip
\noindent
We consider the problem of identifying the CFT's that may
be dual to pure gravity  in three dimensions with
negative cosmological constant.  The $c$-theorem indicates that
three-dimensional pure gravity is consistent only at certain values
of the coupling constant, and the relation to Chern-Simons gauge
theory hints that these may be the values at which the dual CFT can
be holomorphically factorized.  If so, and one takes at face value
the minimum mass of a BTZ black hole, then the energy spectrum of
three-dimensional gravity with negative cosmological constant can be
determined exactly.  At the most negative possible value of the
cosmological constant, the dual CFT is very likely the monster
theory of Frenkel, Lepowsky, and Meurman.  The monster theory may be
the first in a discrete series of CFT's that are dual to
three-dimensional gravity.  The partition function of the second
theory in the sequence can be determined on a hyperelliptic Riemann
surface of any genus.   We also make a similar analysis of supergravity.
 \vskip .5cm
\noindent\Date{June, 2007} \listtoc \writetoc

\newsec{Introduction}
\seclab\intro

\nref\ark{H. Leutwyler, ``A 2+1 Dimensional Model For The Quantum Theory Of Gravity,''
Nuovo Cim. {\bf 42A} (1966) 159.}%
\nref\bark{E. J. Martinec, ``Soluble Systems In Quantum Gravity,''
Phys. Rev. {\bf D30} (1984) 1198.}%
\nref\deser{S. Deser, R. Jackiw, and G. 't Hooft,
``Three-Dimensional Einstein Gravity: Dynamics of Flat Space,''
Annals Phys. {\bf 152}
(1984) 220.}%
 Three-dimensional pure quantum gravity, with the
Einstein-Hilbert action \eqn\harrigo{I={1\over 16\pi G}\int
d^3x\sqrt g\left(R+{2\over \ell^2}\right),} has been studied from
many points of view (see \refs{\ark -\deser} for some early
developments and \ref\carlip{S. Carlip, ``Conformal Field Theory,
$(2+1)$-Dimensional Gravity, and the BTZ Black Hole,''
gr-qc/0503022.} for a recent review and references), but its
status is fundamentally unclear. The present paper is devoted to a
tentative attempt to reconsider it. Before giving an overview of
the paper, we begin with an introduction to the problem.

The first thought about this theory is that at the classical level
it is ``trivial,'' in the sense that there are no gravitational
waves, and any two solutions are equivalent locally. So perhaps it
might be tractable quantum mechanically.

A second thought is that despite being ``trivial,'' the theory
actually is unrenormalizable by power counting, since the
gravitational constant $G$ has dimensions of length.  So perhaps the
quantum theory does not exist.

The claim about unrenormalizability, however, is fallacious,
precisely because the classical theory is trivial.  In three
dimensions, the Riemann tensor $R_{ijkl}$ can be expressed in
terms of the Ricci tensor $R_{ij}$. In the case of pure gravity,
the equations of motion express the Ricci tensor as a constant
times the metric.  So finally, any possible counterterm can be
reduced to a multiple of $\int d^3x \sqrt g $ and is equivalent
on-shell to a renormalization of the cosmological constant, which
is parametrized in \harrigo\ via the parameter $\ell^2$. A
counterterm that vanishes on shell can be removed by a local
redefinition of the metric tensor $g$ (of the general form
$g_{ij}\to g_{ij}+aR_{ij}+\dots$, where $a$ is a constant and the
ellipses refer to local terms of higher order). So a more precise
statement is that any divergences in perturbation theory can be
removed by a field redefinition and a renormalization of $\ell^2$.

\subsec{Relation To Gauge Theory}\subseclab\relation

\nref\Towns{A. Ach\'{u}carro and P. Townsend,  ``A Chern-Simons Action for Three-Dimensional
anti-De Sitter Supergravity Theories,'' Phys. Lett. {\bf B180} (1986) 89.}%
\nref\Witten{E. Witten, ``(2+1)-Dimensional Gravity as an Exactly
Soluble System,''
Nucl. Phys. {\bf B311} (1988) 46.}%
 The claim just made is valid regardless of how one
formulates perturbation theory.  But actually, there is a natural
formulation in which no field redefinition or renormalization is
needed.  This comes from the fact that classically,
$2+1$-dimensional pure gravity can be expressed in terms of gauge
theory.  The spin connection $\omega$ is an $SO(2,1)$ gauge field
(or an $SO(3)$ gauge field in the case of Euclidean signature). It
can be combined with the ``vierbein'' $e$ to make a gauge field of
the group $SO(2,2)$ if the cosmological constant is negative (and
a similar group if the cosmological constant is zero or positive).
We simply combine $\omega$ and $e$ to a $4\times 4$ matrix $A$ of
one-forms: \eqn\matro{A=\left(\matrix{ \omega & e/\ell \cr -e/\ell
& 0 \cr}\right).} Here $\omega$ fills out a $3\times 3$ block,
while $e$ occupies the last row and column.  As long as $e$ is
invertible, the usual transformations of $e$ and $\omega$ under
infinitesimal local Lorentz transformations and diffeomorphisms
combine together into gauge transformations of $A$. This statement
actually has a close analog in any spacetime dimension $d$, with
$SO(d-1,2)$ replacing $SO(2,2)$.  What is special in $d=3$ is that
\refs{\Towns,\Witten} it is also possible to write the action in a
gauge-invariant form. Indeed the usual Einstein-Hilbert action
\harrigo\ is equivalent to a Chern-Simons Lagrangian\foot{Here
$\tr^*$ denotes an invariant quadratic form on the Lie algebra of
$SO(2,2)$, defined by $\tr^*\,ab=\tr\,\,a \star b$, where $\tr$ is
the trace in the four-dimensional representation and $\star$ is
the Hodge star, $(\star b)_{ij}={1\over 2}\epsilon_{ijkl}b^{kl}$.}
for the gauge field $A$: \eqn\equito{I={k\over
4\pi}\int\,\tr^*\,\left(A\wedge dA+{2\over 3}A\wedge A\wedge
A\right).}  In dimensions other than three, it is not possible to
similarly replace the Einstein-Hilbert action with a
gauge-invariant action for gauge fields.\foot{In $d=4$, it is
possible to write the Hamiltonian constraints of General
Relativity in terms of gauge fields \ref\asht{A. Ashtekar, ``New
Variables For Classical And Quantum Gravity,'' Phys. Rev. Lett.
{\bf 57} (1986) 2244-2247.}. This has been taken as the starting
point of loop quantum gravity.}

In the gauge theory description, perturbation theory is
renormalizable by power counting, and is actually finite, because
there are no possible local counterterms. The Chern-Simons
functional itself is the only gauge-invariant action that can be
written in terms of $A$ alone without a metric tensor; as it is not
the integral of a gauge-invariant local density, it will not appear
as a counterterm in perturbation theory. The cosmological constant
cannot be renormalized, since in the gauge theory description, it is
a structure constant of the gauge group.

As we have already remarked, the Chern-Simons description of
three-dimensional gravity is valid when the vierbein is invertible.
This is so for a classical solution, so it is true if one is
sufficiently close to a classical solution.  Perturbation theory,
starting with a classical solution, will not take us out of the
region in which the vierbein is invertible, so the Chern-Simons
description of three-dimensional gravity is valid perturbatively.
The fact that, in this formulation, the perturbation expansion of
three-dimensional gravity is actually finite can reasonably be taken
as a hint that the corresponding quantum theory really does exist.

However, nonperturbatively, the relation between three-dimensional
gravity and Chern-Simons gauge theory is unclear. For one thing, in
Chern-Simons theory, nonperturbatively the vierbein may cease to be
invertible. For example, there is a classical solution with
$A=\omega=e=0$.  The viewpoint in \Witten\ was that such
non-geometrical configurations must be included to make sense of
three-dimensional quantum gravity nonperturbatively.  But it has has
been pointed out  (notably by N. Seiberg) that when we do know how
to make sense of quantum gravity, we take the invertibility of the
vierbein seriously.  For example, in perturbative string theory,
understood as a model of quantum gravity in two spacetime
dimensions, the integration over moduli space of Riemann surfaces
that leads to a sensible theory is derived assuming that the metric
should be non-degenerate.

There are other possible problems in the nonperturbative relation
between three-dimensional  gravity and Chern-Simons theory. The
equivalence between diffeomorphisms and gauge transformations is
limited to diffeomorphisms that are continuously connected to the
identity. However, in gravity, we believe that  more general
diffeomorphisms  (such as modular transformations in perturbative
string theory) play an important role.  These are not naturally
incorporated in the Chern-Simons description.  One can by hand
supplement the gauge theory description by imposing invariance under
disconnected diffeomorphisms, but it is not clear how natural this
is.

Similarly, in quantum gravity, one expects that it is necessary to
sum over the different topologies of spacetime.  Nothing in the
Chern-Simons description requires us to make such a sum.  We can
supplement the Chern-Simons action with an instruction to sum over
three-manifolds, but it is not clear why we should do this.

\nref\teschnerone{J. Teschner, ``On The Relation Between Quantum Liouville Theory And
The Quantum Teichmuller Spaces,'' Int. J. Mod. Phys. {\bf A19S2} (2004) 459-477, hep-th/0303149.}%
 \nref\teschnertwo{J. Teschner, ``From Liouville
Theory To The Quantum Gravity Of Riemann Surfaces'' (International
Congress on
Mathematical Physics 2003), hep-th/0308031.}%
\nref\teschnerplus{J. Teschner, ``A Lecture On The Liouville Vertex
Operators,'' Int. J. Mod. Phys. {\bf A19S2} (2004) 436-458,
hep-th/0303150.}
 \nref\teschner{J. Teschner, ``An Analog Of A
Modular Functor From
Quantized Teichmuller Theory,'' math/0510174.}%

{}From the point of view of the Chern-Simons description, it seems
natural to fix a particular Riemann surface $\Sigma$, say of genus
$g$, and construct a quantum Hilbert space by quantizing the
Chern-Simons gauge fields on $\Sigma$.  (Indeed, there has been
remarkable progress in learning how to do this and to relate the
results to Liouville theory \refs{\teschnerone-\teschner}.) In
quantum gravity, we expect topology-changing processes, such that it
might not be possible to associate a Hilbert space with a particular
spatial manifold.

Regardless of one's opinion of questions such as these, there is a
more serious problem with the idea that gravity and gauge theory
are equivalent non-perturbatively in three dimensions.  Some years
after the gauge/gravity relation was suggested, it was discovered
by Ba\~{n}ados, Teitelboim, and Zanelli \ref\ban{M. Ba\~{n}ados,
C. Teitelboim, and J. Zanelli, ``The Black Hole In
Three-Dimensional Spacetime,'' Phys. Rev. Lett. {\bf 69} (1992)
1849-1851, hep-th/9204099.} that in three-dimensional gravity with
negative cosmological constant, there are black hole solutions.
The existence of these objects, generally called BTZ black holes,
is surprising given that the classical theory is ``trivial.''
 \nref\strom{A. Strominger, ``Black Hole Entropy
{}From Near Horizon Microstates,'' JHEP 9802:009 (1998),
hep-th/9712251.}
\nref\birm{D. Birmingham, I. Sachs, and S. Sen, ``Entropy Of Three-Dimensional Black Holes
In String Theory,'' Phys. Lett. {\bf B424} (1998) 275-280, hep-th/9801019.}%
Subsequent work \refs{\strom,\birm} has made it clear that three-dimensional black
holes should be taken seriously, particularly in the context of
the AdS/CFT correspondence \ref\malda{J. Maldacena, ``The Large
$N$ Limit Of Superconformal Field Theories And Supergravity,''
Adv. Theor. Math. Phys. {\bf 2} (1998) 231-252, hep-th/9711200.}.

The BTZ black hole has a horizon of positive length and a
corresponding Bekenstein-Hawking entropy.  If, therefore,
three-dimensional gravity does correspond to a quantum theory, this
theory should have  a huge degeneracy of black hole states.  It
seems unlikely that this degeneracy can be understood in
Chern-Simons gauge theory, because this essentially topological
theory has too few degrees of freedom.  However, some interesting
attempts have been made; for a review, see \carlip.

The existence of the BTZ black hole makes three-dimensional gravity
a much more exciting problem.  This might be our best chance for a
solvable model with quantum black holes.  Surely in 3+1 dimensions,
the existence of gravitational waves with their nonlinear interactions
means that one cannot hope for
an exact solution of any system that includes quantum
gravity.\foot{An exact solution or at least an illuminating
description of the appropriate Hamiltonian for near-extremal black
holes interacting with external massless particles is conceivable.}
There might be an exact solution of a 1+1-dimensional model with
black holes (interesting attempts have been made \ref\stromcallan{C.
G. Callan, Jr., S. B. Giddings, J. A. Harvey, and A. Strominger,
``Evanescent Black Holes,'' Phys. Rev. {\bf D45} (1992) 1005-1009,
hep-th/9111056.}), but such a model is likely to be much less
realistic than three-dimensional pure gravity. For example, in 1+1
dimensions, the horizon of a black hole just consists of two points,
so there is no good analog of the area of the black hole horizon.

\subsec{What To Aim For}

So we would like to solve three-dimensional pure quantum gravity.
But what would it mean to solve it?

First of all, we will only consider the case that the cosmological
constant $\Lambda$ is negative. This is the only case in which we
know what it would mean to solve the theory.

Currently, there is some suspicion (for example, see \ref\suss{N. Goheer, M. Kleban,
and L. Susskind, ``The Trouble With de Sitter Space,'' JHEP 0307:066
(2003), hep-th/0212209.}) that quantum gravity with $\Lambda>0$ does
not exist nonperturbatively, in any dimension. One reason is that
it does not appear possible with $\Lambda>0$ to define precise
observables, at least none \ref\otherwitt{E. Witten, ``Quantum
Gravity In de Sitter Space,'' hep-th/0106109.}   that can be
measured  by an observer in the spacetime.\foot{At least
perturbatively, the de Sitter/CFT correspondence gives observables
that can be measured
 by an observer who looks at the whole universe from
the outside
\nref\twostrom{A. Strominger, ``The dS/CFT
Correspondence,'' JHEP 0110:034 (2001), hep-th/0106113.}%
\refs{\twostrom,\otherwitt}.  These observables characterize the
wavefunction of the ground state.}
 This is
natural if  a world with positive $\Lambda$ (like the one we may
be living in) is always at best metastable -- as is indeed the
case for known embeddings of de Sitter space in string theory
\ref\kklt{S. Kachru, R. Kallosh, A. Linde, and S. Trivedi, ``de
Sitter Vacua In String Theory,'' Phys. Rev. D68:046005,2003,
hep-th/0301240.}. If that is so, then pure gravity with
$\Lambda>0$ does not really make sense as an exact theory in its
own right but (like an unstable particle) must be studied as part
of a larger system. There may be many choices of the larger system
(for example, many embeddings in string theory) and it may be
unrealistic to expect any of them to be soluble.

Whether that is the right interpretation or not, we cannot in this
paper attempt to solve three-dimensional gravity with $\Lambda>0$,
since, not knowing how to define any mathematically precise
observables, we do not know what to try to calculate.

For $\Lambda=0$, above three dimensions there is a precise
observable in quantum gravity, the $S$-matrix.  However, in the
three-dimensional case, there is no $S$-matrix in the usual sense,
since in any state with nonzero energy, the spacetime is only
locally asymptotic to Minkowski space at infinity \deser. More
relevantly for our purposes, in three-dimensional pure gravity
with $\Lambda=0$, there is no $S$-matrix since there are no
particles that can be scattered. There are no gravitons in three
dimensions, and there are also no black holes unless $\Lambda<0$.
So again, we do not have a clear picture of what we would aim for
to solve three-dimensional gravity with zero cosmological
constant.

With negative cosmological constant, there is an analog, and in
fact a much richer analog, of the $S$-matrix, namely the dual
conformal field theory (CFT).  This is of course a {\it
two}-dimensional CFT, defined on the asymptotic boundary of
spacetime. Not only does AdS/CFT duality make sense in three
dimensions, but in fact one of the precursors of the AdS/CFT
correspondence was the discovery by Brown and Henneaux
\ref\brownhen{J. D. Brown and M. Henneaux, ``Central Charges In
The Canonical Realization Of Asymptotical Symmetries: An Example
{}From Three-Dimensional Gravity,'' Commun. Math. Phys. {\bf 104}
(1986) 207-226.} of an asymptotic Virasoro algebra in
three-dimensional gravity. They considered three-dimensional
gravity with negative cosmological constant possibly coupled to
additional fields.
 The action is \eqn\harrigon{I={1\over 16\pi G}\int d^3x\sqrt g\left(R+{2\over
\ell^2}+\dots\right),} where the ellipses reflect the contributions
of other fields.  Their main result is that the physical Hilbert
space obtained in quantizing this theory (in an asymptotically Anti
de Sitter or AdS spacetime) has an action of left- and right-moving
 Virasoro algebras with $c_L=c_R=3\ell/2G$.  In our modern
understanding \malda, this is part of a much richer structure -- the
boundary conformal field theory.

What it means to solve pure quantum gravity with $\Lambda<0$ is to
find this dual conformal field theory.  We focus on the case
$\Lambda<0$ because this is the only case in which we would know
what it means to solve the theory.  Luckily, and perhaps not
coincidentally, this is also the case that has black holes.

\subsec{A Non-Classical Restriction}\subseclab\nonclassical

This formulation of what we aim to do makes it clear that we must
anticipate a restriction that is rather surprising from a
classical point of view.  In contemplating the classical action
\harrigon, it appears that the dimensionless ratio $\ell/G$ is a
free parameter. But the formula for the central charge
$c_L=c_R=3\ell/2G$ shows that this cannot be the case.  According
to the Zamolodchikov $c$-theorem \ref\zam{A. B. Zamolodchikov,
``Irreversibility of the Flux of the Renormalization Group in a 2D
Field Theory,'' JETP Lett. {\bf 43} (1986) 430.}, in any
continuously varying family of conformal field theories in $1+1$
dimensions, the central charge $c$ is constant. More generally,
the same is true for the left- and right-moving central charges
$c_L$ and $c_R$.

So the central charges of the dual CFT cannot depend on a
continuously variable parameter $\ell/G$.  It must be \otherwitt\ that the
theory only makes sense for specific values of $\ell/G$.

Of course, the $c$-theorem has an important technical assumption:
the theory must have a normalizable and $SL(2,\R)\times
SL(2,\R)$-invariant ground state. (The two factors of $SL(2,\R)$
are for left- and right-moving boundary excitations.) This
condition is obeyed by three-dimensional gravity, with Anti de
Sitter space being the classical approximation to the vacuum.  The
desired $SL(2,\R)\times SL(2,\R)$ symmetry is simply the classical
$SO(2,2)$ symmetry of three-dimensional AdS space.

The statement that $\ell/G$ cannot be continuously varied is not
limited to pure gravity.  It holds for the same reason in any
theory of three-dimensional gravity plus matter that has a
sensible AdS vacuum.  For example, in the string theory models
whose CFT duals are known,   $\ell/G$     is expressed in terms of
integer-valued fluxes; this gives a direct explanation of why it
cannot be varied continuously.

\subsec{Plan Of This Paper}\subseclab\plan

Now we can describe the plan of this paper.

We aim to solve three-dimensional gravity with negative $\Lambda$,
at some distinguished values of $\ell/G$ at which it makes sense.

We do not have any rigorous way to determine the right values.
However, in section 2, taking at face value the Chern-Simons
description of three-dimensional gravity, we will use it to motivate
 certain values of $\ell/G$.  The values that emerge -- with the
help of a small sleight of hand in the choice of the gauge group for
the Chern-Simons theory -- are interesting.  They are the values at
which $c_L$ and $c_R$ are integer multiples of 24, and complete
holomorphic factorization of the dual CFT is conceivable.

Not wishing to look a gift horse in the mouth, we will assume that
these are the right values to consider.  Relying on holomorphic
factorization, to describe the solution of the theory, we must
describe a sequence of holomorphic CFT's with $c=24 k$,
$k=1,2,3,\dots$.

For $k=1$, it is believed \ref\schell{A. N. Schellekens,
``Meromorphic $c=24$ Conformal Field Theories,'' Commun. Math. Phys.
{\bf 153} (1993)159-186, hep-th/9205072.} that there are precisely
71 holomorphic CFT's with the relevant central charge $c=24$.  Of
these theories, 70 have some form of Kac-Moody or current algebra
symmetry extending the conformal symmetry.  In the AdS/CFT
correspondence, these theories are dual to three-dimensional
theories describing gravity plus additional gauge fields (with
Chern-Simons interactions).  To describe pure gravity, we need a
holomorphic CFT with $c=24$ and no Kac-Moody symmetry.

\def\MM{\Bbb{M}}
Such a model was constructed nearly twenty-five years ago by
Frenkel, Lepowsky, and Meurman \ref\flm{I. B. Frenkel, J. Lepowsky,
and A. Meurman, ``A Natural Representation of the Fischer-Griess
Monster With the Modular Function $J$ As Character,'' Proc. Natl.
Acad. Sci. USA {\bf 81} (1984) 3256-3260.}, who also conjectured its
uniqueness. The motivation for constructing the model was that it
admits as a group of symmetries the Fischer-Griess monster group
$\Bbb{M}$ -- the largest of the sporadic finite groups. (The link
between the monster and conformal field theory was suggested by
developments springing from an observation by McKay relating the
monster to the $j$-function, as we explain more fully in section
\bosecase.) Arguably, the FLM model is the most natural known
structure with $\Bbb{M}$ symmetry. A detailed and elegant
description is in the book \ref\flmtwo{I. B. Frenkel, J. Lepowsky,
and A. Meurman, {\it Vertex Operator Algebras And The Monster}
(Academic Press, Boston, 1988).}; for short summaries, see
\nref\dix{L. Dixon, P. Ginsparg, and J. Harvey, ``Beauty And The
Beast: Superconformal Symmetry In A Monster Module,'' Commun. Math.
Phys.
{\bf 119} (1988) 221-241.}%
\nref\dolgo{L. Dolan, P. Goddard, and P.  Montague, ``Conformal Field Theory, Triality,
and the Monster Group,'' Phys. Lett. {\bf B236} (1990) 165-172.}%
\refs{\dix,\dolgo},  and for subsequent developments and surveys,
see
\nref\borch{R. Borcherds, ``Automorphic Forms
And Lie Algebras,'' in {\it Current Developments In Mathematics}
(International Press, Boston, 1997).}%
\nref\mckay{J. McKay, ``The Essentials of Monstrous Moonshine,''  {\it
Groups and Combinatorics - in Memory of M. Suzuki}, Adv. Studies in Pure Math. {\bf 32} (2001)
347-343.}%
\nref\gannon{T. Gannon, ``Moonshine Beyond The Monster: The Bridge Connecting Algebra,
Modular Forms, and Physics'' (Cambridge University Press, 2006).}%
\refs{\borch - \gannon}.
Assuming the (unproved)
uniqueness conjecture of FLM, we
propose that their model must give the CFT that is dual to
three-dimensional gravity at $c=24$.

For $c=24k$, $k>1$, we need an analog of requiring that there is
no Kac-Moody symmetry.  A plausible analog, expressing the idea
that we aim to describe pure gravity, is that there should be no
primary fields of low dimension other than the identity.  A small
calculation shows that at $c=24k$, the lowest dimension of a
primary other than the identity cannot be greater than $k+1$, and
if we assume that this dimension is precisely $k+1$, then the
partition function is uniquely determined.  Conformal field
theories with this property were first investigated by H\"ohn in
\nref\hoehn{G. H\"ohn, ``Selbstduale Vertexoperatorsuperalgebren
und das Babymonster,''
 Ph.D. thesis  (Bonn 1995),
 Bonner Mathematische Schriften 286 (1996), 1-85, arXiv:0706.0236.}
\refs{\hoehn,\hoehntwo} and have been called extremal CFT's;
 see also
 \nref\hoehntwo{G. H\"ohn, ``Conformal Designs based on Vertex Operator
 Algebras,'' arXiv:math/0701626.}%
 \nref\kep{M. Jankiewicz and T. Kephart, ``Modular Invariants And
 Fischer-Griess Monster,'' arXiv:math-ph/0608001.}%
 \refs{\kep}.

 It is not known\foot{H\"ohn's
definition of an extremal CFT allowed holomorphic factorization up
to a phase, so that $c$ may be a multiple of 8, not 24.  As a
result, he discussed several examples of extremal theories that we
will not consider here. These examples have $k$ non-integral and
less than 2.} if extremal CFT's exist for $k>1$. If such a CFT
 does exist, it is an attractive candidate for the dual of
 three-dimensional gravity at the appropriate value of the
 cosmological constant.   The primaries of dimension $k+1$ and above
 would be interpreted as operators that create black holes.  The
 dimension $k+1$ agrees well with the minimum mass of a BTZ black
 hole.  This statement may sound like magic, since the value $k+1$
 is determined from holomorphy and modular invariance without
 mentioning black holes; but the result is not so surprising if one
 is familiar with previous results on the AdS/CFT correspondence in
 three dimensions \ref\farey{R. Dijkgraaf, J. Maldacena, G. Moore, and E. Verlinde,
 ``A Black Hole Farey Tail,'' hep-th/0005003.}.

 Section 3 of this paper is devoted to describing the partition function of
 an extremal CFT and discussing how such a theory could be related
 to three-dimensional gravity.
 In both sections 2 and 3, we consider also the case of
 three-dimensional supergravity.  More precisely, we consider only
 minimal supergravity, corresponding to $N=1$ superconformal
 symmetry for the boundary CFT.  In this case, holomorphic
 factorization is conceivable at $c=12k^*$, $k^*=1,2,3,\dots$.
 Here there is a little ambiguity about exactly what we should mean
 by an extremal superconformal field theory (SCFT), but
 pragmatically, there are good candidates at $k^*=1,2$.
 The $k^*=1$ theory was
 constructed by Frenkel, Lepowsky, and Meurman, who also conjectured
 its uniqueness.  Its discrete symmetries were understood only
 recently in work by Duncan \ref\duncan{J. F. Duncan,
 ``Super-Moonshine For Conway's Largest Sporadic Group,''
 arXiv:math/0502267.}.  For $k^*=2$, the extremal SCFT was constructed by Dixon, Ginsparg, and
 Harvey \dix, by
 modifying the orbifold projection that had been used \flm\ in
 constructing the $k=1$ extremal CFT.
Interestingly, the extremal SCFT's with $k^*=1,2$ both
admit\foot{The fact that they have essentially the same symmetry
group was pointed out by J. Duncan, who also suggested the
identification of the $k^*=2$ theory.} an action of very large discrete groups
related to the Conway group.  This
is a further indication that unusual discrete groups are relevant to
three-dimensional gravity and supergravity.  In fact, we find some
hints that supergravity may have monster symmetry at $k^*=4$ and baby monster symmetry at
$k^*=6$.

Regrettably, we do not know how to construct new examples of
extremal conformal or superconformal field theories.  Section 4 is
devoted to a calculation that aims to give modest support to the
idea that new extremal theories do exist.  We consider an extremal
CFT with $k=2$ and show that its partition function can be
uniquely determined on a hyperelliptic Riemann surface of any
genus (including, for example, any Riemann surface of genus 2).
The fact that a partition function with the right properties
exists and is unique for any genus is hopefully a hint that an
extremal $k=2$ CFT does exist.

We make at each stage the most optimistic possible assumption.
Decisive arguments in favor of the proposals made here
are still lacking.   The literature on three-dimensional
gravity is filled with claims (including some by the present author
\Witten) that in hindsight seem less than fully satisfactory.
Hopefully, future work will clarify things.

\vskip .2 cm Advice by J. Maldacena was essential at the outset of
this work. I also wish to thank J. Duncan, G. H\"ohn, G. Nebe, and J.
Teschner for descriptions of their work and helpful advice; T. Gannon, R. Griess, J.
Lepowsky, and A. Ryba for correspondence about the monster group and related matters;
and many colleagues at the IAS and elsewhere, especially A. Maloney,
G. Moore, and S. Shenker, for helpful comments.

\newsec{Gauge Theory And The Value Of $c$}\seclab\cval

The goal of the present section is to determine what values of the
cosmological constant, or equivalently of the central charge $c$
of the boundary CFT, are suggested by the relation between
three-dimensional gravity and Chern-Simons gauge theory.

Before proceeding to any calculation, we will dispose of a few
preliminary points.  The first is that \ref\destemj{S. Deser, R.
Jackiw, and S. Templeton, ``Topologically Massive Gauge
Theories,'' Annals Phys. {\bf 140} (1982) 372-411.}, as long as
the three-dimensional spacetime is oriented, as we will assume in
this paper, three-dimensional gravity can be generalized to
include an additional interaction, the Chern-Simons functional of
the spin connection $\omega$: \eqn\bogus{\Delta_0 I={k'\over
4\pi}\int_W \tr\,\left(\omega\wedge d\omega+{2\over
3}\omega\wedge\omega\wedge\omega\right).} Here we think of
$\omega$ as an $SO(2,1)$ gauge field (or an $SO(3)$ gauge field in
the case of Euclidean signature).  Also, $\tr$ is the trace in the
three-dimensional representation of $SO(2,1)$, and $k'$ is
quantized for topological reasons (the precise normalization
depends on some assumptions and is discussed in sections
\quantcoup\ and \interpretation). Equivalently, instead of
$\omega$, we could use the $SO(2,2)$ gauge field $A$ introduced in
eqn. \matro, and add to the action a term of the form
\eqn\ogus{\Delta I={k'\over 4\pi}\int_M \tr\,\left(A\wedge
dA+{2\over 3}A\wedge A\wedge A\right),} where now $\tr$ is the
trace in the four-dimensional representation of $SO(2,2)$.
Provided that the conventional Einstein action \harrigo\ is also
present,  it does not matter which form of the gravitational
Chern-Simons interaction we use, since they lead to equivalent
theories. If one adds\foot{This operation is invariant under
diffeomorphisms and local Lorentz transformations because $\omega$
and $e$ transform in the same way under local Lorentz
transformations -- a statement that holds precisely in three
spacetime dimensions.} to $\omega$ a multiple of
 $e$, the Einstein action \harrigo\ transforms in a way that
 cancels  the $e$-dependent part of \ogus, reducing it
 to \bogus\ (while modifying the parameters in the Einstein action).

 For our purposes, the
 $SO(2,2)$-invariant form \ogus\ is more useful.  This way of
 writing the Chern-Simons functional places it precisely in parallel
 with the Einstein-Hilbert action, which as in \equito\ can
 similarly be expressed as a Chern-Simons interaction, defined with
 a different quadratic form.  We will use the fact that all
 interactions can be written as Chern-Simons interactions to
 constrain the proper quantization of all dimensionless
 parameters, including $\ell/G$.

 We start with the fact that the group $SO(2,2)$ is locally equivalent to
 $SO(2,1)\times SO(2,1)$.  Moreover, we will in performing the
 computation assume to start with that $SO(2,1)\times SO(2,1)$ is the right global
 form of the gauge group.  (Then we consider covering groups\foot{For an early treatment of
 coverings in the context of Chern-Simons theory with compact gauge group, see
 \ref\ms{G. Moore and N. Seiberg, ``Taming The Conformal Zoo,'' Phys. Lett. {\bf B220}
 (1989) 422.}} that are only locally
 isomorphic to $SO(2,1)\times SO(2,1)$.)
 Thus, by taking suitable linear combinations of $\omega$ and $e$,
 we will obtain a pair of $SO(2,1)$ gauge
 fields $A_L$ and $A_R$.  These have Chern-Simons interactions
 \eqn\zormo{I={k_L\over 4\pi}\int \tr\,\left(A_L\wedge dA_L+{2\over
 3}A_L\wedge A_L\wedge A_L\right)-{k_R\over 4\pi}\int \tr\,\left(A_R\wedge dA_R+{2\over
 3}A_R\wedge A_R\wedge A_R\right).}
 Both $k_L$ and $k_R$ are integers for topological reasons, and this
 will lead to a quantization of the ratio $G/\ell$ that appears in the Einstein-Hilbert
 action, as well as the
 gravitational Chern-Simons coupling \ogus.  The minus sign
 multiplying the last term in \zormo\ is convenient; it will ensure
 that $k_L$ and $k_R$ are both positive.

\subsec{Quantization Of Parameters}\subseclab\quantcoup

For completeness, we begin by reviewing the quantization of the
Chern-Simons coupling in gauge theory.  The basic case to consider
is that the gauge group is $U(1)$.  The gauge field $A$ is a
connection on a complex line bundle ${\cal L}$ over a
three-manifold $W$, which for simplicity we will assume to have no
boundary. Naively speaking, the Chern-Simons action is
\eqn\toldo{I={k\over 2\pi}\int_W A\wedge dA} with some coefficient
$k$. If the line bundle ${\cal L}$ is trivial, then we can
interpret $A$ as a one-form, and $I$ is well-defined as a
real-valued functional.  If this were the general situation, there
would be no need to quantize $k$.

However, in general ${\cal L}$ is non-trivial, $A$ has Dirac
string singularities, and the formula \toldo\ is not really
well-defined as written.  To do better, we pick a four-manifold
$M$ of boundary $W$ and such that ${\cal L}$ extends over $M$.
Such an $M$ always exists.  Then we pick an extension of ${\cal
L}$ and $A$ over $M$, and  replace the definition \toldo\ with
\eqn\boldo{I_M={k\over 2\pi}\int_M F\wedge F,} where $F=dA$ is the
curvature.  Now there is no Dirac string singularity, and the
definition of $I_M$ makes sense.  But $I_M$ does depend on $M$
(and on the chosen extension of ${\cal L}$, though we do not
indicate this in the notation).  To quantify the dependence on
$M$, we consider two different four-manifolds $M$ and $M'$ with
boundary $W$ and chosen extensions of $\cal L$.  We can build a
four-manifold $X$ with no boundary by  gluing together $M$ and
$M'$ along $W$, with opposite orientation for $M'$ so that they
fit smoothly along their common boundary.  Then we get
\eqn\zolg{I_M-I_{M'}={k\over 2\pi}\int_X F\wedge F.}

Now, on the closed four-manifold $X$, the quantity $\int_X F\wedge
F/(2\pi)^2$ represents $\int_X c_1({\cal L})^2$ (here $c_1$ is the
first Chern class) and so is an integer.  In quantum mechanics,
the action function $I$ should be defined modulo $2\pi$ (so that
$\exp(iI)$, which appears in the path integral, is single-valued).
Requiring $I_M-I_{M'}$ to be an integer multiple of $2\pi$, we
learn that $k$ must be an integer.  This is the quantization of
the Chern-Simons coupling for $U(1)$ gauge theory.\foot{There is a
refinement if the three-manifold $W$ is endowed with a spin
structure. In this case, $k$ can be a half-integer, as explained
in \ref\witgu{E. Witten,``$SL(2,\Z)$ Action On Three-Dimensional
Conformal Field Theories With Abelian Symmetry,''
hep-th/0307041.}.
 This refinement is physically realized in the quantum Hall effect with filling fraction 1.
 That effect can be described by an electromagnetic Chern-Simons coupling with $k=1/2$; the half-integral
 value is consistent because the microscopic theory has fermions and so requires a spin structure.}

Now let us move on to the case of gauge group $SO(2,1)$.  The
group $SO(2,1)$ is contractible onto its maximal compact subgroup
$SO(2)$, which is isomorphic to $U(1)$.  So quantization of the
Chern-Simons coupling for an $SO(2,1)$ gauge field can be deduced
immediately from the result for $U(1)$.  Let $A$ be an $SO(2,1)$
gauge field and define the Chern-Simons coupling
\eqn\rewo{I={k\over 4\pi}\int_W\tr \,\left(A\wedge dA+{2\over
3}A\wedge A\wedge A\right),} where $\tr $ is the trace in the
three-dimensional representation of $SO(2,1)$. Then, in order for
$I$ to be part of the action of a quantum theory, $k$ must be an
integer. The reason that the factor $1/2\pi$ in \boldo\ has been
replaced by $1/4\pi$ in \rewo\ is simply that, when we identify
$U(1)$ with $SO(2)$ and then embed it in $SO(2,1)$, the trace
gives a factor of 2.

\bigskip\noindent{\it Coverings}

So we have obtained the appropriate quantization of the Chern-Simons
coupling for gauge group $SO(2,1)$.  However, this is not quite the
whole story, because $SO(2,1)$ is not simply-connected.  As it is
contractible to $SO(2)\cong U(1)$, it has the same fundamental group
as $U(1)$, namely $\Z$.  Hence it is possible, for every positive
integer $n$, to take an $n$-fold cover of $SO(2,1)$.  The most
familiar of these is the two-fold cover, $SL(2,\R)$.  In addition,
$SO(2,1)$ has a simply-connected universal cover.

We want to work out the quantization of the Chern-Simons
interaction if $SO(2,1)$ is replaced by one of these covering
groups.  Again, it is convenient to start with $U(1)$.  To say
that the gauge group of an abelian gauge theory is $U(1)$ rather
than $\R$ means precisely that the possible electric charges form
a lattice, generated by a fundamental charge that we call ``charge
1.'' Dually, the magnetic fluxes are quantized, with $\int_C
F/2\pi\in\Z$ for any two-cycle $C$.  Replacing $U(1)$ by an
$n$-fold cover means that the electric charges take values in
$n^{-1}\Z$, and dually, the magnetic fluxes are divisible by $n$,
$\int_C F/2\pi \in n\Z$.  As a result, for a four-manifold $X$, we
have $\int_C F\wedge F/(2\pi)^2\in n^2\Z$.  So, in requiring that
the Chern-Simons function \toldo\ should be well-defined modulo
$2\pi$, we require that $k\in n^{-2}\Z$.  This is the appropriate
result for an $n$-fold cover. In the case of the universal cover,
with $U(1)$ replaced by $\R$, the magnetic fluxes vanish and there
is no topological restriction on $k$.

These statements carry over immediately to covers of $SO(2,1)$,
whose covers are all contractible to corresponding covers of $U(1)$.
So for an $n$-fold cover of $SO(2,1)$, we require \eqn\yoldo{k\in
n^{-2}\Z,} and for the universal cover of $SO(2,1)$, $k$ is
arbitrary and can vary continuously.

\bigskip\noindent{\it Diagonal Covers}

There is more to say, because three-dimensional gravity is
actually related to $SO(2,1)\times SO(2,1)$ gauge theory, not just
to gauge theory with a single $SO(2,1)$. So we should consider
covers of $SO(2,1)\times SO(2,1)$ that do not necessarily come
from separate covers of the two factors.

As $SO(2,1)\times SO(2,1)$ is contractible to $SO(2)\times
SO(2)=U(1)\times U(1)$, we can proceed by first analyzing the
$U(1)\times U(1)$ case.  We consider a $U(1)\times U(1)$ gauge
theory with gauge fields $A,B$ and a Chern-Simons action
\eqn\kipog{I={k_L\over 2\pi}\int_W\,A\wedge dA-{k_R\over
2\pi}\int_W B\wedge dB.}  To define $I$ in the topologically
non-trivial case, we pick a four-manifold $M$ over which
everything extends and define \eqn\nikog{I_M=\int_M\left({k_L\over
2\pi}F_A\wedge F_A-{k_R\over 2\pi}F_B\wedge F_B\right),} where
$F_A$ and $F_B$ are the two curvatures.  This is well-defined mod
$2\pi$ if \eqn\ikog{I_X=\int_X\left({k_L\over 2\pi}F_A\wedge
F_A-{k_R\over 2\pi}F_B\wedge F_B\right)} is a multiple of $2\pi$
for any $U(1)\times U(1)$ gauge field over a closed four-manifold
$X$.

In $U(1)\times U(1)$ gauge theory, the charge lattice is generated
by charges $(1,0)$ and $(0,1)$, and the cohomology classes $x=
F_A/2\pi$ and $y= F_B/2\pi$ are integral. For a cover of
$U(1)\times U(1)$, we want to extend the charge lattice. To keep
things simple, we will consider only the case that will actually
be important in our application: a diagonal cover, in which one
adds the charge vector $(1/n,1/n)$ for some integer $n$.  In this
case, $x$ and $y$ are still integral, and their difference is
divisible by $n$: $x=y+nz$ where $n$ is an integral class. We have
\eqn\ykog{I_X=2\pi(k_L-k_R)\int_X y^2 +2\pi
k_L\int_X(n^2z^2+2nyz). } The condition that this is a multiple of
$2\pi$ for any $X$ and any integral classes $y, z$ is that
\eqn\tomigo{\eqalign{k_L&\in \cases{n^{-1}\Z& if $n$ is odd\cr
(2n)^{-1}\Z & if $n$ is even \cr}\cr
                   k_L-k_R&\in \Z.\cr}}

These are also the restrictions on $k_L$ and $k_R$ if the gauge
group is a diagonal cover of $SO(2,1)\times SO(2,1)$ with action
\zormo. For example, the group $SO(2,2)$ is a double cover of
$SO(2,1)\times SO(2,1)$, and this cover corresponds to the case
$n=2$ of the above discussion. So for $SO(2,2)$ gauge theory, the
appropriate restriction on the Chern-Simons levels is
\eqn\tomigot{\eqalign{k_L&\in {1\over 4}\Z\cr
                   k_L-k_R&\in \Z.\cr}}
For a general $n$-fold diagonal cover, one should use \tomigo.

It is also possible to form a ``universal diagonal cover,''
corresponding roughly to the limit $n\to\infty$ in the above
formulas.  With this gauge group, there is no restriction on $k_L$,
but $k_L-k_R$ is an integer.  In terms of three-dimensional gravity,
to which we return next, this corresponds to letting $\ell/G$ be a
freely variable parameter, while the gravitational Chern-Simons
coupling $k'$ defined in eqn. \ogus\ is an integer.  As explained in
section \nonclassical, although this is the state of affairs
classically in three-dimensional gravity, it cannot be the correct
answer quantum mechanically.

\subsec{Comparison To Three-Dimensional Gravity}

So far, we have understood the appropriate gauge theory
normalizations for the Chern-Simons action
 \eqn\ozormo{\eqalign{I=&k_LI_L+k_RI_R\cr =&{k_L\over 4\pi}\int \tr\,\left(A_L\wedge dA_L+{2\over
 3}A_L\wedge A_L\wedge A_L\right)-{k_R\over 4\pi}\int \tr\,\left(A_R\wedge dA_R+{2\over
 3}A_R\wedge A_R\wedge A_R\right).\cr}}
 Our next step will be to express $A_L$ and $A_R$, which are gauge
 fields of $SO(2,1)\times SO(2,1)$ (or a covering group) in terms
 of gravitational variables, and thereby determine the constraints
 on the gravitational couplings.  We have
 \eqn\boromo{I={k_L+k_R\over 2}\left(I_L-I_R\right)+(k_L-k_R){\left(I_L+I_R\right)\over
 2}.}  The  term in \boromo\ proportional to $I_L-I_R$ will gave
 the Einstein-Hilbert action \harrigo, while the term proportional to $(I_L+I_R)/2$
 is
 equivalent to the gravitational Chern-Simons coupling \ogus\ with
  coefficient $k'=k_L-k_R$.

\def\neg{\negthinspace}
The spin connection $\omega^{ab}=\sum_i dx^i \omega_i^{ab}$ is a
one-form with values in antisymmetric $3\times 3$ matrices.  The
vierbein is conventionally a one-form valued in Lorentz vectors,
$e^a=\sum_i dx^i\,e_i^a$. The metric is expressed in terms of $e$
in the usual way, $g_{ij}dx^i\otimes dx^j=\sum_{ab}\eta_{ab}
e^a\otimes e^b$, where $\eta={\rm diag}(-1,1,1)$ is the Lorentz
metric; and the Riemannian volume form is $d^3x \sqrt g= {1\over
6}\epsilon_{abc}e^a\wedge e^b\wedge e^c$, where $\epsilon_{abc}$
is the antisymmetric tensor with, say, $\epsilon_{012}=1$.  All
this has an obvious analog in any dimension. However, in three
dimensions, a Lorentz vector is equivalent to an antisymmetric
tensor; this is the fact that makes it possible to relate gravity
and gauge theory. It is convenient to introduce $^*\neg
e_{ab}=\epsilon_{abc}e^c$, which is a one-form valued in
antisymmetric matrices, just like $\omega$. We raise and lower
local Lorentz indices with the Lorentz metric $\eta$, so ${1\over
2}\epsilon^{abc}\epsilon_{bcd}=-\delta^a_d$, and $e^c=-{1\over
2}\epsilon^{abc}\,{}^*\neg e_{bc}$.

We can combine $\omega $ and $^*\neg e$ and  set
$A_L=\omega-{}^*\neg e/\ell$, $A_R=\omega+{}^*\neg e/\ell$. A
small computation gives \eqn\oromo{I_L-I_R=-{1\over \pi\ell}\int
\tr \,{}^*\neg e (d\omega+\omega\wedge\omega)-{1\over 3\pi
\ell^3}\int \tr \,({}^*\neg e\wedge{} ^ *\neg e\wedge{} ^*\neg
e).} In terms of the matrix-valued curvature two-form
$R^{ab}=(d\omega+\omega\wedge \omega)^{ab} = \half
\sum_{ij}dx^i\wedge dx_j R_{ij}^{ab}$, where $R_{ij}^{ab}$ is the
Riemann tensor, and the metric tensor $g$, this is equivalent to
\eqn\borox{I_L-I_R= {1\over \pi \ell}\int d^3x \sqrt g
\left(R+{2\over \ell^2}\right).} Remembering the factor of
$(k_L+k_R)/2$ in \boromo, we see that this agrees with the
Einstein-Hilbert action \harrigo\ precisely if
\eqn\orox{k_L+k_R={\ell\over 8G}.}

The central charge of the boundary conformal field theory was
originally computed by Brown and Henneaux \brownhen\ for the case
that the gravitational Chern-Simons coupling $k'=k_L-k_R$
vanishes.  In this case, we set $k=k_L=k_R=\ell/16G$.  The formula
for the central charge is $c=3\ell/2G$, and this leads to $c=24k$.
For the case $k'=0$, the boundary CFT is left-right symmetric,
with $c_L=c_R$, so in fact $c_L=c_R=24k$.

In general, the boundary CFT has left- and right-moving Virasoro
algebras that can be interpreted (for a suitable orientation of
the boundary) as boundary excitations associated with $A_L$ and
$A_R$ respectively.  So the  central charges $c_L$ and $c_R$ are
functions only of $k_L$ and $k_R$, respectively.  Hence the
generalization of the result obtained in the last paragraph is
\eqn\zolgo{(c_L,c_R)=(24k_L,24k_R).}

\subsec{Holomorphic Factorization}\subseclab\comparison

In conformal field theory in two dimensions, the ground state
energy is $-c/24$.  More generally, if there are separate left and
right central charges $c_L$ and $c_R$, the ground state energies
for left- and right-movers are $(-c_L/24,-c_R/24)$.  Modular
invariance says that the difference between the left- and
right-moving ground state energies must be an integer.  In the
above calculation, $(c_L-c_R)/24=k_L-k_R$.  According to \zolgo,
this is an integer provided that the gravitational Chern-Simons
coupling $k'=k_L-k_R$ is integral.

 Holomorphic factorization
requires that the left- and right-moving ground state energies
should be separately integral, so that there is modular invariance
separately for left-moving and right-moving modes of the CFT.
Thus, for holomorphic factorization, $c_L$ and $c_R$ must both be
integer multiples of 24.  This is the only constraint, since
holomorphic CFT's with $c=24$ do exist (and have been classified
\schell\ modulo a conjecture mentioned in section \plan).

According to \zolgo, the condition for $c_L$ and $c_R$ to be
multiples of 24 is precisely that $k_L$ and $k_R$ must be
integers.  As in our discussion of eqn. \rewo, this is the right
condition if the gauge group is precisely $SO(2,1)\times SO(2,1)$
rather than a covering group.

It is possible to give an intuitive explanation of why this is the
right gauge group if the boundary CFT is supposed to be
holomorphically factorized.  The Virasoro algebra is, of course,
infinite-dimensional, but it has a finite-dimensional subalgebra,
generated by $L_{\pm 1}$ and $L_0$, that is a symmetry of the
vacuum. It is customary to refer to the corresponding symmetry
group of the vacuum as $SL(2,\R)$, but in fact, in a holomorphic
CFT, in which all energies are integers, the group that acts
faithfully is really $SO(2,1)$.  So a holomorphically factorized
CFT has symmetry group $SO(2,1)\times SO(2,1)$, and it is natural
that this is the right gauge group in a gauge theory description
of (aspects of) the dual gravitational theory.

Now let us consider some other possible gauge groups.  One
possibility is to take a double cover of each factor of
$SO(2,1)\times SO(2,1)$, taking the gauge group to be
$SL(2,\R)\times SL(2,\R)$.  The appropriate restriction on the
gauge theory couplings was determined in \yoldo\ (where we should
set $n=2$) and is that $k_L$ and $k_R$ take values in ${1\over
4}\Z$.  Hence the central charges $c_L$ and $c_R$ are multiples of
6.  In particular, $SL(2,\R)\times SL(2,\R)$ gauge theory would
allow us to consider values of the couplings that contradict
modular invariance of the boundary CFT.

There is perhaps a more intuitive argument suggesting that
$SL(2,\R)\times SL(2,\R)$ is not the right group to consider. If
the gauge group is $SL(2,\R)\times SL(2,\R)$, then upon taking the
two-dimensional representation of one of the $SL(2,\R)$'s, we get
a two-dimensional real vector bundle over $W$ which generalizes
what in classical geometry is the spin bundle.  Thus, this would
be a theory in which, in the classical limit, $W$ is a spin
manifold, endowed with a distinguished spin structure (or even two
of them). That is appropriate in a theory with fermions, but not,
presumably, in a theory of pure gravity.

 We similarly lose modular invariance and have difficult to interpret geometric
 structures if we consider other non-diagonal covers of $SO(2,1)\times SO(2,1)$.
 So let us discuss the diagonal covers of $SO(2,1)\times
SO(2,1)$ that were considered at the end of section \quantcoup. We
know that the universal diagonal cover is not right, since then
$\ell/G$ could vary continuously.  This leaves the possibility of
an $n$-fold diagonal cover for some $n$. The best-motivated
example is perhaps the two-fold cover $SO(2,2)$, which is the
symmetry of Anti de Sitter spacetime (as opposed to a cover of
that spacetime).  In this case, according to \tomigot, $k_L$ and
$k_R$ can take values in $\Z/4$, as long as $k'=k_L-k_R$ is
integral. For the boundary CFT, this means that $c_L$ and $c_R$
can be multiples of 6 (with their difference a multiple of 24).
For example, let us consider a hypothetical CFT of
$(c_L,c_R)=(6,6)$. The ground state energies are $(-1/4,-1/4)$. As
these values are not integers, the symmetry group of the ground
state is not $SO(2,1)\times SO(2,1)$, but a four-fold diagonal
cover, which is a double cover of the gauge group $SO(2,2)$ that
was assumed. Such a CFT cannot be holomorphically factorized,
since the left- and right-moving ground state energies are not
integers; it cannot even be holomorphically factorized up to a
phase.\foot{The left-moving partition function would have to be
$\Phi/\Delta^{1/4}$, where $\Delta=\eta(q)^{24}$ is the
discriminant, $\eta$ being the Dedekind eta function, and $\Phi$
is a modular form of weight 3. The power of $\Delta$ was
determined to get the right ground state energy; modular
invariance implies that the modular weight of $\Phi$ must equal
that of $\Delta^{1/4}$. As there is no modular form of weight 3,
such a theory does not exist. Even at $(c_L,c_R)=(12,12)$, it is
not possible to have holomorphic factorization up to a phase. To
get modular invariance and the right ground state energy, the
left-moving partition function would have to be
$E_6/\Delta^{1/2}$, where $E_6$ is the Eisenstein series of weight
6.  However, the coefficients in the $q$-expansion of this
function are not positive.}  The structure is considerably more
complicated than in the holomorphically factorized case. Similar
remarks apply to other diagonal covers.

Pragmatically, the most important argument against trying to
describe three-dimensional gravity via a cover of $SO(2,1)\times
SO(2,1)$ may be simply the fact that no good candidates are known.
For example, no especially interesting bosonic CFT (as opposed to
a superconformal field theory) seems to be known at $c=6,12,$ or
18, which are values that we would expect if we use the gauge
group $SO(2,2)$.  By contrast, at $c=24$, which is natural for
$SO(2,1)\times SO(2,1)$, the Frenkel-Lepowsky-Meurman monster
theory is a distinguished candidate, as we mentioned in section
\plan\ and will explain in more detail in section \newbie.

The simplest possible hypothesis is that the right gauge group to
consider in studying three-dimensional pure gravity is
$SO(2,1)\times SO(2,1)$, with integral $k_L$ and $k_R$ and
holomorphic factorization of the boundary CFT. It would be highly
unnatural to overlook the fact that $SO(2,1)\times SO(2,1)$ gauge
theory leads to precisely the values of the central charge at
which the drastic simplification of the boundary CFT known as
holomorphic factorization is conceivable. Moreover, the fact that
the classical action \ozormo\ of the gauge theory is a sum of
decoupled actions for $A_L$ and $A_R $ -- related respectively to
left- and right-moving modes of the boundary CFT -- is a hint of
holomorphic factorization of the boundary theory.

At any rate, regardless of whether they give the whole story, it
does seem well-motivated to look for holomorphically factorized
CFT's with central charges multiples of 24 that are dual to
three-dimensional pure gravity at special values of the
cosmological constant. That will be our main focus in the rest of
this paper.

\subsec{Interpretation}\subseclab\interpretation

A few further words of interpretation seem called for.

We do {\it not} claim that three-dimensional gravity is
equivalent, nonperturbatively, to Chern-Simons gauge theory.  Some
objections to this idea were described in section \relation. We
know that Chern-Simons gauge theory is useful for perturbation
theory, as was explained in that section, and we hope that it is
useful for understanding some nonperturbative questions. We used
the gauge theory approach to get some hints about the right values
of the cosmological constant (or equivalently of the central
charge) simply because it was the only tool available.  We
certainly do not claim to have a solid argument that the values of
$(c_L,c_R)$ suggested by $SO(2,1)\times SO(2,1)$ gauge theory are
the only relevant ones.

Implicit in the gauge theory approach to quantization of the
dimensionless parameter $\ell/G$ is an assumption that at least
some of the non-geometrical states that can be described in gauge
theory make sense.  Indeed, when understood in classical geometry
with  the metric assumed to be smooth and nondegenerate, the
Einstein-Hilbert action \harrigo\ is well-defined as a real-valued
function.  The topological problems that cause it to be
multivalued  when interpreted in gauge theory as a Chern-Simons
function, and lead to quantization of $\ell/G$, depend on allowing
certain  configurations that are natural in gauge theory, but
singular in geometry because the vierbein is not invertible.

Furthermore, holomorphic factorization most likely is possible
only if the path integral includes a sum over some non-geometrical
configurations.  In quantum gravity, we at least expect to sum
over all topologies of a three-manifold $W$, perhaps with some
fixed asymptotic behavior.  The choice of topology should be
expected to affect both the left- and right-movers of the boundary
CFT.  To achieve holomorphic factorization, there must presumably
be some sort of separate topological sum for left-movers and
right-movers. As this does not occur in classical geometry, it
must depend on some sort of non-geometric contributions to the
path integral -- though these contributions may be exponentially
small in many circumstances.

\bigskip\noindent{\it The Gravitational Chern-Simons Action
Reconsidered}

Our discussion of the quantization of the gravitational
Chern-Simons coupling \bogus\ \eqn\zogus{\Delta_0 I={k'\over
4\pi}\int_W \tr\,\left(\omega\wedge d\omega+{2\over
3}\omega\wedge\omega\wedge\omega\right)} has been based entirely
on gauge theory.  We now want to describe the similar but slightly
different answer that would come from classical differential
geometry.  As always, the differences reflect the fact that the
gauge theory analysis allows some non-geometrical configurations.

We first briefly restate the gauge theory analysis. $\omega$ is a
connection on an $SO(2,1)$ or (in Euclidean signature) $SO(3)$
bundle over a three-manifold $W$. As usual, to define $\Delta_0 I$
more precisely, we pick an oriented four-manifold $M$ of boundary
$W$ with an extension of $\omega$ over $W$.  Then we define
\eqn\logus{I_M={k'\over 4\pi}\int_M \tr\,F\wedge F.} If $M$ is
replaced by some other four-manifold $M'$, and $X=M-M'$ is a
four-manifold without boundary obtained by gluing together $M$ and
$M'$, then \eqn\nogus{I_M-I_{M'}={k'\over 4\pi}\int_X\tr\,F\wedge
F=2\pi k'\int_X p_1(F),} where $p_1(F)=(1/8\pi^2)\tr\,F\wedge F$
is the first Pontryagin form. In general, $\int_X p_1(F)$ can be
any integer, so the condition that the indeterminacy in $I_M$ is
an integer multiple of $2\pi$ means simply that $k'$ is an
integer.

This is the right answer if the right thing to do is to simply
think of $\omega$ as an $SO(3)$ or $SO(2,1)$ gauge field, ignoring
its classical relation to gravity.  However, in classical gravity,
one can get a better answer. In classical gravity, $\omega$ is a
connection on the tangent bundle $TW$ of $W$.  Now replace $TW$ by
$TW\oplus \epsilon$, where $\epsilon$ is a trivial real line
bundle.  Then $\omega$ can be regarded as a connection on
$TW\oplus \epsilon$ in an obvious way, and $TW\oplus\epsilon$
extends over $M$ as the tangent bundle of $M$.  With this choice,
\logus\ becomes \eqn\gogus{I_M={k'\over 4\pi}\int_M \tr\, R\wedge
R,} where $R$ is the curvature form of $M$, and \nogus\ becomes
\eqn\otus{I_M-I_{M'}=2\pi k' \int_X p_1(R).} The effect of this is
that instead of the first Pontryagin number of a general bundle
over $X$, as in \nogus, we have here the first Pontryagin number
$p_1(TX)$ of the tangent bundle of $X$. This number is divisible
by 3, because of the signature theorem, which says that for a
four-manifold $X$,  $p_1(TX)/3$ is an integer, the signature of
$X$. Hence, in the gravitational interpretation, the condition on
$k'$ is \eqn\gogolp{k'\in {1\over 3}\Z.}

There is also a variant of this.  If $W$ is a spin manifold and we
are willing to define the gravitational Chern-Simons coupling
\zogus\ in a way that depends on the spin structure of $W$ (this
does not seem natural in ordinary gravity, but it may be natural
in supergravity, which we come to in section \supercase), the
condition on $k'$ can be further relaxed.  In this case, we can
select $M$ so that the chosen spin structure on $W$  extends over
$M$. If $M'$ is another choice with the same property, then
$X=M-M'$ is a spin manifold. But (as follows from the
Atiyah-Singer index theorem for the Dirac equation) the signature
of a four-dimensional spin manifold is divisible by 16.  So in
this situation $p_1(TX)$ is a multiple of 48, and the result for
$k'$ under these assumptions is \eqn\isto{k'\in {1\over 48}\Z.}

\bigskip\noindent{\it The Physical Hilbert Space}

Now we shall discuss the meaning for gravity of the physical
Hilbert space of Chern-Simons gauge theory.

In three-dimensional Chern-Simons gauge theory, one can fix a
Riemann surface $C$ and construct a Hilbert space ${\cal H}_C$ of
physical states obtained by quantizing the given theory on $C$.
This Hilbert space  depends on the Chern-Simons couplings, so when
we want to be more precise, we will call it ${\cal H}_C(k_L,k_R)$.
In \Witten, it was proposed that the physical Hilbert space of
three-dimensional gravity on a Riemann surface $C$ should be
obtained in essentially this way, using $SO(2,1)\times SO(2,1)$
gauge theory (or something similar, depending on whether the
cosmological constant is positive, negative, or zero).  Some
serious objections to the claim that pure gravity and Chern-Simons
gauge theory are equivalent in three dimensions were noted in
section \relation. Yet remarkable progress has made made
\refs{\teschnerone,\teschner} in understanding the quantization of
$SO(2,1)$ gauge theory, and even more remarkably, in relating this
quantization to Liouville theory.  It  therefore seems likely that
the Hilbert space obtained by quantizing Chern-Simons gauge theory
of $SO(2,1)\times SO(2,1)$ means something for three-dimensional
gravity even if the proposal in \Witten\ was premature.  The rest
of this subsection will be devoted to an attempt (not used in the
rest of the paper) to reconcile the different points of view.  See
\ref\carliptwo{S. Carlip ``Quantum Gravity In 2+1 Dimensions: The
Case Of A Closed Universe,'' Living Rev. Rel. {\bf 8} (2005) 1,
gr-qc/0409039.} for some useful background.

We may start by asking what we mean by the  physical Hilbert space
${\cal H}_C$  obtained in quantum gravity by quantizing on a
closed manifold $C$. What type of question is this space supposed
to answer? Quantization on, for example, an asymptotically flat
spacetime leads to a Hilbert space that can be interpreted in a
relatively straightforward way, but the physical meaning of a
Hilbert space obtained by quantizing on a compact spatial manifold
(if such a Hilbert space can be defined at all) is not clear.

\ifig\welco{\bigskip The Hartle-Hawking wavefunction $\Psi$ is
computed by integrating over three-manifolds $W$ with a given
boundary $C$. The dotted line labeled $r$ denotes a path in $W$
connecting two points in $C$.  It is possible to vary $W$ so that
the length of this path goes to zero with no change in the
geometry of $C$. In this limit, $W$ becomes singular even though
$C$ is smooth.  This gives a ``source term'' in the Wheeler-de
Witt equation, because of which the Hartle-Hawking wavefunction
does not obey this equation. } {\epsfxsize=2in\epsfbox{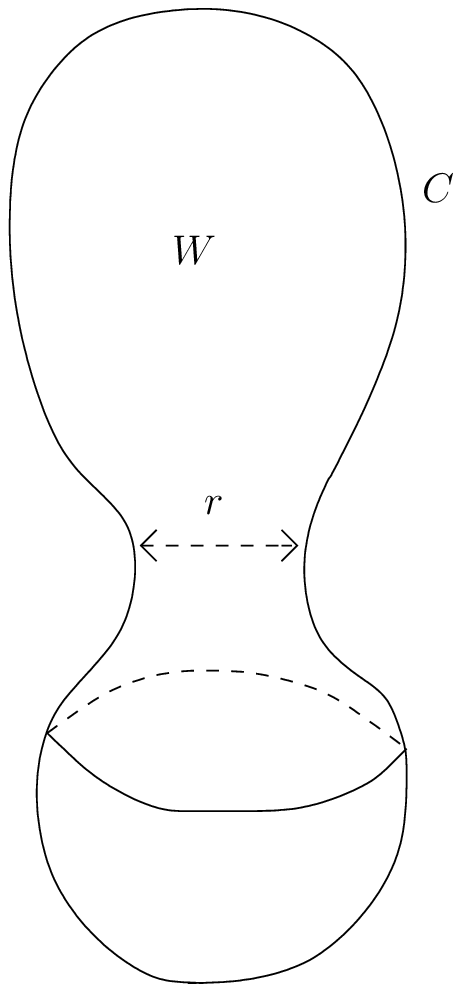}}

One line of thought that is relatively close to working is to
consider the Hartle-Hawking wavefunction \ref\hartle{J. Hartle and
S. B. Hawking, ``Wavefunction Of The Universe,'' Phys. Rev. {\bf
D28} (1983) 2960.} and claim that it is a vector in ${\cal H}_C$.
(The obvious idea that quantum mechanical probabilities would be
calculated in terms of inner products of vectors in a Hilbert
space ${\cal H}_C$ of physical states is afflicted with similar
but more serious problems.) The Hartle-Hawking wavefunction is a
functional of metrics on $C$. For every metric $h$ on $C$, we
define $\Psi(h)$ as the result of performing a path integral over
three-manifolds $W$ whose boundary is $C$ and whose metric $g$
coincides with $h$ on the boundary. Formally, one can try to argue
that $\Psi(h)$ obeys the Wheeler-de Witt equation and thus is a
vector in a Hilbert space ${\cal H}_C$ of solutions of this
equation. Moreover, one can formally match the Wheeler-de Witt
equations of gravity with the conditions for a physical state in
Chern-Simons gauge theory. Though many steps in these arguments
work nicely, one runs into trouble because a Riemann surface can
be immersed, rather than embedded, in a three-manifold, and hence
it is possible for $W$ to degenerate without $C$ degenerating
(\welco).  As a result, the Hartle-Hawking wavefunction does not
obey the Wheeler-de Witt equation and is not a vector in ${\cal
H}_C$.

 In the case of negative cosmological constant, the boundary CFT
gives a sort of cure for the problem with the Hartle-Hawking
wavefunction. Instead of thinking of $C$ as an ordinary boundary
of $W$, we think of it as a conformal boundary at infinity.  The
partition function $\hat \Psi(h)$ of the boundary CFT is defined
by performing the path integral over all choices of $W$ with $C$
as conformal boundary. This is well-behaved, because, with $C$ at
conformal infinity, it is definitely embedded rather than
immersed. Moreover, $\hat\Psi(h)$ is a sort of limiting value of
the Hartle-Hawking wavefunction. Indeed, let $\phi$ be a positive
function on $C$. Then $\hat\Psi(h)$ is essentially\foot{One needs
some renormalization in this limit; the necessary renormalization
reflects the conformal anomaly. Because of this anomaly, $\hat
\Psi(h)$ is not quite a function only of the conformal structure
of $C$ but a function of the metric that transforms with a certain
weight under conformal rescalings. As a result, the precise choice
of $\phi$ matters, but only in a rather simple way.} the limiting
value of $\Psi(e^\phi h)$ as $\phi\to \infty$.

This suggests that we should be able to think of $\hat \Psi(h)$ as a
vector in the Hilbert space ${\cal H}_C$ associated with
three-dimensional gravity and a two-manifold $C$.  This viewpoint
can be explained more fully. The phase space of $SO(2,1)\times
SO(2,1)$ Chern-Simons theory on $C$ is the space of $SO(2,1)\times
SO(2,1)$ flat connections on $C$. The space of $SO(2,1)$ flat
connections on $C$ has several topological components (labeled by
the first Chern class, which comes in because $SO(2,1)$ is
contractible to $SO(2)\cong U(1)$). One of these components, the
only one that can be simply interpreted in terms of classical
gravity with negative cosmological constant,\foot{In the approach to
quantization developed in \refs{\teschnerone,\teschner}, this is the
only component considered.}  is isomorphic to Teichmuller space
$\cal T$. Thus, this component of the classical phase space $\cal M$
is a product of two copies of $\cal T$, parametrized by a pair of
points $\tau, \tau'\in \cal T$.  One can quantize $\cal M$ (or at
least this component of it) naively by using the standard
holomorphic structure of $\cal T$.   If we do this, the wavefunction
of a physical state is a ``function'' of $\tau$ and $\tau'$ that is
holomorphic in $\tau$ and antiholomorphic in $\tau'$.
(Antiholomorphy in one variable reflects the relative minus sign in
the Chern-Simons action \ozormo; we assume that $k_L$ and $k_R$ are
positive.)  Actually, a physical state wavefunction
$\Psi(\tau,\bar\tau\,')$ is not quite a function of $\tau$ and
$\bar\tau\,'$ in the usual sense, but is a form, of weights
determined by $k_L$ and $k_R$. So it takes values in a Hilbert space
${\cal H}_C(k_L,k_R)$ that depends on the Chern-Simons couplings.
Such a wavefunction $\Psi(\tau,\bar\tau\,')$ is determined by its
restriction to the diagonal subspace $\tau=\tau'$. Moreover, if we
want to make a relation to gravity, it is natural to require that
$\Psi$ should be invariant under the diagonal action of the mapping
class group on $\tau$ and $\tau'$; this condition is compatible with
restricting to $\tau=\tau'$.

Similarly, the partition function of a CFT on the Riemann surface
$C$ is a not necessarily holomorphic ``function'' (actually a form
of appropriate weights) $\Psi(\tau,\bar\tau)$.  Being real
analytic, $\Psi$ can be analytically continued to a function
$\Psi(\tau,\bar\tau')$ with $\tau'$ at least slightly away from
$\tau$. It does not seem to be a standard fact\foot{However, G.
Segal has obtained results in this direction. } that $\Psi$
analytically continues to a holomorphic function on ${\cal
T}\times {\cal T}$ (with invariance only under one diagonal copy
of the mapping class group). However, this is true in genus 1,
since the partition function can be defined as $\Tr\, q^{L_0}\bar
q\,'^{\bar L_0}$, where we can take $q$ and $q'$ to be independent
complex variables of modulus less than 1. It seems very plausible
that  the statement is actually true for all values of the genus,
since one can move on Teichmuller space by ``cutting'' on a circle
and inserting $q^{L_0}\bar q\,'^{\bar L_0}$. If so, the partition
function of the CFT can always be interpreted as a vector\foot{It
may be necessary here to extend ${\cal H}_C(k_L,k_R)$ to a space
of forms on ${\cal T}\times {\cal T}$ that are invariant under the
action of the mapping class group but not necessarily
square-integrable.} in the Chern-Simons Hilbert space ${\cal
H}_C(k_L,k_R)$.

If we are given a theory of three-dimensional gravity, possibly
coupled to other fields, the partition function of the dual CFT is
a wavefunction $\Psi(\tau,\bar\tau{}\,')$ which, according to the
conjecture just stated, is a vector in ${\cal H}_C(k_L,k_R)$. Any
gravitational theory of the same central charges leads to another
vector in the same space.

{}From this point of view, it seems that we should not claim, as was
done in \Witten, that ${\cal H}_C(k_L,k_R)$ is a space of physical
states that are physically meaningful in pure three-dimensional
gravity. Rather, a particular bulk gravitational theory, such as
pure gravity, gives rise to a particular dual CFT whose partition
function gives a definite vector in ${\cal H}_C(k_L,k_R)$. Another
gravitational theory, perhaps with matter fields, whose dual CFT has
the same values of the central charges, will lead to a dual
partition function that is another vector in the same space. Thus,
${\cal H}_C(k_L,k_R)$ is in a sense a universal target for
gravitational theories -- with arbitrary matter fields -- of given
central charges.

We have formulated this for a particular Riemann surface $C$, but
in either the gravitational theory or the dual CFT, $C$ can vary
and there is a nice behavior when $C$ degenerates. So it is more
natural to think of this as a structure that is defined for all
Riemann surfaces.  In conformal theory, this perspective is
described in \ref\frsh{D. Friedan and S. Shenker, ``The Analytic
Geometry Of Two-Dimensional Conformal Field Theory,'' Nucl. Phys.
{\bf B281} (1987) 509.}.

\subsec{Analog For Supergravity}\subseclab\supercase

Here we will discuss the extension of some of these ideas to
three-dimensional supergravity.

We consider primarily minimal supergravity, corresponding to the
case that the boundary CFT has $\N=1$ supersymmetry for
left-movers or right-movers or perhaps both.  Thus, for, say, the
left-movers, the Virasoro algebra is replace by an $\N=1$
super-Virasoro algebra.  The symmetry algebra generated by $L_{\pm
1}$ and $L_0$ is extended (in the Neveu-Schwarz sector) to a
superalgebra that also includes the fermionic generators ${\cal G}_{\pm
1/2}$.  This is the Lie superalgebra of the supergroup $OSp(1|2)$,
whose bosonic part is $Sp(2,\R)$, or equivalently $SL(2,\R)$. In
particular, since the operators ${\cal G}_{\pm 1/2}$ transform in the
two-dimensional representation of $SL(2,\R)$, the relevant group
is definitely $SL(2,\R)$ (or possibly a covering of it), not its
quotient $SO(2,1)$.

\def\str{{\rm str}}
For definiteness, consider a two-dimensional CFT with $(0,1)$
supersymmetry, that is, with $\N=1$ supersymmetry for right-movers
and none for left-movers. Then left-movers have an ordinary Virasoro
symmetry and right-movers have an $\N=1$ super-Virasoro symmetry.
Such a theory can be dual to a three-dimensional supergravity
theory, which classically can be described by a Chern-Simons gauge
theory in which the gauge supergroup is $SO(2,1)\times OSp(1|2)$, or
possibly a cover thereof.  For brevity, we will here assume that the
gauge group is precisely $SO(2,1)\times OSp(1|2)$. The action is the
obvious analog of \ozormo:
 \eqn\ozormox{\eqalign{I=&k_LI_L+k_RI_R\cr =&{k_L\over 4\pi}\int \tr\,\left(A_L\wedge dA_L+{2\over
 3}A_L\wedge A_L\wedge A_L\right)-{k_R\over 4\pi}\int \str\,\left(A_R\wedge dA_R+{2\over
 3}A_R\wedge A_R\wedge A_R\right).\cr}}
Here $A_L$ is an $SO(2,1)$ gauge field, $A_R$ is an $OSp(1|2)$
gauge field, and $\str$ is the supertrace in the adjoint
representation of $OSp(1|2)$.

We want to generalize the analysis of section \quantcoup\ to
determine the allowed values of $k_L$ and $k_R$.  There is
actually almost nothing to do.  $A_L$ is simply an $SO(2,1)$ gauge
field, so $k_L$ must be an integer.  As for $A_R$, we can for
topological purposes replace the supergroup $OSp(1|2)$ by its
bosonic reduction $SL(2,\R)$, since the fermionic directions are
infinitesimal and carry no topology.  So we can borrow the result
of \yoldo: \eqn\pyoldo{\eqalign{k_L&\in \Z\cr k_R & \in {1\over
4}\Z.\cr}}

We still have $(c_L,c_R)=(24k_L,24k_R)$, since the Brown-Henneaux
computation of the central charge depends only on the bosonic part
of the action.  So $c_L$ must be a multiple of 24, as before, but
now it seems that $c_R$ should be a multiple of 6.

This is not the result that one might hope for, because
holomorphic factorization in $\N=1$ superconformal field theories
requires that $c_R$ should be a multiple of 12, not 6.  So half of
the seemingly allowed values of $c_R$ are difficult to interpret
in the spirit of this paper.  Replacing $SO(2,1)\times OSp(1|2)$
by a covering group would only make things worse, as in the
bosonic case.

There are many conceivable ways to interpret this result, including
the possibility that our assumptions have been too optimistic.
However, one additional possibility seems worthy of mention here.
Part of the structure of a superconformal field theory is that there
are Ramond-sector vertex operators.  They introduce a ``twist'' in
the supercurrent, which has a monodromy of $-1$ around a point at
which a Ramond vertex operator is inserted. Let us assume that the
superconformal dual of three-dimensional supergravity should be a
theory in which Ramond vertex operators make sense. What is the
gravitational dual of an insertion of such a vertex operator?

\ifig\welcox{\bigskip  $C$ is a Riemann surface with a pair of
insertions of Ramond-sector vertex operators at points labeled $p$ and $p'$. They are connected
by a line $L$ that runs in a three-manifold $W$ whose boundary is
$C$. Supergravity on $W$ contributes to conformal field theory on
$C$. The fermion fields of $OSp(1|2)$ receive a minus sign in
monodromy around $L$. } {\epsfxsize=3in\epsfbox{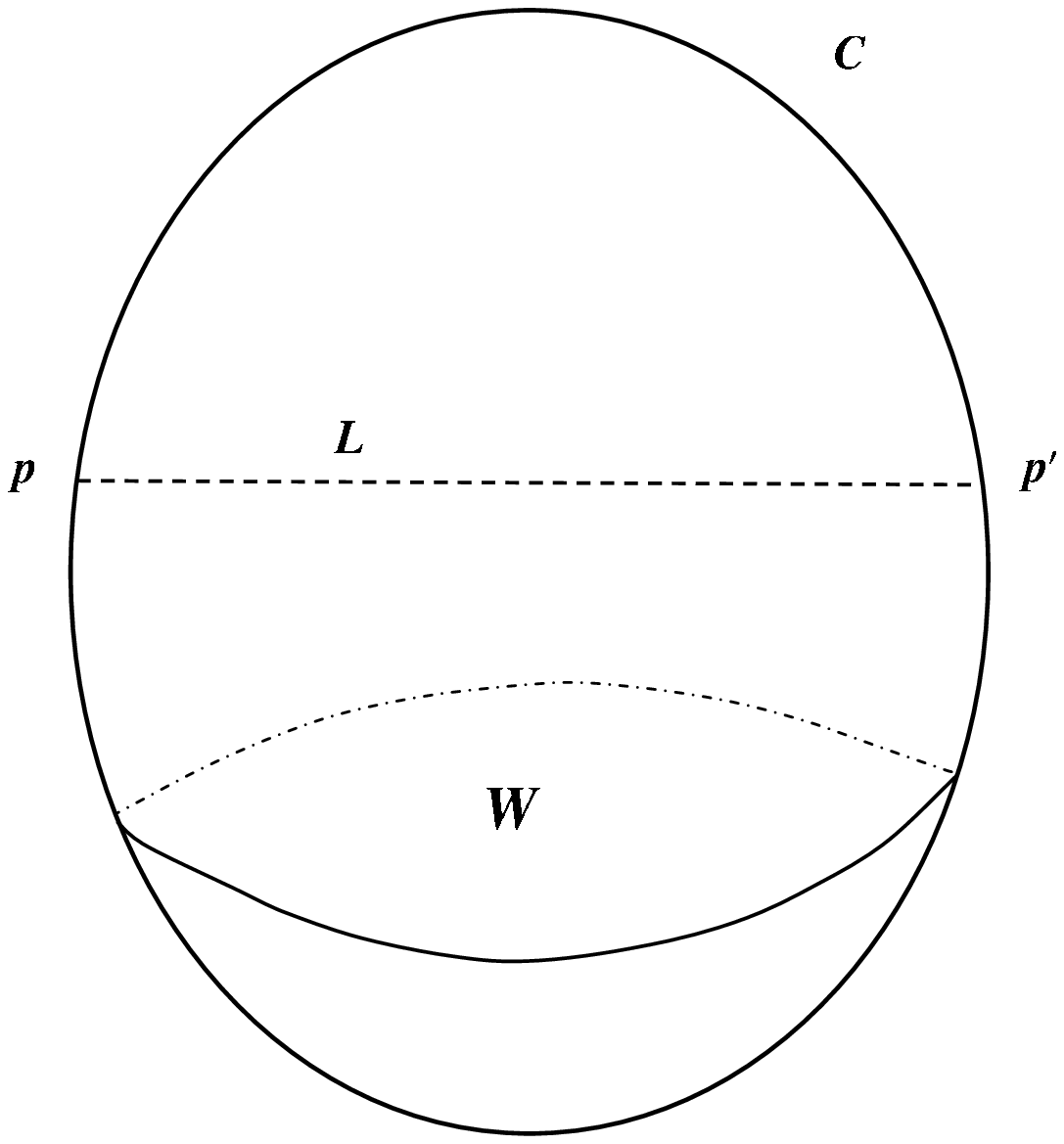}}

As in \welcox, points in the conformal boundary of spacetime at
which a Ramond vertex operator is inserted must be connected in
the bulk by lines -- such as the line labeled $L$ in the figure --
around which the fermionic fields of $OSp(1|2)$ (that is, the
gravitinos) have a monodromy $-1$.  It is plausible to interpret
these lines, which we will call Ramond lines, as world-lines of
Ramond-sector black holes. A spacetime history containing a black
hole trajectory is of course more complicated than a spacetime
with a simple line drawn in it. But for our present purposes,
which are purely topological, the difference may not be important.
There can also be Neveu-Schwarz sector black holes (and black
holes in purely bosonic gravity), but we are about to discuss an
effect which seems  special to Ramond-sector black holes.

So now we have a new problem.  We want to define
\eqn\nolman{I_R={1\over 4\pi}\int_W \str\,\left(A_R\wedge
dA_R+{2\over
 3}A_R\wedge A_R\wedge A_R\right)}
in the presence of a Ramond world-line $L$ on $W$.  In the absence
of the Ramond line, we know that $I_R$ in $OSp(1|2)$ gauge theory is
defined modulo $4\cdot 2\pi$, which is why as stated in \pyoldo,
$k_R$ can be a multiple of $1/4$.  What happens in the presence  of
the Ramond line?

Rather than solving a new topological problem, we can take the
following shortcut.  Let $W'$ be a new three-manifold obtained by
taking a double cover of $W$ branched over the line $L$.  We let
$I_R(W)$ be the action  \nolman\ and $I_R(W')$ be the corresponding
action for the gauge field $A_R$ pulled back to $W'$.  When pulled
back to $W'$, the singularity of $A_R $ along the Ramond line
disappears, so $I_R(W')$ is defined modulo $4\cdot 2\pi$.  There is
no better way to define $I_R(W)$ in the presence of a Ramond line
than to say that $I_R(W)=I_R(W')/2$.  So $I_R(W)$ is defined modulo
$2\cdot 2\pi$.

This means that $k_R$ should be a multiple of $1/2$, not $1/4$. So
in other words, if including Ramond lines is the right thing to
do, we get \eqn\pyoldox{\eqalign{k_L&\in \Z\cr k_R & \in {1\over
2}\Z,\cr}} and hence $c_L$ and $c_R$ are multiples of 24 and 12,
respectively.

Unfortunately, the above ``derivation'' is little more than a
scenario to try to justify the answer that we hoped for.  However, a
good pragmatic reason to focus on the case that  $c_R $ is a
multiple of 12 is that there are interesting candidate
superconformal field theories (SCFT's) in that case, as we discuss
in section \newbie. There are no obvious interesting candidates at
$c_R=6,18$, etc.

In the supersymmetric case it is convenient to express the
Chern-Simons coupling $k$ as $k=k^*/2$, where we will focus on the
case that $k^*$ is an integer.  In terms of $k^*$, the central
charge is $c=12k^*$.

\bigskip\noindent{\it Some Generalizations}

We conclude this section by briefly mentioning some simple
generalizations.

First of all, $(1,1)$ supersymmetry in two dimensions, with $\N=1$
super-Virasoro symmetry for both left-movers and right-movers, is
dual to three-dimensional supergravity theories related to
$OSp(1|2)\times OSp(1|2)$ Chern-Simons gauge theory.  If one wants
the left- or right-movers to have more than $\N=1$ supersymmetry,
one simply replaces $OSp(1|2)$ by an appropriate supergroup with
more fermionic generators.  For example, $OSp(2|2)$ is related to
$\N=2$ supersymmetry, $PSU(2|2)$ is related to what is usually
called $\N=4$ supersymmetry (with the ``small'' $\N=4$
superconformal algebra), and $OSp(r|2)$ with $r>2$ is related to
theories with ``large'' $\N=k$ superconformal algebras. (The most
widely studied case of these algebras is $r=4$; for example, see
\ref\guk{S. Gukov, E. Martinec, G. W. Moore, and A. Strominger,
``An Index for 2-D Field Theories with Large $\N = 4$
Superconformal Symmetry,'' hep-th/0404023.}.)

It is also possible to consider other  extended chiral algebras,
apart from superconformal algebras.  For example, one can start in
three dimensions with an $SL(3,\R)$ Chern-Simons gauge theory, which
plausibly may be related to some sort of three-dimensional $W_3$
gravity theory in much the same (not fully understood) way that
$SL(2,\R)$ Chern-Simons theory is related to ordinary
three-dimensional gravity.  A dual CFT would very likely then have a
$W_3$ chiral algebra.  An analogous statement plausibly holds for
many extended chiral algebras.

\newsec{Partition Functions}\seclab\newbie

In this section, we will determine what we propose to be the exact
spectrum of physical states of three-dimensional gravity or
supergravity with negative cosmological constant, in a spacetime
asymptotic at infinity to Anti de Sitter space.  Equivalently, we
will determine the genus one partition function of the dual CFT.

In all cases, we work at the values of $\ell/G$ at which
holomorphic factorization is possible. These values were related
to gauge theory in section \cval. We assume holomorphic
factorization, and mainly consider only the holomorphic sector of
the theory.  The full partition function is the product of the
function we determine and a similar antiholomorphic function.
These partition functions have been studied before
\refs{\hoehn,\hoehntwo}, with different motivation.

\subsec{The Bosonic Case}\subseclab\bosecase

We begin with the bosonic case.  What are the physical states of
pure gravity in a spacetime asymptotic at infinity to ${\rm AdS}_3$?

Since there are no gravitational waves in the theory, the only state
that is obvious at first sight is the vacuum, corresponding in the
classical limit to Anti de Sitter space.   In a conformal field
theory with central charge $c=24k$, the ground state energy is
$L_0=-c/24=-k$.  The contribution of the ground state
$|\Omega\rangle$ to the partition function $Z(q)=\Tr\,q^{L_0}$ is
therefore $q^{-k}$.

Of course, there is more to the theory than just the ground state.
According to Brown and Henneaux \brownhen, a proper treatment of the
behavior at infinity leads to the construction of a Virasoro algebra
that acts on the physical Hilbert space.  The Virasoro generators
$L_n,\,n\geq -1$ annihilate $|\Omega\rangle$, but by acting with
$L_{-2},L_{-3},\dots$, we can make new states of the general form
$\prod_{n=2}^\infty L_{-n}^{s_n}|\Omega\rangle$, with energy
$-k+\sum_n ns_n$.  (We assume that all but finitely many of the
$s_n$ vanish.)  If these are the only states to consider, then the
partition function would be
\eqn\notorgo{Z_0(q)=q^{-k}\prod_{n=2}^\infty{1\over 1-q^n}.}

This cannot be the complete answer, because the function $Z_0(q)$
is not modular-invariant.  There must be additional states such
that $Z_0(q)$ is completed to a modular-invariant function.

Additional states are expected, because the theory also has BTZ
black holes.  The main reason for writing the present paper, after
all, is to understand the role of the BTZ black holes in the quantum
theory.  We will assume that black holes account for the difference
between the naive partition function $Z_0(q)$ and the exact one
$Z(q)$.  To use this assumption to determine $Z(q)$, we need to know
something about the black holes.

The classical BTZ black hole is characterized by its mass $M$ and
angular momentum $J$.  In terms of the Virasoro generators,
\eqn\poko{\eqalign{M=&{1\over \ell}(L_0+\bar L_0)\cr
                   J=& (L_0-\bar L_0),\cr}}
                   so $L_0=(\ell M+J)/2$, $\bar L_0=(\ell M-J)/2$.
The classical BTZ black hole obeys $M\ell\geq |J|$, or $L_0,\,\bar
L_0\geq 0$.  The BTZ black hole is usually studied in the absence of
the gravitational Chern-Simons coupling, that is for $k_L=k_R=k$.
Its entropy is
$S=\pi(\ell/2G)^{1/2}\left(\sqrt{M\ell-J}+\sqrt{M\ell+J}\right)$.
(This entropy was first expressed in terms of two-dimensional
conformal field theory in \strom.) With $\ell/G=16k$ as in \orox,
this is equivalent to $S=4\pi \sqrt k\left(\sqrt{L_0}+\sqrt{\bar
L_0}\right).$  For the holomorphic sector, the entropy is therefore
\eqn\rogo{S_L=4\pi \sqrt{k_L L_0},} and similarly for the
antiholomorphic sector.

There is no classical BTZ black hole with $L_0<0$, and the entropy
of such a black hole is zero if $L_0=0$.  We will take this as a
suggestion that quantum states corresponding to black holes exist
only if $L_0>0$, that is $L_0\geq 1$.
 This means that the exact
partition function $Z(q)$ should differ from the function $Z_0(q)$
in \notorgo\ by terms of order $q$:
\eqn\zelb{Z(q)=q^{-k}\prod_{n=2}^\infty{1\over 1-q^n}+{\cal O}(q).}

Admittedly, we are here trying to squeeze more information from the classical result
than is justified.
But it turns out that a modular-invariant partition function of this
form exists and is unique.  This result (which is due to H\"ohn \hoehn)
follows from the fact that the
moduli space ${\cal M}_1$ of Riemann surfaces of genus 1 is itself a
Riemann surface of genus 0, in fact parametrized by the
$j$-function.  If $E_4$ and $E_6$ are the usual Eisenstein series of weights 4 and 6,
then $j=1728E_4^3/(E_4^3-E_6^2)$.  Its expansion in powers of $q$ is
 \eqn\gombo{j(q)=q^{-1}+744+196884
q+21493760q^2+864299970q^3+20245856256q^4+\dots.} Actually, it is
more convenient to use the function
\eqn\ombo{J(q)=j(q)-744=q^{-1}+196884 q+21493760q^2+ \dots,} which
likewise parametrizes the moduli space.

The $J$-function has a pole at $q=0$ and no other poles. The
statement that $J$ parametrizes the moduli space means precisely
that any modular-invariant function can be written as a function
of $J$. The partition function $Z(q)$ has a pole at $q=0$, that is
at $J=\infty$. This pole is of order $k$, and the partition
function has no other poles. Any holomorphic function of $J$ that
has no singularity except at $J=\infty$ is a polynomial in $J$. In
the present case, as the pole in $Z(q)$ at $q=0$ is of order $k$,
$Z$ must be a polynomial in $J$ of degree $k$. Thus
\eqn\dombo{Z(q)=\sum_{r=0}^kf_r J^r,} with some coefficients
$f_r$. These $k+1$ coefficients can be adjusted in a unique
fashion to ensure that $Z(q)$ takes the form \zelb, or in other
words to ensure that the terms in $Z(q)$ of order $q^{-n}$,
$n=0,\dots,k$, coincide with the naive function $Z_0(q)$.

When we do this, we get a function that we will call $Z_k(q)$,
$k=1,2,3,\dots$.  This function is our candidate for the generating
function that counts the quantum states of three-dimensional gravity
in a spacetime asymptotic to ${\rm AdS}_3$.  For example, for $k=1$
we have simply $Z_1(q)=J(q)$, and the next few examples are
\eqn\welfo{\eqalign{Z_2(q)=&J(q)^2-393
767\cr =&q^{-2}+1+42987520q+40491909396q^2+\dots\cr Z_3(q)=&J(q)^3-590651
J(q)-64481279\cr
=&q^{-3}+q^{-1}+1+2593096794q+12756091394048q^2+\dots\cr
Z_4(q)=&J(q)^4-787535J(q)^2-8597555039J(q)-644481279\cr=&
q^{-4}+q^{-2}+q^{-1}+2+81026609428q+1604671292452452276q^2+\dots.\cr}}

 Following \hoehn, we refer
to a holomorphic CFT with $c=24k$ and partition function $Z_k(q)$ as
an extremal CFT. According to our proposals, the dual of
three-dimensional gravity should be an extremal CFT.

As was already noted in the introduction, Frenkel, Lepowsky, and
Meurman constructed \refs{\flm,\flmtwo} an extremal CFT with $k=1$,
that is, a  holomorphic CFT with $c=24$ and partition function
$J(q)=Z_1(q)$. They also conjectured its uniqueness.  If that
conjecture as well as the ideas in the present paper are correct,
then the FLM theory must be the dual to quantum gravity for $k=1$.
Unfortunately, as also noted in the introduction, for $k>1$,
extremal CFT's are not known, though their possible existence has
been discussed in the literature \refs{\hoehn-\kep} for reasons not
related to three-dimensional gravity.  Our reasoning in this paper
suggests that such theories should exist and be unique for each $k$.

The main point of the FLM construction was that their theory has as
a group of symmetries the Fischer-Griess monster group $\Bbb{M}$,
the largest of the sporadic finite groups.
 Arguably, the FLM theory
is the most natural known structure with $\Bbb{M}$ symmetry.  The
coefficients in the $q$-expansion of the $J$-function are integers
for number-theoretic reasons, but the FLM construction gave a new
perspective on why they are positive; this is a property of the
partition function of any CFT.  It also gave a new perspective on
why these coefficients are so large; this follows from $\Bbb{M}$
symmetry, since $\Bbb{M}$ does not have small representations.

\nref\thompson{J. Thompson, ``Some Numerology Between the
Fischer-Griess Monster and the Elliptic Modular Function,''  Bull.
London Math. Soc.  11
(1979),  352-353.}%
\nref\connort{J. H. Conway and S. P. Norton, ``Monstrous
Moonshine,'' Bull. London
Math. Soc. {\bf 11} (1979) 308-339.}%
Indeed, the original clue to the FLM construction was the
observation by J. McKay that the first non-trivial coefficient
196884 of the function $J(q)$ is nearly equal to the smallest
dimension 196883 of a non-trivial representation of $\Bbb{M}$.
(This observation was later greatly generalized
\refs{\thompson,\connort}.) The FLM interpretation is that 196884
is the number of operators of dimension 2 in their theory. One of
these operators is the stress tensor, while the other 196883 are
primary fields transforming in the smallest non-trivial
representation of $\Bbb{M}$.

In our interpretation, the 196883 primaries are operators that
(when combined with suitable anti-holomorphic factors) create
black holes. It is illuminating to compare the number 196883 to
the Bekenstein-Hawking formula.  An exact quantum degeneracy of
196883 corresponds to an entropy of $\ln 196883\cong 12.19$.  By
contrast, the Bekenstein-Hawking entropy at $k=1$ and $L_0=1$ is
$4\pi\cong 12.57$.  We should not expect perfect agreement,
because the Bekenstein-Hawking formula is derived in a
semiclassical approximation which is valid for large $k$.

Agreement improves rapidly if one increases $k$.  For example, at
$k=4$, and again taking $L_0=1$, the exact quantum degeneracy of
primary states is 81026609426, according to eqn. \welfo.  (Two of
the states at this level are descendants.)  This corresponds to an
entropy $\ln 81026609426\cong 25.12$, compared to the
Bekenstein-Hawking entropy $8\pi\cong 25.13$.  Shortly we will
compute the entropy in the large $k$ limit.

Our interpretation is that a primary state
$|\Lambda\rangle$  represents a black hole, while a descendant
$\prod_{n=1}^\infty L_{-n}^{s_n}|\Lambda\rangle$ describes a black
hole embellished by boundary excitations.  These boundary
excitations are the closest we can come in $2+1$ dimensions to the
gravitational waves that a black hole can interact with in a larger
number of dimensions.  If this interpretation is correct, then for an exact count of black hole
states, we should count primaries only.  However, as the above examples illustrate, in practice
this issue has only a very slight effect on the black hole degeneracies.
The boundary excitations contribute to the
entropy an amount that is independent of $k$ and thus negligible in
the regime where we compare to the Bekenstein-Hawking entropy, and also, numerically,
negligible even for small $k$. The
separation between the black hole and the boundary excitations is
best-motivated for large $k$.

\bigskip\noindent{\it Alternative Formula}

Now we will present an alternative formula for the partition
functions $Z_k(q)$.  This formula avoids the large coefficients involved in writing the $Z_k$ as
a polynomial in $J$, and will make some properties of the $Z_k$  more manifest.

Let $q=\exp(2\pi i\tau)$ with $\tau$ in the upper half plane.
If  $p$ is a prime number, then the Hecke
operators acting on functions $F(\tau)$ are defined essentially by
\eqn\nub{ T'_pF(\tau)=F(p\tau)+\sum_{b=0}^{p-1}F((\tau+b)/p).}
(What we call $ T'_p$ is $p$ times the Hecke operator $T_p$ as
usually defined. See \ref\koblitz{N. Koblitz, {\it Introduction To Elliptic Curves And Modular
Forms} (Springer-Verlag, 1984).} for an introduction to Hecke operators and \refs{\mckay,\gannon}
for their use in the present subject.)
More generally, for any positive integer $t$, we
define \eqn\zub{
T'_tF(\tau)=\sum_{d|t}\sum_{b=0}^{d-1}F((t\tau+bd)/d^2).}
For $t=0$, we define $T'_0F(\tau)=1$.
If
$F(\tau)$ is modular-invariant, then so is $ T'_tF(\tau)$. The
definition of Hecke operators generalizes naturally to modular forms, but we will not
need this.

An immediate consequence of  the definition is that, for any
$F(\tau)$ that is invariant under $\tau\to\tau+1$, and so has a
Laurent expansion in powers of $q=\exp(2\pi i\tau)$, the
coefficients in the $q$-expansion of $ T'_nF(\tau)$ are linear
combinations with positive integer coefficients of the $q$-expansion
coefficients of $F(\tau)$.  The FLM construction shows that the
$q$-expansion coefficients of the $J$-function are non-negative
integers and the coefficients of positive powers of $q$ are
dimensions of non-trivial (reducible) monster representations.
Hence the same is true of $T'_n J$.

An important special case of how the Hecke operators act on
$q$-expansion coefficients is that if $F(\tau)=q^{-1}+{\cal
O}(q)$, then \eqn\jurot{ T'_nF(\tau)=q^{-n}+{\cal O}(q).}

Now we can give an alternative description of the partition
functions  $Z_k(\tau)$.  We recall that these functions are
uniquely determined by being modular-invariant and agreeing up to
order $q$ with the function
$Z_0=q^{-k}\prod_{n=2}^\infty(1-q^n)^{-1}$.  To find a function
with these properties, we simply expand
\eqn\wyto{Z_0=\sum_{r=-k}^\infty a_r q^r,} and then we let
\eqn\yto{Z_k(\tau)= \sum_{r=0}^ka_{-r}T'_{r}J(\tau).} For
example, \eqn\gyto{Z_2(\tau)=(T'_2+ T'_0)J(\tau)=J(2\tau)+J(\tau/2)+J((\tau+1)/2)+1.}
This representation makes it clear that, just like the $T'_r J$, the functions $Z_k$
share some properties with the $J$-function: the $q$-expansion coefficients are non-negative,
and the coefficients of positive powers of $q$ are dimensions of non-trivial monster representations.

This makes it tempting to speculate that the CFT's dual to three-dimensional gravity have
$\Bbb{M}$ symmetry for all $k$.  If so, the $\Bbb{M}$ symmetry is invisible in classical
General Relativity and acts on the microstates of black holes.  (However, the analogous conjecture
for supergravity, which would involve the Conway group, appears to be untrue, as we will see
in section \extremal.)

As an illustration of the usefulness of the expression for the
partition function in terms of Hecke operators, we will use it to
compare with the Bekenstein-Hawking entropy.  We consider the
semiclassical limit $k,L_0\to\infty$, with $r=L_0/k$ positive and
fixed. We take $r=p/q$ to be a rational number and assume that $k$
is divisible by $q$, so that $L_0$ is an integer $n$.
 If we write
\eqn\bellow{J(q)=\sum_{m=-1}^\infty c_mq^m,} then a formula of
Petersson and Rademacher gives the asymptotic behavior
\eqn\migot{\ln c_m\sim 4\pi\sqrt m-{3\over 4}\ln m-{1\over 2}\ln
2+\dots.} Let\eqn\polito{Z_k(\tau)=\sum_{n=-k}^\infty b_{k,n}q^n.}
We want to determine the behavior of $b_{k,n}$ for large $k$ and $n$
with $r=n/k$ fixed.
To evaluate the partition function using the formula \yto, let us first look at the contribution
from $r=k$, that is $ T'_kJ(\tau)$.  In evaluating $ T'_k
J$ from the definition \zub, the dominant term (for large $k$ and
$n$) is the term with $d=k$.  This contribution to $b_{k,n}$ is  $b_{k,n}^0\sim k c_{kn}$, and
hence, using the Petersson-Rademacher formula, \eqn\olito{\ln
b_{k,n}^0\sim 4\pi\sqrt{kn}+{1\over 4}\ln k -{3\over 4}\ln n-{1\over
2}\ln 2+\dots.} The first term is the Bekenstein-Hawking entropy
\rogo.  The additional terms in \olito, as well as the
remaining contributions in \yto\ with $r<k$ that we have omitted in deriving \olito,
do not modify
the Bekenstein-Hawking formula in the limit of large $k$, fixed $n/k$,
but give interesting subleading corrections that will be studied elsewhere \ref\malwi{A. Maloney
and E. Witten, to appear.}.

Another method to get similar results (expressing black hole degeneracies in terms of
coefficients of the singular part of the partition function) is
 the Farey tail expansion \farey.  See also
\ref\birmsen{D. Birmingham and S. Sen, ``Exact Black Hole Entropy Bound In Conformal Field Theory,''
hep-th/0008051.}.

\bigskip\noindent{\it The FLM Construction}

Consider a holomorphic CFT with $c=24$.  If the number of primary
fields of dimension 1 is $s$, then the expansion of the partition
function near $q=0$ begins $Z(q)=q^{-1}+s+{\cal O}(q)$, and
modular invariance implies that the genus 1 partition function is
precisely $Z(q)=J(q)+s=q^{-1}+s+196884 q+\dots$.  Of the 196884
fields of dimension 2, one is a descendant of the identity and $s$
are descendants of dimension 1 primaries. Hence there are
$196883-s$ primary fields of dimension 2, too few to furnish a
non-trivial representation of the monster group unless $s=0$.

For this reason, Frenkel, Lepowsky, and Meurman \flm\ aimed to
construct a holomorphic CFT of $c=24$ with no primary fields of
dimension 1, hoping that it would have monster symmetry.  Their
construction was made as follows.  The Leech lattice is an even,
unimodular lattice of rank 24 with no vector of length squared
less than 4. One can construct a holomorphic theory of $c=24$ by
compactifying  $24$ chiral bosons $X_i$, $i=1,\dots,24$ using any
even unimodular lattice.  If one uses the Leech lattice, then
because it contains no vector of length squared less then 4, a
primary field of the form $\exp(ip\cdot X)$ has dimension at least
2, as desired. However, this lattice theory has 24 primary fields
$\partial X_i$ of dimension 1. Aiming to eliminate those fields,
FLM considered an orbifold\foot{Their work preceded the general
study of orbifolds in string theory \ref\dixop{L. Dixon, J. A.
Harvey, C. Vafa, and E. Witten, ``Strings On Orbifolds,'' Nucl.
Phys. {\bf B261} (1985) 678-686.}, though some related stringy
constructions, such as the Ramond and Neveu-Schwarz sectors of
string theory, were already known.} by the $\Z_2$ symmetry $X_i\to
-X_i$ which eliminates the dimension 1 primaries. As it turns out,
this particular orbifold preserves modular invariance, does not
add dimension 1 primaries in\ twisted sectors, and gives the
hoped-for theory with monster symmetry.

FLM conjectured that the model they constructed is the unique
holomorphic CFT with partition function $J$. If this conjecture is
correct, it implies that many other constructions give the same
theory. There are \ref\tuitetwo{M. P. Tuite, ``On The Relation
Between Generalized Moonshine And The Uniqueness Of The Moonshine
Module,'' Commun. Math. Phys. {\bf 166} (1995) 495-532,
hep-th/9305057. } many orbifolds of the Leech lattice theory that
give holomorphic CFT's with no primary of dimension 1. According to
the FLM uniqueness conjecture, all of these theories are isomorphic
to their monster theory. Moreover, the Leech lattice theory itself
can be obtained as an orbifold theory starting from another even
unimodular lattice of rank 24, so other lattices can be used as
starting points as well.

Aiming to imitate the FLM construction for $k=2$, one might start
with an even unimodular rank 48 lattice with no vector of length squared less than 6.
Three such lattices are known; two are described in \ref\consloan{J.
Conway and N. A. Sloane, {\it Sphere Packings, Lattices, and Groups}
(Grundlehren der Mathematischen Wissenschaften, 1998).} and a third
has been constructed more recently \ref\nebe{G. Nebe, ``Some
Cyclo-Quaternionic Lattices,'' J. Algebra {\bf 199} (1998)
472-298.}.  (There is no reason to believe that these are the only
three rank 48 lattices with this property; there may be a vast
number of them.)   In the theory of 48 free chiral bosons $X_i$
compactified using such a lattice, a primary field $\exp(ip\cdot X)$
has dimension at least 3, but there are dimension 1 primaries
$\partial X_i$ and dimension 2 primaries $\partial X_i\partial
X_j-{1\over 48}\delta_{ij}
\partial X\cdot\partial X$. A simple $\Z_2$ orbifold eliminates
the dimension 1 primaries, as in the FLM case, but not the dimension
2 primaries.  One may attempt to find a more complicated orbifolding
construction to remove the unwanted primaries.  Though all three
lattices have interesting discrete symmetry groups, it appears that
there is no anomaly-free subgroup that removes all dimension 2
primaries.\foot{I am grateful for assistance from G. Nebe in
investigating this question.}  Many other orbifolds can be
considered, such as the symmetric product of $k$ copies of the $k=1$
monster theory, but it appears difficult to remove all primaries of
low dimension.

Optimistically speaking, this situation might be compared to current
algebra of a simply-laced compact Lie group $G$.  At level 1, one
has the Frenkel-Kac-Segal construction of current algebra of $G$ via
free bosons.  At higher integer level, the theory still exists, but
generically has no equally straightforward realization in terms of
free fields. Perhaps the situation is somewhat similar for extremal
CFT's.

\bigskip\noindent{\it Further Remarks}

We have emphasized here the partition function, because it is what
we can determine for general $k$.  However, if one can actually
describe the relevant CFT -- as conjecturally we can at $k=1$ via
the FLM construction -- this gives much more than a partition function.
In this case, by computing matrix elements of primary fields, we get
a detailed description of the black hole quantum mechanics.

Above $2+1$ dimensions, a black hole can form out of radiation and
ordinary matter.  In pure gravity in $2+1$ dimensions, the only
thing that a black hole can form from is, roughly speaking, smaller
black holes.  To be more precise, let $\Phi_i,\,i=1,\dots,w$ be
primary fields of dimension not much greater than $k$ that
individually could, in acting on the vacuum, create black holes of
fairly small mass. A product of such fields acting on the vacuum at
prescribed points \eqn\welp{\prod_{i=1}^w\Phi_i(z_i)|\Psi\rangle}
may create a black hole of large mass, that is, a primary state of
large energy or a descendant of such a state. Evaluating such matrix
elements is the closest analog we can find, in the present model, to
describing the formation of a large mass black hole from matter.
Unfortunately, at the moment, we can perform such computations only
for $k=1$, since for other values of $k$ we do not know the CFT.
Ideally, one would like to describe the CFT for all $k$ and
investigate the semiclassical limit of large $k$.

If our hypotheses are correct, the following may be the most
significant difference between three-dimensional pure gravity and a
more realistic theory of black holes in $3+1$ dimensions. In the
present model, as the holomorphic fields are all conserved currents, and the
energy levels are integers, the dynamics is
integrable and periodic on a short time scale.  This is certainly
not expected for black holes in general. An embedding of
three-dimensional gravity in a larger system, such as a string
theory, would be expected to give a non-integrable deformation of
the model.

\bigskip\noindent{\it A Conundrum}

We have here considered the holomorphic sector of a CFT, but the CFT
dual of three-dimensional gravity has both holomorphic and
antiholomorphic degrees of freedom.  Let us therefore discuss a
puzzle (stressed by J. Maldacena) that arises when one combines the
holomorphic and antiholomorphic degrees of freedom.  We let
$|\Omega\rangle$, $|\tilde\Omega\rangle$ be the ground states in the
holomorphic and antiholomorphic sectors, and let $|\Phi\rangle$ and
$|\tilde\Phi\rangle$ denote primary states other than the ground
state.  The state $|\Omega\rangle\otimes |\tilde\Omega\rangle$ and
its descendants correspond, according to our picture, to Anti de
Sitter space and its boundary excitations.  A state
$|\Phi\rangle\otimes |\tilde\Phi\rangle$, or a descendant thereof,
corresponds to a BTZ black hole, perhaps with boundary excitations.
But what do we make of states $|\Omega\rangle\otimes
|\tilde\Phi\rangle$ or $|\Phi\rangle\otimes |\tilde\Omega\rangle$,
and their descendants? Such states are trying to be Anti de Sitter
space for holomorphic variables and black holes for antiholomorphic
variables, or vice-versa.

In classical three-dimensional gravity, there is no satisfactory solution
that has this interpretation.  We can, however, see what is involved
in trying to make one.  First of all, three-dimensional Anti de
Sitter space ${\rm AdS}_3$ is simply the universal cover of the
$SL(2,\R)$ group manifold, with the symmetry $SL(2,\R)\times
SL(2,\R)$ (or rather a cover thereof) coming from the left and right
action of $SL(2,\R)$ on itself. The conformal boundary of the
universal cover of ${\rm AdS}_3$ is a cylinder $\Bbb{R}\times S^1$.
We can parametrize it by $-\infty<\tau<\infty$, $0\leq \sigma\leq
2\pi$, with conformal structure described by the metric
$ds^2=d\tau^2-d\sigma^2$. The BTZ black hole is the quotient of (a cover of)
$SL(2,\R)$ by a subgroup $\Z\subset SL(2,\R)\times SL(2,\R)$ that is
not simply a subgroup of one factor
or the other.  This quotient has conformal boundary that is again a
cylinder, conformally equivalent to the boundary of Anti de Sitter
space itself. That is why the BTZ black hole can be regarded as an
excitation of Anti de Sitter space.

To describe a state $|\Omega\rangle\otimes |\Phi\rangle$ (or its
parity conjugate) at the classical level, we want a solution of
Einstein's equations that is invariant under precisely one of the
two factors of the $SL(2,\R)\times SL(2,\R)$ symmetry. This will reflect the fact that
$|\Omega\rangle$
is $SL(2,\R)$-invariant and $|\Phi\rangle$ is not. A space with
just one of the $SL(2,\R)$ symmetries is ${\rm AdS}_3/ \Z$, where $\Z$ is a subgroup of one
$SL(2,\R)$ factor.  The quotient ${\rm AdS}_3/\Z$ is a perfectly
good solution of Einstein's equations in bulk, but its asymptotic
behavior at infinity does not quite agree with that of ${\rm
AdS}_3$.  The boundary at infinity is a cylinder, but the compact
direction in the cylinder is lightlike rather than spacelike.  For
our approach to three-dimensional gravity in the present paper to be
correct, something like this quotient ${\rm AdS}_3/\Z$ has to make
sense at the quantum level, perhaps because of a small quantum
correction that makes the compact direction in the cylinder
effectively spacelike.  Note that the metric
$ds_\epsilon^2=d\tau^2-\epsilon d\sigma^2$ induces on the cylinder a
conformal structure that is independent of $\epsilon$ up to
isomorphism as long as $\epsilon>0$.

What we have described is an analog for black holes of a statement
made in section \interpretation: to ensure holomorphic
factorization, the sum over topologies probably must be extended to
include configurations that are difficult to interpret classically.
In the present case, the analog of the sum over topologies is the
sum over states with or without the presence of a black hole.

\subsec{The Supersymmetric Case}\subseclab\supercase

Now we consider the analog for supergravity.  The analysis is a
little more complicated and in some ways the results are less
satisfactory.

A rather similar problem has been treated by H\"ohn \hoehn.  He considered not a superconformal
field theory, but a more general holomorphic theory with bosonic operators of integral
dimension and fermionic operators of half-integral dimension.
In that case, the partition function can have a pole at the Ramond cusp, and is given by a Laurent
series in $j_\theta= K-24$ rather than a polynomial.

We continue to assume holomorphic factorization, and we assume that
the holomorphic part of the boundary CFT is an ${\cal N}=1$
superconformal field theory (SCFT).  This means (in whatever not
well understood sense gravity is related to gauge theory) that the
gauge group of the bulk theory has a factor $OSp(1|2)$.  From a
classical point of view, the $OSp(1|2)$ bundle over a three-manifold
$W$ endows $W$ with a spin structure.  In the AdS/CFT
correspondence, $W$ has for conformal boundary a Riemann surface
$C$, and the spin structure on $W$ determines one on $C$. One is
instructed in the AdS/CFT correspondence to specify $C$, including
its spin structure, and sum over all choices of $W$. (Depending on
the theory, the antiholomorphic degrees of freedom may themselves be
supersymmetric and endowed with a choice of spin structure.)  We
will assume that it is physically appropriate to
specify the spin structure on $C$ rather than summing over it (just
as one specifies the complex structure of $C$), though at the end of section \extremal,
we discuss what happens if one sums over spin structures.

\def\H{{\cal H}}
\def\HNS{{\cal H}_{\rm NS}}
\def\HR{{\cal H}_{\rm R}}
\ifig\spinst{\bigskip We represent a genus 1 Riemann surface $C$
as the quotient of the complex plane by the lattice generated by
complex numbers  $1$ and $\tau$, where $\tau$ lies in the upper
half plane $\eusm H$. There are four possible spin structures,
classified by whether the fermions are periodic or antiperiodic
around the ``horizontal'' and ``vertical'' cycles. This is
indicated by labeling the horizontal and vertical directions with
a $+$ sign for periodicity or a $-$ sign for antiperiodicity. }
{\epsfxsize=4in\epsfbox{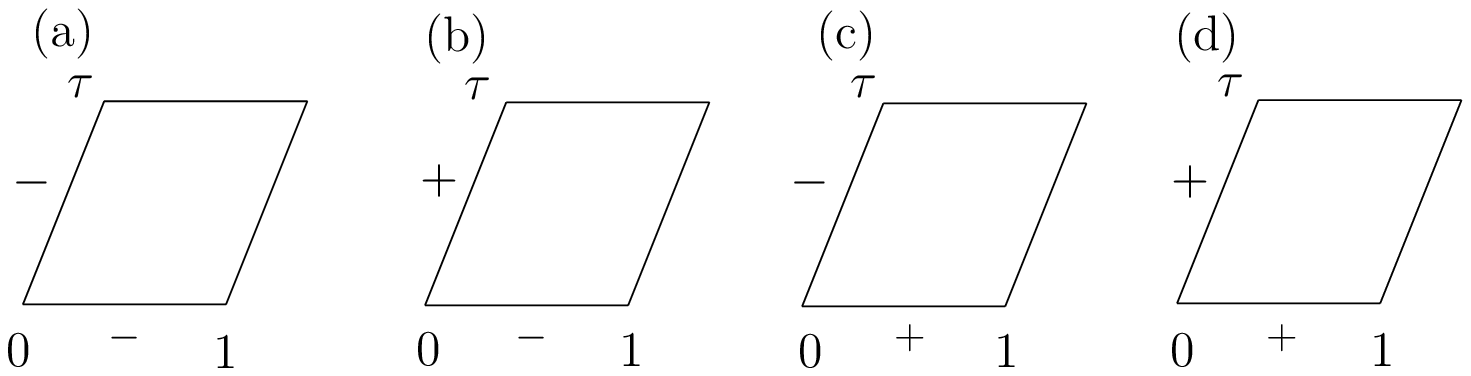}}

We focus on the case that $C$
has genus 1, in which case there are four possible spin structures
(\spinst).  The path integrals for these four spin structures can
be interpreted in terms of traces in two Hilbert spaces, known as
the Neveu-Schwarz (NS) and Ramond (R) Hilbert spaces $\H_{\rm NS}$
and $\H_{\rm R}$, respectively.

Traces in $\HNS $ are constructed using the spin structures, shown
in (a) and (b) in the figure, in which fermions are antiperiodic in
the horizontal direction. If additionally the fermions are
antiperiodic in the vertical direction, as in (a), then the path
integral computes \eqn\itcom{F(\tau) =\Tr_{\HNS}\,q^{L_0},} where
$q=\exp(2\pi i\tau)$.  We call this the NS partition function. If
they are periodic in the vertical direction, as in (b), the path
integral computes instead
\eqn\witcom{G(\tau)=\Tr_{\HNS}(-1)^Fq^{L_0},} where $(-1)^F$ is the
operator that is $+1$ on bosonic states, including the ground state
of the NS sector, and $-1$ on fermionic states.

Traces in $\HR$ are computed using spin structures, shown in (c) and
(d) in the figure, in which fermions are periodic in the horizontal
direction.  If the fermions are antiperiodic in the vertical
direction, the path integral computes an ordinary trace
\eqn\ubitcom{H(\tau)=\Tr_{\HR}\,q^{L_0}.} If fermions are periodic
in both directions, the path integral gives
\eqn\ycombo{\chi=\Tr_{\HR}\,(-1)^Fq^{L_0}=\Tr\,(-1)^F.}

Both $H(\tau)$ and $\chi$ are severely constrained by the fact that
in the R sector, one of the superconformal generators, which we will
${\cal G}_0$, commutes with $L_0$ and obeys ${\cal G}_0^2=L_0$. This
implies first of all that $L_0\geq 0$ in the R sector, so $H(\tau)$
has no pole at $q=0$. Further, the action of ${\cal G}_0$ implies
that states of $L_0>0$ are paired between bosons and fermions.  They
cancel out of $\chi$ and make equal contributions to $H(\tau)$.  It
follows that $\chi$ is simply an integer, the trace of the operator
$(-1)^{F}$ in the subspace with $L_0=0$ (or alternatively the index
of the operator ${\cal G}_0$ mapping from bosonic states to
fermionic ones). Moreover, if we expand $H(\tau)$ in powers of $q$
\eqn\yerox{H(\tau)=\sum_{n=0}^\infty h_nq^n,} then all coefficients
$h_n$ are even except possibly $h_0$.  In this expansion, only
integral powers of $q$ appear, since all fields obey integral
boundary conditions in the R sector.

\def\k{k^*}
In the NS sector, the ground state energy is $-c/24$. In the
holomorphically factorized case that we focus on, $c=12k^*$ for an
integer $k^*$, so the ground state energy is $-k^*/2$. Since
fermions in the NS sector are antiperiodic in the spatial direction,
NS excitations may have either integer or half-integer energy above
the ground state. The partition function therefore takes the general
form
\eqn\feudo{F(\tau)=q^{-\k/2}(1+aq^{1/2}+bq+cq^{3/2}+\dots)=\sum_{m\in
\Z/2,\,m\geq -k^*/2}
                               f_mq^m.}
In this expansion, the states in which $m+k/2$ is an integer are
bosons, while the states in which $m+k/2$ is half-integral are
fermions.  That is so because, in the NS sector, bosonic fields are
periodic and have integral excitation energies, while fermionic
fields are half-integral and have half-integral excitation energies.
$G(\tau)$ can therefore very explicitly be expressed in terms of the
same coefficients: \eqn\eudo{G(\tau)=\sum_{m\in \Z/2,\,m\geq
-k^*/2}(-1)^{2m+\k}
                               f_mq^m.}  We have simply included a
                               factor that is $+1$ on bosons and
                               $-1$ on fermions.
A more succinct way to say the same thing is that
\eqn\budo{G(\tau)=(-1)^{\k} F(\tau+1).}

Thus, $G$ is simply determined in terms of $F$.  The same is true
for $H$.  Consider the modular transformation $\tau\to -1/\tau$,
which has the effect of exchanging the horizontal and vertical
directions in \spinst\ (with a reversal of orientation for one).
This exchanges spin structures (b) and (c), as a result of which
\eqn\bludo{ H(\tau)=G(-1/\tau).}
We can combine \budo\ and \bludo\ to
get \eqn\ludo{H(\tau)=(-1)^{\k}F(-1/\tau+1).}  An important
special case of this is \eqn\udo{\lim_{\tau\to
i\infty}H(\tau)=(-1)^{\k}F(1).} The limit on the left hand side
exists, because $H$ has no pole at $q=0$. The limit is just the
coefficient $h_0$ in \yerox: \eqn\gudo{h_0=(-1)^{\k}F(1).} $h_0$
is the number of Ramond states of zero energy.

According to the above formulas,  everything (except an integer
$\chi$, the supersymmetric index) can  be expressed in terms of $F$.
So it is useful to understand the modular properties of $F$. $F$ is
not invariant under $\tau\to \tau+1$, as we have already seen. But,
since all energies take values in $\Z/2$, $F$ is invariant under
$\tau\to \tau+2$. In addition, since the modular transformation
$\tau\to-1/\tau$ maps spin structure (a) itself, $F$ is invariant
under $\tau\to -1/\tau$.  The two transformations $\tau\to\tau+2$
and $\tau\to -1/\tau$ generate a subgroup of $SL(2,\Z)$ that
consists of $2\times 2$ integral unimodular matrices
\eqn\busno{\left(\matrix{a& b \cr c& d\cr}\right)} that are
congruent mod 2 to one of \eqn\usno{\left(\matrix{1 & 0 \cr 0 &
1\cr}\right),~~ \left(\matrix{0 & -1 \cr 1 & 0\cr}\right).} These
matrices of course act on $\tau$ in the usual fashion, $\tau\to
(a\tau+b)/(c\tau+d)$.  This group is sometimes called $\Gamma_\theta$;
it is conjugate to the group $\Gamma_0(2)$ characterized by
requiring $b$ to be even.  (A third conjugate group is characterized
by the condition that $c$ should be even.  Each of these three
groups can be defined as the subgroup of $SL(2,\Z)$ that leaves
fixed one of the even spin structures, that is, one of the first
three shown in \spinst.)

\ifig\pinst{\bigskip (a) A fundamental domain for the action of
$SL(2,\Z)$ on the upper half plane.  By the action of $\tau\to
\tau+1$, one can take $|{\rm Re}\,\tau|\leq 1/2$, and by the action
of $\tau\to -1/\tau$, one can take $|\tau|\geq 1$.  The $SL(2,\Z)$
action identifies the left and right hand halves of the boundary of
the fundamental domain, making what topologically is a Riemann
surface of genus zero with one missing point at $\tau=i\infty$. This
point is called the cusp. (b) The analogous fundamental domain for
the action of $\Gamma_\theta$.  Here the symmetry
$\tau\to\tau+2$ lets us reduce to $|{\rm Re}\,\tau|\leq 1$, and
$\tau\to -1/\tau$ still lets us reduce to $|\tau|\geq 1$.  The group
action still identifies the left and right hand halves of the boundary of the
fundamental domain, and the result is a Riemann surface of genus
zero, now with two points omitted.  The missing points are the
Neveu-Schwarz cusp at $\tau=i\infty$ and the Ramond cusp at
$\tau=\pm 1$ (those two points  are equivalent under
$\tau\to\tau+2$). The Ramond cusp is missing in the quotient of the
upper half plane by $\Gamma_\theta$ because the points $\tau=\pm
1$ are not in the upper half plane, but on its boundary.}
{\epsfxsize=4in\epsfbox{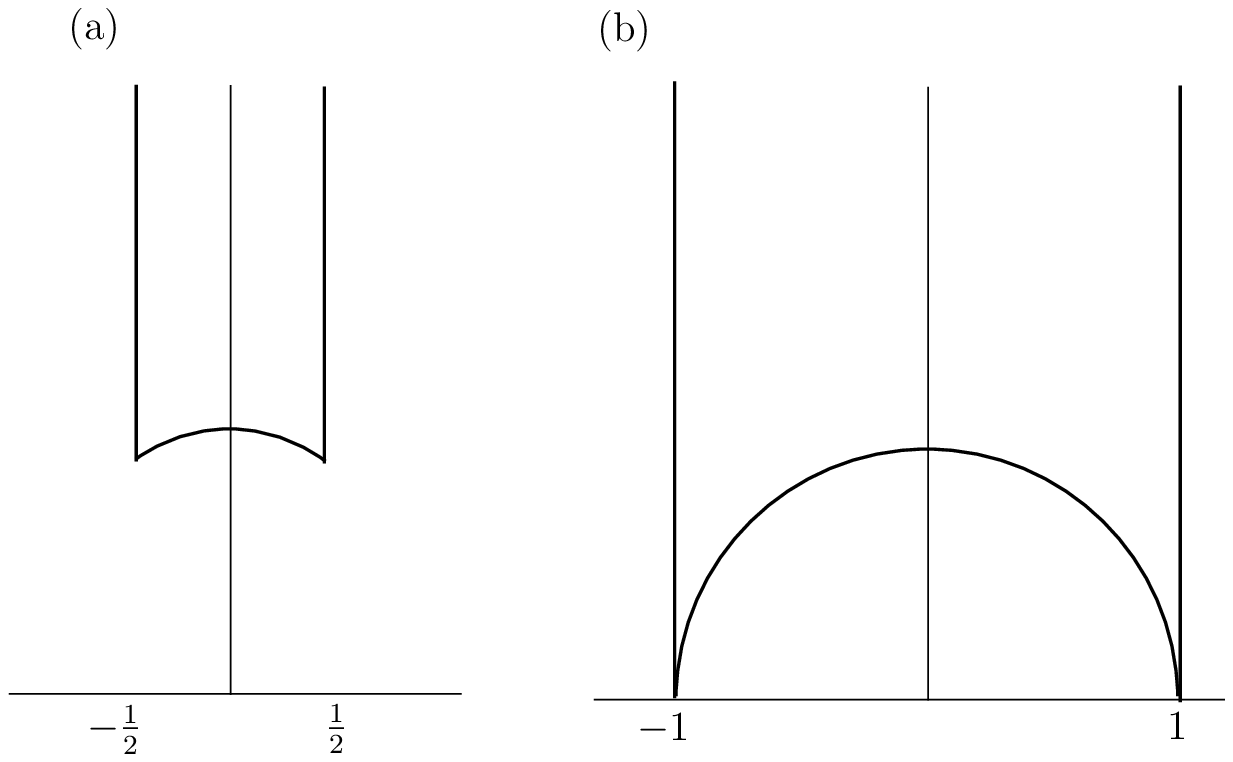}}

The quotient of the upper half plane $\eusm H$ by $SL(2,\Z)$ is a
curve of genus zero with one point missing.  This can be seen by studying the fundamental
domain (\pinst(a)).  The missing point is called the ``cusp'' at $\tau=i\infty$.
It is because the quotient has genus zero that Riemann surfaces of genus 1 can
be parametrized by a single holomorphic function, namely
$J(\tau)$.
As we have seen in section \bosecase, this is useful
because it means that the constraints of modular invariance are
equivalent to the statement that the partition function is a
function of $J$.
 Similarly, the quotient of the upper half plane $\eusm H$ by
$\Gamma_\theta$ is again a curve of genus zero, this time with
two missing points or cusps, as explained in \pinst(b).   This means
that there is a function $K(\tau)$ such that any
$\Gamma_\theta$-invariant function $F$ can be expressed as a
function of $K$.

The two cusps of $\eusm H/\Gamma_\theta$ are at $\tau\to i\infty$ and $\tau=1$, and
correspond respectively to the NS sector and  the R sector.  For
$\tau\to i\infty$, the NS partition function $F(\tau)$ has an
expansion in terms of $L_0$ eigenvalues in the NS sector.
Likewise, the Ramond partition function $H(\tau)$ can be expanded for
$\tau\to i\infty$ in $L_0$ eigenvalues in the Ramond sector; but
according to \ludo, the behavior of $H(\tau)$ for $\tau$ near
$i\infty$ is the same as the behavior of $F(\tau)$ for $\tau$ near
1.  So in terms of the function $F$, the behavior at the two cusps is determined by the low-lying
spectrum in the NS and Ramond sectors, respectively.

We therefore call these the NS and R cusps. The natural
uniformizing parameter at the NS cusp is $q^{1/2}=\exp(i\pi\tau)$,
where $\tau$ is the argument of the function $F$ in \feudo.  The
right parameter is $\exp(i\pi\tau)$ since the symmetry is
$\tau\to\tau+2$.  At the R cusp, the natural parameter is
$q=\exp(2\pi i \tau)$, where now $\tau$ is the argument of $H$.

  To get a clear picture of the nature of a holomorphic
function $F$ on $\eusm H/\Gamma_\theta$, it is necessary to
describe the behavior near the cusps, or equivalently to describe
what happens when one tries to extend $F$ to a function on the
compactification of $\eusm H/ \Gamma_\theta$ that is obtained
by adding the cusps.  This compactification is a space $\eusm
Y\cong \Bbb{CP}^1$.

We can explicitly describe a function $K$ that
parametrizes $\eusm H/\Gamma_\theta$ and has a pole only at the
NS cusp. This can be done in several ways.  One formula is
\eqn\oneform{K(\tau)={\Delta(\tau)^2\over \Delta(2\tau)\Delta(\tau/2)}-24,} where
$\Delta=q\prod_{n=1}^{24}(1-q^n)^{24}$ is the discriminant, a
modular form of weight 24.  In \oneform, we have subtracted the
constant 24 so that the expansion of $K(\tau)$ in powers of
$q^{1/2}$ has no constant term:
\eqn\umbob{\eqalign{K(\tau)\,=&\,\,q^{-1/2} + 276 q^{1/2} + 2048 q +
11202 q^{3/2} + 49152 q^2 +
 184024 q^{5/2} + 614400 q^3\cr & + 1881471 q^{7/2} + 5373952 q^4 +
 14478180 q^{9/2} +\dots.\cr}}
This is analogous to the definition of the $J$-function without a
constant term.  Another formula for $K$ is
\eqn\tumbob{K={q^{-1/2}\over
2}\left(\prod_{n=1}^\infty(1+q^{n-1/2})^{24}+\prod_{n=1}^\infty(1-q^{n-1/2})^{24}\right)
+2048q\prod_{n=1}^\infty(1+q^n)^{24}.}   The product formula in
\oneform, since it converges for all $|q|<1$, shows that $K$ is
non-singular as a function on $\eusm H$.  As for the behavior at the
cusps, either formula shows that $K$ has a simple pole at the NS
cusp, that is, it behaves for $q\to 0$ as $q^{-1/2}$.  $K$ is
regular at the Ramond cusp; in fact \eqn\uzzo{K(\tau=1)=-24.  }
This statement is equivalent to the statement  that $K+24=
{\Delta(\tau)^2/\Delta(2\tau)\Delta(\tau/2)}$  vanishes at $\tau=1$.  Indeed,
that function has a pole at the NS cusp, so it must have a zero somewhere.  Its representation
as a convergent infinite product shows that it is nonzero for $0<|q|<1$, so the zero is at the
Ramond cusp.
We give
another explanation of \uzzo\ in section \extremal\ in analyzing the
$\k=1$ model. The fact that the holomorphic function $K$ on $\eusm
Y$ has only one pole, which is of first order, gives another way to
prove that $\eusm Y$ is of genus zero.

Now let us consider the Neveu-Schwarz partition function $F$ of a
holomorphic SCFT.  Any $\Gamma_\theta$-invariant function $F$ on
$\eusm H$  can be written as a function of $K$.  The function $F$
arising in a holomorphic SCFT is actually {\it polynomial} in $K$.
Indeed, since the definition of $F$ as $\Tr\,q^{L_0}$ is convergent
for $0<|q|<1$,  $F$ is regular as a function on $\eusm H$. So the
only poles of $F$ are at cusps, but the formula \gudo, which
reflects the fact that the ground state energy in the Ramond sector
is zero, says that $F$ has no pole at the Ramond cusp. So the only
pole of $F$ is at the Neveu-Schwarz cusp, that is at $K=\infty$.
Consequently, in any holomorphic SCFT, the Neveu-Schwarz partition
function $F$ is a polynomial in $K$.

The degree of this polynomial is precisely $\k$, since $F\sim
q^{-\k/2}$ for $q\to 0$.  So  \eqn\zirox{F=\sum_{r=0}^{k^*} f_r
K^r.} Thus, $F$ depends on $k^*+1$ coefficients.

Now, for the case of an SCFT that is dual to three-dimensional
supergravity, what would we expect $F$ to be?  To follow the same
logic as in the bosonic case, we would argue that since the
minimum energy of  a BTZ black hole is $L_0=0$, and the entropy
vanishes at $L_0=0$, primary states corresponding to black holes
should be absent for $L_0\leq 0$.  Equivalently, primary fields
other than the identity should be absent for dimension less than
$(\k+1)/2$. Either statement would mean that up to terms of order
$q^{1/2}$, $F$ would coincide with the naive function
\eqn\yeroto{F_0(\k)=q^{-\k/2}\prod_{n=2}^\infty{1+q^{n-1/2}\over
1-q^n}} that counts superconformal descendants of the identity.

For each positive integer $\k$, there is a uniquely determined
function $F_{\k}$ that is a polynomial in $K$ and coincides with
$F_0(\k)$ up to order $q^{1/2}$.  This is a natural analog of the
partition function $Z_k$ that we defined for an extremal CFT without
supersymmetry.

However, the same logic that makes us think that there should be no
NS primaries of $L_0\leq 0$ also would make us believe that there
should be no Ramond primaries of $L_0=0$.  (There are automatically
no Ramond states of $L_0<0$, since $L_0={\cal G}_0^2$.)  In either
the NS or Ramond sector, the classical black hole entropy vanishes
for $L_0 \leq 0$.

The number of Ramond primaries of $L_0=0$ in a theory with
Neveu-Schwarz partition function $F_{\k}$ is, according to \gudo,
$h_0=(-1)^{\k}F_{\k}(1)$.  Let us call this number $\beta_{\k}$.
$\beta_{\k}$ is uniquely determined for each $\k$.  A practical
way to determine it, using \uzzo, is to write
\eqn\tyrx{\beta_{\k}=(-1)^{\k}F_{\k}(1)=(-1)^{\k}\sum_{r=0}^{\k}f_rK(1)^r=
(-1)^{\k}\sum_{r=0}^{\k}f_r(-24)^r.}
So one first evaluates the coefficients $f_r$ in \zirox\ to ensure
that there are no NS primary states with $L_0\leq 0$, and then one
evaluates the sum in \tyrx.

The first ten values of $\beta_{\k}$ are given in Table 1.  The
first observation is that $\beta_{\k}$ never vanishes in this range.
Consequently, it is impossible to assume that there are no primary
operators other than the identity of dimension less than
$(k^*+1)/2$.  If we assume that there are no such primaries in the
NS sector, then there are primaries of dimension $\k/2$ in the
Ramond sector.

\smallskip
$$
\centerline{\vbox{\offinterlineskip
\def\tablerule{\noalign{\hrule}}
\halign to 4.5truein{\tabskip=1em plus 2em#\hfil&\vrule height12pt
depth5pt#&#\hfil&\vrule height12pt depth5pt#&#\hfil\tabskip=0pt\cr
{\hfill $\k$}&&{\hfill $\beta_{\k}$}\cr
\tablerule
{\hfill $1$}&&{\hfill 24}\cr
{\hfill $2$}&&{\hfill 24}\cr
{\hfill $3$}&&{\hfill 95}\cr
{\hfill $4$}&&{\hfill 1}\cr
{\hfill $5$}&&{\hfill 143}\cr
{\hfill $6$}&&{\hfill 1}\cr
{\hfill $7$}&&{\hfill 262}\cr
{\hfill $8$}&&{\hfill -213}\cr
{\hfill $9$}&&{\hfill 453}\cr
{\hfill $10$}&&{\hfill -261}\cr}}}
$$
\bigskip
\centerline{\vbox{\hsize=5.1truein\baselineskip=12pt \noindent
Table 1. Values of the index $\beta_{\k}$ for $\k=1,\dots,10$.
}}\vskip .5cm

The good news is that the numbers in the table are rather small
compared to black hole multiplicities that arise in the classically
allowed region $L_0>0$.  So an optimistic view is to interpret the
numbers in the table as quantum corrections.  For example, at
$\k=9$, where $\beta$ takes the relatively large value 453, the
multiplicity of the lowest mass classically allowed black hole,
namely $L_0=1/2$ in the NS sector, turns out to be 135149371 if the
partition function $F_9$ can be trusted. (As usual, we count only
primaries, although this involves only a small correction.) The
lowest classically allowed black hole in the Ramond sector, at
$L_0=1$, has multiplicity 381161020987.

However, it may not seem logical to assume that there are no NS
primaries at dimension $\k/2$ and allow Ramond primaries of that
dimension.  Moreover, it really does not make sense to do this,
since some values of $\beta$ in the table are negative.

So we retreat from claiming that there are no NS primary fields of
dimension less than $(\k+1)/2$, and instead we consider the
hypothesis that there are no such primaries of dimension less than
$\k/2$.  In this case, we are free to add an integer $s$ to
$F_{\k}$.  If we do so, the number of NS primaries of dimension
$\k/2$ becomes $f_0=s$, and the number $h_0$ of Ramond primaries of
that dimension changes by $(-1)^{\k}s$.  The combination
\eqn\tolgo{\beta_{\k}=h_0-(-1)^{\k}f_0} is unchanged, and so is
still given by the numbers in the table.

Unfortunately, then, we have no way to determine the numbers $f_0$
and $h_0$ separately, only the ``index'' $\beta$.  We also have
little insight about the more conventional index $\chi$ defined in
\ycombo.  Both of these quantities characterize the primary states
of $L_0=0$.

As in the bosonic case, one can construct useful formulas for the functions $F_{\k}$ in terms of
Hecke operators acting on $F_1$.  However, we will omit this.

\bigskip\noindent{\it Optimistic Conjecture}

We define an extremal SCFT to be one with no primary fields other
than the identity of dimension less than $k^*/2$. Equivalently, the
NS partition function is equal up to an additive integer to the function $F_{k^*}$ that was
defined above.

Generalizing what we have said about the bosonic case, the most
optimistic conjecture we can propose is that an extremal SCFT exists
and is unique for every positive integer $k^*$, and is dual to
three-dimensional supergravity.   The best evidence we can offer is
that extremal SCFT's in this sense do exist for $\k =1,2$ and there are results
about uniqueness \refs{\hoehn,\duncan} at least for $\k=1$.

In addition, as we will see, the functions $F_{\k}$ with $\k=3,4$
have interesting properties suggestive of the existence of actual
theories.

These matters are discussed in section \extremal.  In addition, in
the appendix, we describe the functions $F_{\k}$ for $5\leq \k\leq
10$.

\def\NSZ{{\rm NS}_0}
\def\RZ{{\rm R}_0}
\subsec{Extremal SCFT's With Small $\k$}\subseclab\extremal

\bigskip\noindent{$k^*$=1}

The first construction of an extremal SCFT was made by FLM at
$k^*=1$, that is $c=12$. They considered eight free bosons $X_i$
compactified using the $E_8$ root lattice, combined with eight free
fermions $\psi_i$ to achieve superconformal symmetry.  This theory
has NS primary fields of dimension 1/2, namely the $\psi_i$.  To
eliminate these fields, they considered a $\Z_2$ orbifold, dividing
by the operation that acts as $-1$ on all $X_i$ and $\psi_i$.  They
conjectured that this construction gave the unique SCFT that has $c=12$
and no NS primary of dimension 1/2.

While the construction is simple, it has the drawback of not making
manifest the global symmetry of the model.  This was remedied much
more recently by Duncan \duncan, who described the same model in
another way.  In this construction, one begins with 24 free fermions
$\lambda_i$, $i=1,\dots,24$, forming again a system with $c=12$.  In
quantization, one can require the $\lambda_i$ to be either
antiperiodic or periodic around the spatial direction, giving what
we will call $\NSZ$ and $\RZ$ sectors.  (We reserve the name NS and
R for another construction that will appear shortly.)

With 24 fermions, the ground state energy in the $\NSZ$ sector is
$-1/2$, while in the $\RZ$ sector, the ground state energy is $+1$.
The minimum dimension of a spin field is the difference between
these two energies, or $3/2$.

The model has an $O(24)$ symmetry that rotates the fields
$\lambda_i$. The ground state in the $\RZ$ sector is highly
degenerate, because of zero modes of the fields $\lambda_i$. It
consists of   $2^{12}=4096$ states, transforming in the spinor
representation of  $O(24)$ (or rather its double cover).

The spin fields of lowest dimension are therefore 4096 fields
$W_\alpha$ of dimension 3/2 transforming as spinors.  Dimension 3/2
is the correct dimension for a supercurrent, but a generic linear
combination $W=\sum_\alpha \epsilon^\alpha W_\alpha$, with
$c$-number coefficients $\epsilon^\alpha$, does not generate a
superconformal algebra.  Schematically, and without worrying about
the precise coefficients, the operator product $W\cdot W$ gives
\eqn\omigolt{W(x)W(0)\sim {\bar\epsilon\epsilon\over x^3}+{\bar
\epsilon \epsilon\,T\over
x}+{\bar\epsilon\Gamma^{ij}\epsilon\,\lambda_i\lambda_j\over
x^2}+{\bar\epsilon
\Gamma^{ijkl}\epsilon\,\lambda_i\lambda_j\lambda_k\lambda_l\over
x}+{\rm regular}.}
(Here $T$ is the stress tensor and $\Gamma_i$ are the gamma matrices of $O(24)$.)
 The condition for $W$ to generate a
superconformal algebra is that the last two terms should be absent.
This is equivalent to
\eqn\zomigolt{\bar\epsilon\Gamma^{ij}\epsilon=\bar\epsilon\Gamma^{ijkl}\epsilon=0.}
  It is shown in
\duncan\ that a spinor $\epsilon$ obeying these conditions exists
and, if normalized to have $\bar\epsilon\epsilon=1$, is unique up to
an $O(24)$ transformation.  Another important fact is that any such
$\epsilon$ has one definite $SO(24)$ chirality or the other.

\def\Coo{{\rm Co}_0}
\def\Co{{\rm Co}_1}
Any choice of a solution of \zomigolt\  turns the theory into an
$\N=1$ superconformal field theory, with $W$ as the supercurrent.
However, as $\epsilon$ is not $O(24)$-invariant, the theory regarded
as an $\N=1$ SCFT does not have $O(24)$ symmetry.  Rather, the
choice of $\epsilon$ breaks $O(24)$ symmetry to a group that is known as the Conway
group $\Coo$. It is the symmetry group of the Leech lattice, and is a double cover
of a sporadic finite group $\Co$.
Thus, as a superconformal field theory, this model has symmetry
$\Coo$, not $O(24)$.

With this choice of superconformal algebra, we need to understand
what are the Neveu-Schwarz and Ramond vertex operators.  A Ramond
vertex operator $\cal O$ has a square root singularity in the
presence of the supercurrent $W$: \eqn\toglo{W(x){\cal O}(x')\sim
{{\cal O}'\over (x-x')^{n-1/2}}} for some integer $n$ and some
operator ${\cal O}'$. With $W$ understood as a spin operator with
respect to the original fermions $\lambda$, those fermions have
precisely this property. So they are Ramond fields.  This enables us
to analyze all of the states in the
original $\NSZ$ sector. The ground state corresponds to the identity
operator, whose OPE with $W$  of course has no branch cut, so it is
an NS rather than Ramond operator.  The operators in the $\NSZ$
sector that have no branch cut with $W$ are those that are products of
an  even number of $\lambda$'s and their derivatives.
The partition function that counts the corresponding states is \eqn\hsot{{q^{-1/2}\over
2}\left(\prod_{n=1}^\infty(1+q^{n-1/2})^{24}+\prod_{n=1}^\infty(1-q^{n-1/2})^{24}\right).}
We recognize this as part of the formula \tumbob\ for the function
$K$.  In the $\RZ$ sector, of the 4096 ground states, half have one
chirality or fermion number and half have the other. Any excitation
at all can be combined with a  ground state of properly chosen
chirality to get an operator that either does or does not have a
branch cut with $W$, as desired. So for each $L_0$ eigenvalue in the
$\RZ$ sector, precisely half the states contribute to the NS sector
and half to the R sector.  The contribution of $\RZ$ states to the
NS sector is therefore
\eqn\zimbo{2048q\prod_{n=1}^\infty(1+q^n)^{24}.} Adding up \hsot\
and \zimbo, we see that the total partition function $F_1$ of the NS
sector in this model is precisely what we have called $K$:
\eqn\polko{F_1=K={q^{-1/2}\over
2}\left(\prod_{n=1}^\infty(1+q^{n-1/2})^{24}+
\prod_{n=1}^\infty(1-q^{n-1/2})^{24}\right)+2048q\prod_{n=1}^\infty(1+q^n)^{24}.}

We can similarly compute the Ramond partition function $H_1$ of this
model.  The contribution of the $\NSZ$ sector is obtained from
\hsot\ by changing a sign so as to project onto states of odd
fermion number, rather than even fermion number.  And the
contribution of the $\RZ$ sector is the same as \zimbo.  So
\eqn\yero{\eqalign{H_1&={q^{-1/2}\over
2}\left(\prod_{n=1}^\infty(1+q^{n-1/2})^{24}-\prod_{n=1}^\infty(1-q^{n-1/2})^{24}\right)+
2048q\prod_{n=1}^\infty(1+q^n)^{24}\cr &= 24 + 4096 q + 98304 q^2 +
1228800 q^3 + 10747904 q^4+\dots.\cr}} Except for states of $L_0=0$,
the global supercharge ${\cal G}_0$ of the Ramond sector exchanges
the part of the Ramond sector coming from $\NSZ$ with the part
coming from $\RZ$.  This implies that we can alternatively write
\eqn\gero{H_1=24+4096q\prod_{n=1}^\infty(1+q^n)^{24},} where we have
removed the $\NSZ$ contribution except for the ground states, and
doubled the $\RZ$ contribution. The equivalence of these formulas is
not very obvious.

 A consequence of the above formulas
is that $h_0$, the number of Ramond states of $L_0=0$, is equal to
24. This gives another explanation for \uzzo.
Another consequence of
\gero\ and \polko\ is that for $L_0>0$, the coefficients $h_n$ in
the $q$-expansion of $H_1$ are precisely twice the corresponding
coefficients $f_n$ in the expansion of $F_1$: \eqn\bero{h_n=2f_n.}
(Here $h_n$ is defined for integer $n$, while $f_n$ is defined for integer or  half-integer
$n$, so this formula involves all coefficients $h_n$ but only half
of the $f_n$.) This is highly exceptional for SCFT's and reflects
the particular way that this one was constructed.  The relation
\bero\ is equivalent to
\eqn\overo{0=F_1(\tau)+F_1(\tau+1)-H_1(\tau)=F_1(\tau)-G_1(\tau+1)-H_1(\tau).}
Although disguised in our way of presenting it, this is a standard
relation in string theory.  In the FLM construction of the model via
8 free bosons and  8 free fermions, \overo\ amounts to the statement
of Gliozzi, Olive, and Scherk \ref\gso{F. Gliozzi, D. I. Olive, and
J. Scherk, ``Supersymmetry, Supergravity Theories, And The Dual
Spinor Model,'' Nucl. Phys. {\bf B122} (1977) 253-290.} that  the
Ramond-Neveu-Schwarz model has equal partition function in the
Ramond and Neveu-Schwarz sectors, because of spacetime
supersymmetry.

The factor of 2 in \bero\ is consistent with asymptotic equality
between the numbers of NS and Ramond black holes because it only
involves half of the coefficients of $F_1$.  $H_1$ has only half as
many coefficients as $F_1$, but they are twice as big.

As we have explained, the NS partition function $F_{k^*}$ of an
extremal SCFT with any $k^*$ is a polynomial in $K$ or
equivalently in $F_1$, \eqn\yert{F_{\k}=f(F_1),} for some
polynomial $f$. From \ludo, it follows that the Ramond partition
function $H_{k^*}$ of an extremal SCFT can be obtained from $H_1$ using the same polynomial:
\eqn\pelk{H_{\k}=(-1)^{\k}f(-H_1).}

\bigskip\noindent{\it A Model With $\k=2$}

What in our language is an extremal SCFT of $\k=2$ was constructed \dix\
by Dixon, Ginsparg, and Harvey (DGH), relatively soon after the
work of FLM, by twisting the FLM monster construction in a somewhat
similar fashion. Note that an extremal SCFT of $\k=2$ has $c=24$,
like the FLM monster theory.

We recall that the starting point of the FLM construction is 24 free
bosons $X_i$ that are compactified via the Leech lattice.  Then one
performs a $\Z_2$ orbifold, dividing by the symmetry $X_i\to -X_i$.
To construct an orbifold, one first constructs untwisted and twisted
sectors (by quantizing fields $X_i(\sigma)$ that are assumed to be
periodic or antiperiodic functions of $\sigma$) and then one
projects both sectors onto their $\Z_2$-invariant subspaces.

DGH observed that the ground state energy in the untwisted sector is
$-1$, while the ground state energy in the twisted sector is $+1/2$.
So a twist operator  of lowest energy has dimension $3/2$, the right
dimension for a supercurrent.  They went on to show that it is
possible to pick a twist operator ${\cal S}$ that generates a
superconformal algebra.

In the FLM construction, the field ${\cal S}$ is not present, since
it is odd rather than even under $\Z_2$, and is projected out when
one forms the orbifold theory.  However, DGH showed that it is
possible to obtain an ${\cal N}=1$ SCFT by modifying the usual
orbifold projection.  They defined the NS sector to consist of
fields that do not have a cut in their OPE with ${\cal S}$, while the Ramond sector
consists of fields that do have such a cut. Concretely, the NS sector consists
of the
$\Z_2$-even part of the untwisted sector of the orbifold plus the part of
the twisted sector that transforms under $\Z_2$ as does ${\cal S}$
(so that ${\cal S}$, in particular, is an NS
field), while the Ramond sector consists of the $\Z_2$-odd part of
the untwisted sector plus the $\Z_2$-even part of the twisted
sector.

{}From this information, one can of course compute the NS and
Ramond partition functions $F_2$ and $H_2$.  From our point of
view, of course, $F_2$ is a polynomial in $F_1=K$, chosen so that
$F_2=q^{-1}+{\cal O}(q^{1/2})$.  This gives $F_2=K^2-552$, and
similarly, in view of \pelk, $H_2=H_1^2-552$.  From this, we get
\eqn\tyzo{\eqalign{F_2 & =q^{-1} + 4096 q^{1/2} + 98580 q +
1228800 q^{3/2} + 10745856 q^2 +
 74244096 q^{5/2} + 432155586 q^3+\dots\cr
                    H_2 & =24 + 196608 q + 21495808 q^2 + 864288768 q^3+\dots.\cr }}
The number of states of $L_0=0$ is 24 in the Ramond sector and 0
in the NS sector.

The global symmetry of the DGH model is closely related to
 the symmetry group $\Coo$ of the Leech lattice (it is an extension of this by a finite abelian
 group that involves the momenta of the Leech lattice).
 Since the $k^*=1$
model also has $\Coo$ symmetry, one might optimistically conjecture
that this group is relevant to supergravity at all $\k$.  However,
results below about $\k=3$ indicate that this is not the case.

\bigskip\noindent{\it Partition Function With $\k=3$}

Unfortunately, this is the last case for which a suitable SCFT is
known, but we can of course determine the appropriate NS and
Ramond partition functions for all $\k$, at least modulo an
additive integer. We consider $\k=3,4$ here because they turn out
to have unusual properties, and relegate the further cases
$\k=5,6,\dots,10$ to an appendix.

For $\k=3$, we have \eqn\orbox{F_3=K^3-828K-6143.} This leads to
\eqn\oxlo{\eqalign{F_3&= q^{-3/2} + 1 + 33606 q^{1/2} + 1843200 q
+ 43434816 q^{3/2} +
 648216576 q^2\cr& + 7171304841 q^{5/2} + 63903727616 q^3 +\dots        \cr
                   H_3&=   95 + 3686400 q + 1296433152 q^2 + 127807455232 q^3+\dots.\cr}}
Here one finds an identity just like \bero: the expansion
coefficients $h_n$ of $H_3$ are related for $n>0$ to analogous
coefficients $f_n$ of $F_3$ by \eqn\tomo{h_n=2f_n.} This expresses
the fact that \eqn\boxo{F_1^3(\tau)+F_1^3(\tau+1)=H_1^3(\tau)-1536,}
along with the similar linear relation \overo.

It seems that the identity $h_n=2f_n$, $n>0$, does not hold for any
values of $\k$ except 1 and 3.  In general, the differences
$h_n-2f_n$ are relatively small, reflecting the fact that
asymptotically an NS or Ramond black hole has the same entropy, but
they are not zero. The fact that these degeneracies are actually
equal for $\k=3$ may be a clue to the construction of this model.

One important point about the $\k=3$ model is that it seems very
unlikely to have $\Coo$ or $\Co$ symmetry.  Indeed the leading coefficient
95 of $H_3$ is not in any economical way the dimension of a representation
of $\Coo$.  (The dimensions of the first few irreducible representations of this
group are 1, 24, 276, 299, and 1771; for $\Co$, one must omit the number 24 from this list.)
 Adding an integer $n$ to $F_3$ and therefore
subtracting $n$ from $H_3$, does not help; we cannot pick $n$ so
that $n$ and $95-n$ are both dimensions of  $\Coo$
representations in an economical fashion. So it appears that,
despite the tempting evidence from the cases $\k=1,2$, the group
$\Coo$ is probably not a general symmetry of three-dimensional
supergravity.

\bigskip\noindent{\it Monster Symmetry For $\k=4$?}

However, in the next case, $\k=4$, there may well be a much larger
discrete symmetry group, namely the monster group $\MM$.

For $\k=4$, we have \eqn\nongo{F_4=K^4 - 1104K^2 - 8191K +
107545.} This leads to the NS and Ramond partition functions
\eqn\onongo{\eqalign{F_4&=q^{-2} + q^{-1/2} + 1 + 196884 q^{1/2} +
21493760 q +
 864299970 q^{3/2} + 20246053140 q^2 \cr &+ 333202640600 q^{5/2} +
 4252023300096 q^3 + 44656994071935 q^{7/2} + 401490908149760 q^4+\dots \cr H_4&=1
 + 42987520 q + 40491712512 q^2 + 8504046600192 q^3 +
 802981773312000 q^4+\dots .\cr}}
 The first non-trivial coefficient in $F_4$ is the famous number
 $196884=196883+1$ whose appearance in the $J$ function gave the
 original hint of a connection between $\MM$ and modular functions
 and ultimately conformal field theory.  Moreover, the leading
 coefficient in $H_4$ is 1, which is compatible with any
 assumption about an automorphism group, since it is the
 dimension of an irreducible representation, namely the trivial
 one.  The contrasts with what we found for $\k=3$, where the
 leading coefficient, namely $95$, makes it difficult to postulate
 a large discrete symmetry.

 To explore further the hypothesis of $\MM$ symmetry, we attempt to
 express the coefficients in $F_4$ and $H_4$ in terms of
 dimensions of representations of $\MM$.  The dimensions
 $d_1,d_2,\dots,d_{12}$ of the first 12 monster representations, which we call $R_i$,
 $i=1,\dots,12$,
  are given
 in the table.

\bigskip
\centerline{\vbox{\offinterlineskip
\def\tablerule{\noalign{\hrule}}
\halign to 4.5truein{\tabskip=1em plus 2em#\hfil&\vrule height12pt
depth5pt#&#\hfil&\vrule height12pt depth5pt#&#\hfil\tabskip=0pt\cr
 \tablerule
$d_1$&& 1 \cr $d_2$ && 196883\cr
 $d_3$ &&
21296876  \cr  $d_4$ && 842609326 \cr $d_5$ && 18538750076 \cr $d_6$
&& 19360062527 \cr $d_7$ && 293553734298 \cr $d_8$ && 3879214937598
\cr $d_9$ && 36173193327999 \cr $d_{10}$ && 125510727015275 \cr
$d_{11}$ && 190292345709543 \cr $d_{12}$ && 222879856734249
\cr}}}\bigskip \centerline{ \vbox{\hsize=5.1truein\baselineskip=12pt
\noindent Table 2. Presented here from \ref\atlas{J. H. Conway, R.
T. Curtis, S. P. Norton, R. A. Parker, and R. A. Wilson, {\it Atlas
Of Finite Groups} (Oxford University Press, 1985).} are the
dimensions $d_i$ of the $i^{th}$ irreducible representation of the
monster group $\MM$, for $i=1,\dots, 12$. We denote as $R_i$ the
representation of dimension $d_i$.}}\bigskip

A little experimentation soon shows that the nontrivial coefficients in the NS partition function
$F_4$ can be nicely written in terms of the $d_i$.  The first 8 nontrivial coefficients
$f_{1/2},f_1,
\dots, f_4$ are
\eqn\toxico{\eqalign{196884         &= d_1+d_2               \cr
                      21493760 &= d_1+d_2+d_3\cr
                      864299970 & = 2d_1+2d_2+d_3+d_4 \cr
                      20246053140&=3d_1+4d_2+2d_3+d_4+d_6\cr
                      333202640600&=4d_1+5d_2+3d_3+2d_4+d_5+d_6+d_7\cr
                      4252023300096&=5d_1+7d_2+4d_3+4d_4+2d_5+2d_6+d_7+d_8\cr
                      44656994071935&=7d_1+11d_2+7d_3+6d_4+3d_5+4d_6+2d_7+2d_8+d_9\cr
                      401490908149760&=10d_1+16d_2+12d_3+9d_4+5d_5+7d_6+4d_7+4d_8+d_9+d_{10}+d_{12}.
                       \cr}}
The coefficients on the right hand side
are rather small given the numbers involved, and as we will explain shortly,
they do not grow much faster than is required by superconformal symmetry.  The first three numbers
here equal the first three non-trivial coefficients of the $J$ function, but afterwards the two
series diverge.

The coefficients $h_n$ in the $q$-expansion of $H_4$ can similarly
be expressed in terms of dimensions of monster representations.  In
doing so, however, the following is a useful shortcut and also
possibly a clue to constructing the model.  First of all, in
contrast to the case of $\k=1,3$, it is not true that $h_n-2f_n=0$
for $\k=4$ and all $n>0$. But some of these differences vanish, and
they are all relatively small.  As we have already expressed the
$f_n$ for small $n$ in terms of monster dimensions in \toxico, it
suffices now to find similar expressions for $h_n-2f_n$.  Actually,
as the $h_n$ are all even for $n>0$, we prefer to divide by 2, and
we find \eqn\oxico{\eqalign{{h_1\over 2}-f_1&=0 \cr
                    {h_2\over 2}-f_2&=-f_{1/2}\cr
                    {h_3\over 2}-f_3&=0 \cr
                    {h_4\over 2}-f_4&=-f_1\cr
                    {h_5\over 2}-f_5&=0.\cr}}
When this is combined with \toxico, we see that the $h_i/2$, $i\leq 5$
 are linear combinations of
the $d_i$ with small positive integer coefficients.
So the initial coefficients of the Ramond partition function are compatible with $\MM$ symmetry, and
in addition many of the differences $h_n/2-f_n$ vanish.

T. Gannon has given a simple explanation of these results by showing
that one can express $F_4$ and $H_4$ in terms of the $J$-function
via $F_4(\tau)=J(2\tau)+J(\tau/2)+1$,
$H_4(\tau)=J(\tau/2)+J((\tau+1)/2)+1$.  Hence $F_4$ and $H_4$
inherit a relation to the monster from $J$.  The fact that the formulas
have such a direct explanation
may lessen the case for a new SCFT with monster symmetry.

Given that the above formulas exist,
they are not quite unique, because there are some linear relations\foot{These relations
can be explained as follows.   The symmetric part of $R_2\otimes R_2$ decomposes as ${\rm Sym}^2\,R_2\cong R_1\oplus R_2
\oplus R_4\oplus R_5$, and the antisymmetric part decomposes as $\wedge^2\,R_2\cong R_3\oplus R_6$.
On the other hand, ${\rm dim}({\rm Sym}^2\,R_2)-{\rm dim}(\wedge^2\,R_2)={\rm dim}\,R_2=d_2$.
Taken together, these facts imply the first relation in \herot, and the second follows similarly
by considering the decomposition of $R_3\otimes R_3$.} among the $d_i$ with small coefficients:
\eqn\herot{\eqalign{d_1 + d_4 + d_5 &= d_3 + d_6 \cr
                     d_3+ d_8 + d_{12} & = d_2+d_7+d_9+d_{11}.\cr}}
We have made some choices to ensure that the above formulas are compatible with superconformal
symmetry, in the following sense.  The first formula in \toxico\ indicates that at $L_0=1/2$,
 the NS sector has 196883 primary states $|\rho_i\rangle$
transforming in the representation $R_2$.  It follows that at
$L_0=1$, there are descendants ${\cal G}_{-1/2}|\rho_i\rangle$
transforming in the same representation.  This continues at higher
levels; the number of copies of $R_2$ appearing as descendants of
the primary states $|\rho_i\rangle$ at $L_0=s/2$ is
$1,1,1,2,3,4,5,7$ for $s=1,2,\dots,8$.  If we count only primary
states, we find, assuming that \toxico\ is the right decomposition,
that primary states in the representation $R_2$ occur at
$L_0=1/2,3/2,2,7/2,4$, each time with multiplicity 1.  Similar
remarks apply to other representations. A special case is that, for
the range of $L_0$ considered in \toxico,
 most of the states that transform in the trivial representation
 $R_1$ are actually descendants of the identity.  If \toxico\ is the right
 decomposition, then the first
 primary field that is $\MM$-invariant is
 at $L_0=5/2$.  If we were to rewrite \toxico\ in terms of primary
 states only, the coefficients would be even smaller (but still nonnegative),
 strengthening the case for monster symmetry.

In this particular example, we cannot add an integer to $F_4$ without spoiling the hypothesis
of monster symmetry.  If there were an NS primary with $L_0=2$, it would have a descendant
at $L_0=5/2$, and the number of primaries at $L_0=5/2$ would be less than 196883.

In the appendix, we present somewhat similar though less extensive evidence for baby monster
symmetry at $k^*=6$.

\bigskip\noindent{\it Sum Over Spin Structures}

Given an SCFT with $c$ an integer multiple\foot{If $c$ is an odd multiple of 12, the sum over
spin structures is modular-invariant but projects out the NS ground state and does not give
a CFT in the usual sense.  This is actually important in superstring theory, where, with $c=12$
in the light-cone treatment, the NS ground state is a ``tachyon'' and is removed in the sum over
spin structures.} of 24,
by summing over spin structures, one can make an ordinary bosonic CFT.  The sum over spin structures
projects both the NS and Ramond sectors onto  bosonic states of integer dimension,
as is appropriate for a bosonic CFT.

For $c$ to be a multiple of 24, $k^*$ must be even.  It was already shown by DGH that the sum over
spin structures,
applied to their model which in our language is $k^*=2$, gives the FLM model at $k=1$.  A natural
question (raised by J. Duncan) is what this operation does at higher even values of $k^*$.

The sum over spin structures removes fermionic operators such as the supercurrent ${\cal S}$, of
dimension 3/2.  However, it leaves operators such as ${\cal S}\partial{\cal S}$ that are even in
${\cal S}$.  This operator has dimension 4, and is a Virasoro primary (though, of course, it is
a descendant in the ${\cal N}=1$ super-Virasoro algebra).  A bosonic extremal CFT of $k\geq 4$
should not have a primary of dimension 4.  Hence, the ``sum over spin structures'' operation
applied to an extremal SCFT of $k^*\geq 8$ does not give an extremal CFT.  However, it does give
an extremal CFT if applied to a theory of $k^*=4$ or 6.

Hence, if extremal SCFT's exist with $k^*=4,6$, they can be used to generate extremal CFT's of
$k=2,3$.  It is interesting that $k^*=4,6$ are precisely the values at which extremal SCFT's may have
monster or baby monster symmetry.  (The baby monster group is the centralizer of an involution in
the monster, and hence it is conceivable for an SCFT at $k^*=6$ to have baby monster symmetry
while the CFT at $k=3$ has monster symmetry.  The framework for this is described in \dix.)

Part of what we have said can be understood in another way.
Supergravity and gravity are different in the
semiclassical limit, and one cannot be obtained from the other by merely
summing over spin structures.
So whatever is the SCFT dual of supergravity (even if our assumptions are too optimistic
and it is not extremal),
the sum over spin
structures applied to this SCFT cannot give, for arbitrarily large $k^*$, the CFT that is dual
to gravity.

\def\W{{\cal W}}
\def\V{{\cal V}}
\def\SV{{\rm Sym}^2{\cal V}}
\def\SW{{\rm Sym}^2{\cal W}}
\def\E{{\cal E}}
\newsec{$k=2$ Partition Function On A Hyperelliptic Riemann Surface}\seclab\confalg

The most important question raised by this paper is certainly
whether appropriate CFT's and SCFT's exist, beyond the few examples
that are known.

In this section, we will perform a small computation that aims to
give a hint that this is the case, at least for the bosonic theory
with $k=2$ and hence $c=48$.  We will show that the partition
function of such a theory on a hyperelliptic Riemann surface of
any genus can be determined in a unique and consistent way.  This
includes, for example, any Riemann surface of genus 2.  The fact
that we get a unique and consistent result in this situation gives
some encouragement for believing that the $k=2$ model does exist
and is unique.

In this paper, we merely demonstrate the consistency of an
algorithm for determining the partition function.  Hopefully it
will be possible in future work to get explicit formulas, at least
for genus 2.  Our method also works for $k=1$, though here the
ability in principle to determine the partition function comes as
no surprise, since the model has been constructed explicitly \flm\
and in fact the genus 2 partition function has been computed
\ref\tuite{M. P. Tuite, ``Genus Two Meromorphic Conformal Field
Theory,'' math.qa/9910136.} by a quite different method.

Perhaps it would help orient the reader to compare what we will do
to another possible approach to determining the partition function.
Genus 2 partition functions were determined in \tuite\ for a variety
of holomorphic CFT's with $c=24$.   The basic method was to express
the partition function in terms of a Siegel modular form, which
depends on only finitely many coefficients, and determine the
coefficients by considering the behavior when the Riemann surface
$C$ degenerates. There are two types of degeneration ($C$ can break
up into two genus 1 curves joined at a point, or can reduce to a
single genus 1 curve with two points glued together). The partition
function can be determined from the behavior at just one
degeneration, and there is a problem of consistency to show that one
gets the same result either way. One could attempt to demonstrate
the consistency by studying the appropriate Siegel modular forms,
but we prefer to prove consistency by establishing the associativity
of a certain operator product algebra.  One advantage is that this
method can be used to determine the partition function on a
hyperelliptic Riemann surface of any genus.  On the other hand, if
one could overcome the technicalities in genus 2 (see
\ref\tuitetwo{G. Mason and M. P. Tuite, ``On Genus Two Riemann
Surfaces Formed From Sewn Tori,'' Commun. Math. Phys. {\bf 270}
(2007) 587-634, math.qa/0603088.} for one approach), one could
possibly compute the genus 2 partition function of an extremal CFT
for all $k$.

\subsec{Twist Fields}

 A hyperelliptic Riemann
surface $C$ is a double cover of the complex plane, for example a
double cover of the complex $x$-plane, which we will call $C_0$,
described by an equation
\eqn\doofus{y^2=\prod_{i=1}^{2g+2}(x-e_i).} The $2:1$ cover $C\to
C_0$ is branched at the points $e_1,\dots,e_{2g+2}$. $C$ is smooth
if the $e_i$ are distinct, and has genus $g$ if the number of
branch points is precisely $2g+2$.

\nref\zam{Al. B. Zamolodchikov,  ``Correlation Functions Of Spin
Operators In The Ashkin-Teller Model And The Scalar Field On A
Hyperelliptic Surface,''  Soviet Phys. JETP  {\bf 63}  (1986) 1061-1066.}%
\nref\kni{V. G. Knizhnik, ``Analytic Fields On Riemann Surfaces,''
Commun. Math. Phys. {\bf 112} (1987) 567.}%
\nref\dixfms{L. Dixon, D. Friedan, E. Martinec, and S. Shenker,
``The Conformal Field Theory Of Orbifolds,'' Nucl. Phys. {\bf B282} (1987) 13-73.}%
\nref\vafham{S. Hamidi and C. Vafa, ``Interactions On Orbifolds,''
Nucl. Phys. {\bf B279}
(1987) 465.}%
 Our
approach to determining the partition function on such a Riemann
surface is based on an old idea \refs{\zam-\vafham}. The partition
function of a conformal field theory $\W$ on the hyperelliptic
Riemann surface $C$ can be determined by computing, in a doubled
theory, the correlation function of $2g+2$ copies of a ``twist
field'' $\E$, inserted at the points $e_1,\dots,e_{2g-2}$ in $C_0$.

Consider any CFT $\W$ of central charge $c$.   Away from branch
points, the theory $\W$ on the double cover $C$ looks locally like
the theory $\W\times \W$ on $C_0$.  Here we have one copy of $\W$
for each of the two branches of $C\to C_0$.

In going around a branch point, the two copies of $\W$ are
exchanged.  So a more complete description is to say that the theory
on $C_0$ is $\SW$, the symmetric product of two copies of $\W$.  The
symmetric product theory is an orbifold of the product theory
$\W\times \W$ in which one divides by the $\Z_2$ symmetry that
exchanges the two branches.

\def\E{{\cal E}}

Let us describe what kind of operators the symmetric product
theory has.  First, there are ``untwisted'' operators.  These are
simply operators of the theory $\W\times \W$ that are invariant
under the exchange of the two factors.  For example, let $T^+$ and
$T^-$ be the stress tensors of the two factors.  Any local
operator constructed as a polynomial in $T^+$ and $T^-$ and their
derivatives and invariant under the exchange $T^+\leftrightarrow
T^-$ gives an operator in the symmetric product theory. Operators
of this particular type generate a chiral algebra that we will
call $\SV$, where $\V$ denotes the chiral algebra generated by a
single stress tensor.

In addition, there are ``twisted sector'' operators.  These
operators correspond (in the operator/state correspondence of CFT)
to states obtained by quantizing the theory $\W\times \W$ on a
circle $S^1$, in such a way that the two copies of $\W$ are
exchanged in monodromy around the circle.  These states are the
states of a {\it single} copy of $\W$ on a circle of twice the
circumference.  The energies of the states in the twisted sector
of the $\SW$ theory are therefore precisely one-half the energies
of the original $\W$ theory (in conformal field theory, doubling
the circumference of the circle divides the Hamiltonian by two).

Operators related to states in the twisted sector are called twist
fields.  The dimension of a twist field is the same as the
difference in energy between the corresponding twisted sector
state and the untwisted ground state.  The ground state energy in
the untwisted sector is $-2\cdot c/24$ (where the 2 comes from the
two copies of $\W$) and the ground state energy in the twisted
sector is $-(1/2)\cdot c/24$ (where the factor of $1/2$ was
explained in the last paragraph).  The difference is
$d_\E=(c/24)(-1/2-(-2))=(3/2)c/24$, and this is the dimension of
the twist field $\E$ of lowest energy. For our application, we are
interested in the case that $c/24$ is an integer $k$, and hence
$d_\E=3k/2$.

Now we can explain how the partition function of the theory $\W$
on the hyperelliptic Riemann surface $C$ defined in eqn. \doofus\
can be interpreted as a genus zero correlation function in the
theory $\SW$. From the standpoint of the $\SW$ theory on the
$x$-plane, the role of the branch points is just to exchange the
two copies of $\W$, via a twist field.  On the double cover, there
is no operator insertion  at the branch points except the
identity; the identity operator at a branch point corresponds in
the downstairs description to the ground state in the twisted
sector and hence to the twist operator $\E$ of lowest dimension.
So the partition function on the double cover can be expressed in
terms of  the correlation function $\big\langle
\E(e_1)\E(e_2)\dots \E(e_{2g+2})\big\rangle$ on the $x$-plane.
(The precise statement involves the conformal anomaly; this is
deferred to section \role.)

\def\U{{\cal U}}\def\O{{\cal O}}
To compute this correlation function, we need to understand the
chiral algebra that is obtained by extending the symmetric product
$\SV$ of two Virasoro algebras by the field $\E$. For this, we
need to know what primary fields (of $\SV$) appear in the operator
product $\E\cdot \E$.  To calculate the product $\E \cdot \E$, we
need to look at the behavior when a pair of branch points approach
each other.  This situation is described locally by an equation
\eqn\lofus{y^2=(x-e)(x-e'),} and we are interested in the behavior
for $e\to e'$.  In that limit, the equation becomes $y^2=(x-e)^2$,
and the surface breaks up into two branches $C_\pm:y=\pm (x-e)$.
Each branch is a copy of the complex plane. We can express the
product $\E(e)\cdot \E(e')$ for $e\to e'$ as a sum of operators of
the form  $\U_+\otimes \U_-$, where $\U_\pm$ is an operator in the
original CFT $\W$ on the branch $C_\pm$.  Moreover, the $\U_\pm$
are descendants of some Virasoro primary fields $\O_\pm$.

\ifig\coxie{\bigskip (a) The equation $y^2=(x-e)(x-e')$ describes
two branches connected by a tube.  The tube collapses to a point
as $e\to e'$. (b) Topologically, the tube is equivalent to a
cylinder.  Propagation along the tube multiplies any energy
eigenstate by a $c$-number factor. If a primary field $\CO$ flows
in from the past, the same field flows out in the future.
Alternatively, if we think of all fields as flowing out, then the
two fields emerging at the two ends of the cylinder are
conjugate.} {\epsfxsize=4.5in\epsfbox{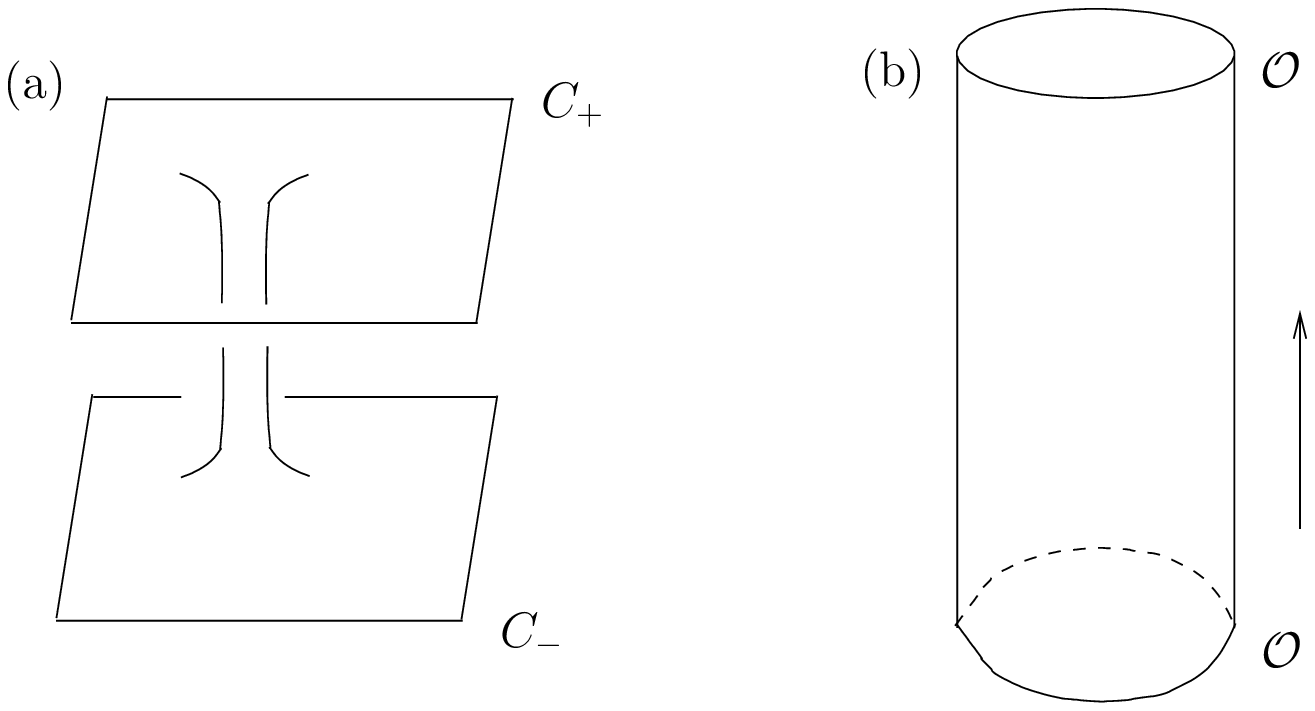}} The decomposition
of $\E(e)\cdot \E(e')$ as a sum of descendants of operators
$\CO_+\otimes \CO_-$ can be made in such a way that the operator
$\CO_+$ is conjugate to $\CO_-$, meaning that the two-point
function $\langle\CO_+\CO_-\rangle$ (in the original theory $\W$)
is nonzero in genus zero. To see this, go back to the smooth
double cover with $e\not= e'$. For $e$ near $e'$, the two branches
are smoothly connected by a tube (\coxie a). The surface thus has
the topology of a cylinder (\coxie b). Propagation along this
cylinder is the ``identity operator'' on primary fields. So if a
given primary state flows in at the bottom, it or  flows out at
the top. When we write $\E(e)\cdot \E(e')$ as a sum of descendants
of operators $\CO_+\otimes \CO_-$, the convention is that both
$\CO_+$ and $\CO_-$ are outgoing, so they are conjugate, rather
than being equal.  At any rate, the important conclusion for us is
that $\CO_+$ and $\CO_-$ have the same dimension.

According to our interpretation of three-dimensional gravity, it
is described by a conformal field theory with $c=24k$ and no
primary field other than the identity of dimension less than
$k+1$. Hence, a primary operator
$\CO_+\otimes \,\CO_-$ of the symmetric product theory, where $\CO_+$ and $\CO_-$ have the same
dimension, is either the identity operator or has dimension at
least $2(k+1)$.

To understand the chiral algebra or superalgebra\foot{As $\E$ has
dimension $3k/2$, it is a fermionic operator if $k$ is odd.}
 generated by $\E$ together with $\SV$, we need to know all primary
fields that appear with singular coefficients in the operator
product $\E(e)\cdot \E(e')$. Since $\E$ has dimension $3k/2$, an
operator that will contribute a singularity must have dimension
less than $3k$ and hence at most $3k-1$.   On the other hand, we
have just seen that all operators appearing in the OPE are either
descendants of the identity or have dimension at least $2(k+1)$.

For $k=1$ or 2, we have $3k-1<2(k+1)$.  This leads to a drastic
simplification, which is the reason that we will restrict
ourselves here to $k=1,2$: singularities in the product
$\E(e)\cdot \E(e')$ come only from the identity operator and its
$\SV$ descendants. Therefore, to understand correlation functions
of $\E$, we need only understand the chiral algebra that is
obtained by adding to $\SV$ the primary field $\E$ of dimension
$3k/2$ with the OPE schematically \eqn\gryt{\E\cdot \E\sim 1+{\rm
descendants}.}

This is a chiral algebra that can be described in closed form and
explicitly proved to obey the Jacobi identity.  The only $\SV$
primaries are 1 and $\E$.  The only genus zero three point
functions of primaries are $\langle 1 \cdot 1\cdot 1 \rangle$,
which is trivial, and $\langle 1\cdot  \E\cdot \E\rangle$, which
is almost trivial in the sense that it reduces to a two-point
function.  To understand correlation functions of descendants, it
suffices to understand the module for $\SV$ consisting of $\E$ and
its descendants.   One way to construct this module explicitly is
to observe that on the double cover of the $x$-plane, $\E$ simply
corresponds to the identity operator and $\SV$ to the ordinary
Virasoro algebra; so this module for $\SV$ can be deduced from the
identity module for $\V$.

To get beyond the three-point function, the key step is to show that the chiral algebra obtained by
adding $\E$ to $\SV$ is consistent, that is, that the Jacobi
identity is obeyed.  If so, this chiral algebra determines all genus
zero correlation functions of $\E$, and hence determines the
partition function of our CFT on a general hyperelliptic Riemann
surface.

In such a problem of trying to extend a known chiral algebra $\SV$ by additional primary fields $\E$
(of integer or half-integer dimension, in which case one will get an extended
chiral algebra or a superalgebra),
the  Jacobi identity is equivalent \ref\zam{A. B. Zamolodchikov, ``Infinite Additional
Symmetries In Two-Dimensional Conformal Quantum Field Theory,'' Theor.
Math. Phys. {\bf 65} (1985) 1205-1213.}
to the statement that there exists
a four-point function of the additional primary fields that has the
appropriate singularities determined by the operator product
expansion in all channels. (Such a function automatically has the
right symmetries.)  In the case at hand, since the only primary field that we are
trying to add is $\E$, the only non-trivial primary four-point function that we have to consider
is $\big\langle \E(e_1)\E(e_2) \E(e_3) \E(e_4)\big\rangle$. In section
\determin, we explicitly construct this four-point function and show
that it behaves correctly in all channels.

In the following sense, this result is not at all surprising. The
four-point function $\big\langle
\E(e_1)\E(e_2)\E(e_3)\E(e_4)\big\rangle$ is essentially the
partition function of the underlying CFT on the hyperelliptic
Riemann surface $y^2=(x-e_1)(x-e_2)(x-e_3)(x-e_4)$. That surface
has genus 1, and we already know from section
\newbie\ that for every $k$ there is a unique, natural genus 1 partition
function with the properties we want. In section \revisit, we show
explicitly, for $k=1,2,$ that the four-point function $\big\langle
\E(e_1)\E(e_2)\E(e_3)\E(e_4)\big\rangle$ determined from the OPE's
agrees with the genus 1 partition function as described in section
\newbie.

\subsec{The $\E\cdot\E$ Operator Product}\subseclab\details

As a first step in that direction, we will calculate the details of
the $\E\cdot \E$ operator product. For this, we start with a double
cover $C$ of the $x$-plane branched at $e$ and $e'=-e$, and so
described by an equation $y^2=x^2-e^2$. If $u=x+y$, $v=x-y$, then
the equation is $uv=e^2$. The two branches $C_+$ and $C_-$
correspond respectively to $u\to\infty,\,v\to 0 $ and $u\to
0,\,v\to\infty$.

The path integral over $C$ gives a quantum state $\Psi$ in the
theory $\W\times \W$, that is, one copy of $\W$ for each branch.
This state is invariant under exchange of the two branches by the
symmetry $y\to -y$, so it is really a state in the symmetric
product theory $\SW$.  We are really only interested in the part
of $\Psi$ proportional to the vacuum state and its descendants. As
we have seen, this part suffices to describe the desired chiral
algebra if $k\leq 2$.

\ifig\coxit{\bigskip A contour $\cal S$ that wraps once around the
hole can be deformed to the contour $\cal S_+$ in the upper branch
or to the contour $\cal S_-$ in the lower branch.}
{\epsfxsize=3in\epsfbox{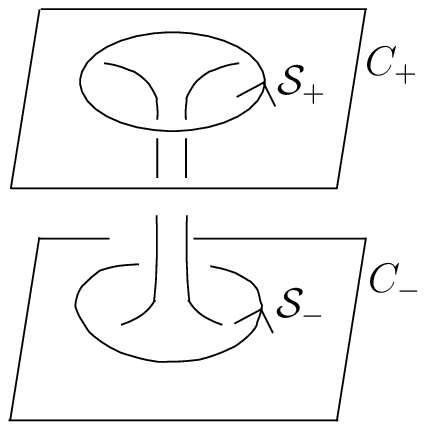}}
\def\CC{{\cal S}}
We will determine $\Psi$ by using the fact that certain elements
in the product $\V\times \V$ of two Virasoro algebras annihilate
$\Psi$. This is so because there are globally-defined holomorphic
vector fields on $C$, of the form
$V_n=2^{-n}u^{n+1}d/du=-2^{-n}(e^2/v)^nv d/dv$.  Let $\CC$ be a
contour on the surface $C$ that wraps once around the ``hole.'' If
$T$ is the stress tensor, the contour integral $\int_\CC V_nT$ can
be regarded, for any $n$, as an operator acting on the state
$\Psi$. This operator is invariant under deformation of the
contour. It can be deformed (\coxit) to a contour $\CC_+$ in the
upper branch $C_+$ or a contour $\CC_-$ in the lower branch $C_-$.
So we have for all $n$
\eqn\zelg{\left(\int_{\CC_+}V_nT-\int_{\CC_-}V_nT\right)\Psi = 0.
}

We want to express the two contour integrals that appear here in
terms of Virasoro generators on the two branches.  To do this, we
simply express $V_n$ as a vector field on the branch $C_+$ or
$C_-$, either of which we identify with the $x$-plane.  For
example, on the branch $C_+$, we write explicitly
$y=\sqrt{1-e^2/x^2}=1-e^2/2x^2-e^4/8x^4+{\cal O}(e^6)$.  We have
carried the expansion far enough to determine (for $k\leq 2$) all
singular terms in the product $\E(e)\cdot \E(-e)$.  So
\eqn\hocal{V_n=x^{n+1}\left(1-(n+2){e^2\over 4x^2}+{e^4\over
x^4}\left(n^2+n-4\over 32\right)+{\cal O}(e^6)\right){d\over dx}.}
This means, if we ignore the conformal anomaly for the moment,
that $\int_{\CC_+}V_nT$ corresponds, on the branch $C_+$, to the
operator \eqn\telmoxo{Q^+_n=L^+_n-{n+2\over
4}e^2L^+_{n-2}+\left(n^2+n-4\over 32\right)e^4L^+_{n-4}+{\cal
O}(e^6).} As a check, one can verify that
$[Q^+_n,Q^+_m]=(n-m)Q^+_{n+m}$.  Similarly, $\int_{\CC_-}V_nT$
corresponds on the branch $C_-$ to the operator
\eqn\zelmoxo{Q^-_n=\left({e^2\over
4}\right)^n\left(L^-_{-n}-e^2\left({-n+2\over
4}\right)L^-_{-n-2}+e^4\left(n^2-n-4\over
32\right)L^-_{-n-4}+{\cal O}(e^6)\right).}

The state $\Psi$ is determined for each value of $e$ (up to
multiplication by a complex scalar) by the condition that $\hat
Q_n\Psi=0$, where $\hat Q_n=Q_n^+-Q_n^-$. However, because of the
Virasoro anomaly some $c$-number terms must be added to the above
formulas, reflecting the conformal anomaly in the mapping from $u$
to $x$. There is no such $c$-number contribution to $\hat Q_0$,
since it cancels between the two branches.  The $c$-numbers in $\hat
Q_n$ for other $n$ can be conveniently determined by requiring that
$[\hat Q_n,\hat Q_m]=(n-m)\hat Q_{n+m}$.  For our purposes, the only
formulas we need are \eqn\zeldo{\eqalign{\hat Q_0& =
\left(L^+_0-{e^2\over 2}L^+_{-2}-{e^4\over
8}L^-_{-4}\right)-\left(L^+_0-{e^2\over 2}L^-_{-2}-{e^4\over
8}L^-_{-4}\right)+\dots \cr \hat Q_1& =\left(L^+_1-{3e^2\over 4}
L^+_{-1}-{e^4\over 16} L^+_{-3}\right)-{e^2\over 4}\left(
L^-_{-1}-{e^2\over 4} L^-_{-3}\right)+\dots \cr \hat Q_2& =
L_2^+-e^2L_0^+-3ke^2+{e^4\over 16}L^+_{-2}-{e^4\over
16}L^-_{-2}+\dots.\cr}} Terms of order $e^6$ have been omitted. The
constant in $\hat Q_2$ was obtained from $[\hat
Q_0,\hat Q_2]=-2\hat Q_2$.

By requiring that $\hat Q_m\Psi=0$ for $m=0,1,2$, and that $\Psi$
converges to the Fock vacuum $|\Omega\rangle$ for $e\to 0$, we now
find $\Psi$ to be \eqn\gesho{\eqalign{\Psi(e)=&\left(1+{e^2\over
4}(L^+_{-2}+L^-_{-2})+{e^4\over 32}(L^+_{-4}+L^-_{-4})\right. \cr &
\left.+{e^4\over 32}(L^+_{-2}+L^-_{-2})^2+{e^4\over
192k}L^+_{-2}L^-_{-2} +\dots\right)|\Omega\rangle.\cr}}

We can immediately use this formula to determine the singular part
of the $\E\cdot \E$ OPE.  We normalize $\E$ so that the most
singular term is $\E(x/2)\E(-x/2)\sim 1/x^{3k}$.  Then, setting
$e=x/2$, we get \eqn\mimico{\eqalign{\E(x/2)\E(-x/2)\sim {1\over
x^{3k}}&\left(1+{x^2\over 16}T+{x^4\over
2^{10}}\partial^2T\right.\cr &\left.+{x^4\over 2^9}T\star
T+{x^4\over 3k\cdot 2^{10}}T^+T^-\right)+{\cal O}(x^{-3k+6}).\cr}}
All we have done is to write, on the right hand side, the operator
that corresponds to the state $\Psi(e)/x^{3k}$, for $e=x/2$. (The
factor of $1/x^{3k}$ must be supplied by hand, since in defining
$\Psi(e)$, we just normalized it so that the coefficient of the Fock
vacuum is 1.)  Also, $T^+$ and $T^-$ are the stress tensors on the
branches $C_+$ and $C_-$, respectively.  $T=T^++T^-$ is the
diagonal or total stress tensor, and similarly $L_n=L_n^++L_n^-$.
In addition, we use the fact that
in the correspondence between operators and states, the states
$L^\pm_{-2}|\Omega\rangle$ correspond to the operators $T^\pm$,
while $L^\pm_{-4}|\Omega\rangle$ correspond to $\half
\partial^2 T^\pm$.  Finally, we have written $T\star T$
for the operator corresponding to the state
$L_{-2}^2|\Omega\rangle$.
$T\star T(0)$ can be obtained as the term of order $x^0$ in the
operator product $T(x)\cdot T(0)$.

\subsec{Determining The Four-Point Function}\subseclab\determin

Our goal is to use the explicit formula \mimico\ to show that the
chiral algebra obtained by adjoining $\E$ to $\SV$ obeys the Jacobi
identity.  The Jacobi identity is equivalent to the existence of a
four-point function for primary fields with the right symmetry
properties and the right OPE singularities in all channels.  It is
enough to consider primary fields with respect to the $\SV$ algebra,
which we already know to exist.  The only non-trivial case is the
four-point function of $\E$.

For $k=1$, the singular part of the $\E\cdot \E$ operator product
only involves the diagonal stress tensor $T=T^++T^-$. This is a
substantial simplification; the extended chiral superalgebra
obtained by incorporating $\E$ can be viewed an extension of an
ordinary Virasoro algebra $\V$, rather than a symmetric product
$\SV$. In fact, this algebra is a familiar one, an ${\cal N}=1$
super Virasoro algebra: \eqn\zolbo{\E(x/2)\E(-x/2)\sim {1\over
x^3}\left(1+{x^2\over 16}T\right)+{\cal O}(x).} The stress tensor
$T$ has $c=48$ (as it is the diagonal stress tensor in the
symmetric product of two copies of a $c=24$ theory).  A
superconformal field theory with this central charge can be
constructed from 32 chiral multiplets consisting of bosonic and
fermionic fields $\phi_i$ and $\psi_i$, where $\langle
\phi_i(x)\phi_j(0)\rangle=-\delta_{ij}\ln (x)$, and $\langle
\psi_i(x)\psi_j(0)\rangle =\delta_{ij}/x$, for $i,j=1,\dots,32$.
In terms of these fields, we can take \eqn\ocolbo{\E={i\over
4\sqrt 2}\sum_{k=1}^{32}\psi_k\partial\phi_k.} With this free
field realization, all correlation functions of $\E$ can be
explicitly computed.  For example, the four-point function is
\eqn\tofog{\eqalign{\big\langle
\E(e_1)\E(e_2)\E(e_3)\E(e_4)\big\rangle=\prod_{i=1}^4(de_i)^{3/2}&\biggl(\left({1\over
(e_1-e_2)^3(e_3-e_4)^3}+{\rm cyclic}\right)\biggr.\cr
&\biggl.+{3\over 32}{1\over \prod_{i<j}(e_i-e_j)}\biggr),\cr}}
where additional terms obtained by cyclic permutations of
$e_2,e_3,e_4$ are to be added in the first term on the right. We
will eventually compare this formula to the known genus 1
partition function of the monster theory of $k=1$.

In \tofog, we included a factor of $(de)^{3/2}$ for each operator
$\E(e)$ to reflect the fact that $\E$ has dimension $3/2$.  Thus
the correlation ``function'' $\big\langle
\E(e_1)\E(e_2)\E(e_3)\E(e_4)\big\rangle$ is most naturally
understood as a $3/2$-differential in each variable.  This factor
is often omitted in writing such formulas, and we will do so as an
abbreviation except when the factor is important.

For $k=2$, the singular part of $\E\cdot \E$ cannot be expressed
in terms of the diagonal stress tensor $T$ only, and there is no
way to avoid using $\SV$.  Correspondingly, the chiral algebra for
$k=2$ seems to be unfamiliar, and it is considerably harder to get
the analog of \tofog\ for $k=2$.  Since we do not have an explicit
realization of the algebra analogous to  \ocolbo, we cannot
compute the four-point function directly, and instead we will
determine it by requiring the appropriate OPE singularities.

The function $\big\langle
\E(e_1)\E(e_2)\E(e_3)\E(e_4)\big\rangle$, with $e_1,\dots,e_4\in
C_0\cong \Bbb{CP}^1$, must be symmetric in all arguments and must
have the appropriate OPE singularities in all channels.  In
addition, it is highly constrained by invariance under the action
of $SL(2,\C)$ on $C_0$. The most general ``function,'' actually a
cubic differential in each variable, that is symmetric,
$SL(2,\C)$-invariant, and consistent with at least the leading OPE
singularity, is \eqn\helpfun{\eqalign{\big\langle
\E(e_1)\E(e_2)\E(e_3)\E(e_4)\big\rangle=&
\prod_{i=1}^4(de_i)^3\Biggl(\left({1\over
(e_1-e_2)^6(e_3-e_4)^6}+{\rm cyclic}\right)\Biggr.\cr &+\left.
A\biggl({1\over(e_1-e_2)^4(e_3-e_4)^4}\prod_{i=1,2,\,j=3,4}{1\over
e_i-e_j}+{\rm cyclic}\biggr)\right.\cr&\left. +B{1\over
\prod_{i<j}(e_i-e_j)^2}.\right).\cr}} Cyclic permutations of
$e_2,e_3,e_4$ are to be added as shown, and the coefficients $A$
and $B$ must be determined to ensure that the OPE singularities
are correct.

After a detailed calculation, one finds that the right values are
\eqn\nurtyo{A={3\over 16},~B={12k+1\over 2^{16}}} where of course
$k=2$. To get these formulas, one may for instance set
$e_1=-e_2=e$, and use \mimico\ to determine the singular behavior
for $e\to 0$.   In this way, one expresses the singular behavior
of the four-point function in terms of three-point functions
$\big\langle{\cal X}(0)\E(e_3)\E(e_4)\big\rangle$, where ${\cal
X}$ is one of the descendants of the identity that appear in
\mimico.

The coefficient $A$ appears in the coefficient of $1/e^4$ for
$e\to 0$, and in determining it, the only important choices of
${\cal X}$ are $1$ and $T$.  The three-point functions that we
need are \eqn\urtyo{\eqalign{\big\langle 1\cdot
\E(z)\E(w)\big\rangle & ={1\over (z-w)^6}\cr \big\langle
T(x)\E(z)\E(w)\big\rangle & = {3\over (z-x)^2(w-x)^2(z-w)^4}.\cr}}
The first is immediate from the way we have normalized $\E$. The
second, since the three fields involved are quasi-primary fields
(they transform as primaries under $SL(2,\C)$) is determined by
$SL(2,\C)$ invariance up to a constant multiple.  The constant can
be determined using the $T\cdot \E$ OPE (which reflects the fact
that $\E$ is a primary of dimension $3=3k/2$) or the $\E\cdot \E$
OPE.

To determine $B$, we need three more three-point functions, namely
\eqn\gelbo{\eqalign{\big\langle\partial^2T(0)\E(z)\E(w)\big\rangle&
=\left({1\over z^2}+{1\over w^2}\right){18\over
z^2w^2(z-w)^4}+{24\over z^3w^3(z-w)^4}\cr\big\langle T\star
T(0)\E(z)\E(w)\big\rangle &=\left({5\over z^2}+{5\over
w^2}-{4\over zw}\right){3\over z^2w^2(z-w)^4}\cr\big\langle
T^+T^-(0)\E(z)\E(w)\big\rangle& = \left({9k^2\over 16}+{3k\over
64}\right){1\over z^4w^4(z-w)^2} .\cr}} The first of these results
is obtained simply by differentiating the second formula in
\urtyo.

To get the second, we use the fact that \eqn\elbox{T\star
T(0)={\rm Res}_{x=0}{dx\,T(x)\over x}T(0).} Now consider the
expression \eqn\melgo{\Big\langle {dx\,T(x)\over
x}T(0)\E(z)\E(w)\Big\rangle.} We view this as a differential form
on the complex $x$-plane, with $z$ and $w$ kept fixed.  As such,
it has poles precisely at $x=0,z,w$.  The residue at $x=0$ is the
desired three-point function $\big\langle T\star
T(0)\E(z)\E(w)\big\rangle$.  The residues at at $x=z,w$ can be
computed in terms of the correlator $\big\langle
T(0)\E(z)\E(w)\big\rangle$ by using the fact that $\E$ is a
primary of dimension 3, which determines the singular behavior of
the product $T\cdot \E$.  The sum of the residues must vanish, and
this gives our result.

Finally, to get the last formula in \gelbo, we note first that
since the operators $T^+T^-$ and $\E$ are quasiprimary fields of
dimension 4 and 3, we have \eqn\rexox{\big\langle
T^+T^-(x)\E(z)\E(w)\big\rangle={\theta\over
(x-z)^4(x-w)^4(z-w)^2},} for some constant $\theta$.  We have
\eqn\ruffox{\lim_{w\to\infty}{\big\langle
T^+T^-(x)\E(0)\E(w)\big\rangle\over \big\langle
\E(0)\E(w)\big\rangle}={\theta\over x^4}.} This is supposed to be
computed in the theory $\SW$ on the $x$-plane, with insertions of
$\E$ at $0,w$.  This is equivalent to a computation in the theory
$\W$ on a double cover of the $x$-plane branched at $0,w$.  In the
limit $w\to\infty$, the double cover is described by the equation
$y^2=x$, and it is convenient to transfer the computation to the
$y$-plane.

We will now write the stress tensor explicitly as $T_{xx}$ or
$T_{yy}$ depending on which local parameter is used in defining it.
Allowing for the conformal anomaly, the relation between the stress
tensor $T_{xx}$ on the $x$-plane and the corresponding stress tensor
$T_{yy}$ on the $y$-plane is \eqn\tolto{\left({\partial
y\over\partial x}\right)^2T_{yy}=T_{xx}-{c\over 12}\{y,x\},} where
$\{y,x\}$ denotes the Schwarzian derivative of these two functions.
With $y^2=x$, we get $\{y,x\}=3/8x^2$.  In the case of interest,
$c=24k$ so \eqn\tefflo{T_{xx}={1\over 4x}T_{yy}+{3k\over 4x^2}.}
This formula enables us to express $T^+$ in terms of
$T_{yy}|_{y=\sqrt x}$ and $T^-$ in terms of $T_{yy}|_{y=-\sqrt x}$.
The limit in \ruffox\ becomes \eqn\verfo{\Big\langle\left({1\over
4x}T_{yy}|_{y=\sqrt x}+{3k\over 4x^2}\right)\left({1\over
4x}T_{yy}|_{y=-\sqrt x}+{3k\over 4x^2}\right)\Big\rangle.} (The idea
is that passing to the $y$-plane eliminates the twist operators $\E$
from the formalism, and replaces $T^\pm(x)$ by a stress tensor
evaluated at $y=\pm\sqrt x$.) Using the fact that in conformal field
theory on the $y$-plane, $\langle T_{yy}\rangle=0$ and $\big\langle
T_{yy}(y_1)T_{yy}(y_2)\big\rangle =c/2(y_1-y_2)^4$ with $c=24k$, the
correlation function in \verfo\ can be evaluated to give
$\theta/x^4$ with $\theta=9k^2/16+3k/64$. This gives the last
formula in \gelbo.

With the aid of the three-point functions in \urtyo\ and \gelbo\ and
the operator product formula \mimico, one can verify that the
candidate four-point function \helpfun\ has the right singularities
precisely if the coefficients $A$ and $B$ are as claimed in \nurtyo.

Our derivation has in fact been somewhat redundant, since the
formulas for three-point functions can largely be deduced from the
formula for the $\E\cdot\E$ operator product, and vice-versa.

\subsec{The Genus One Partition Function
Revisited}\subseclab\revisit

We will now explicitly use our formulas for the four-point
function of the twist field $\E$ to recover the partition
functions of a $k=1$ or $k=2$ extremal CFT on a Riemann surface of
genus 1.

First we must describe the necessary formalism more precisely. The
partition function $Z_C(e_1,e_2,\dots,e_{2g+2})$ of any conformal
field theory $\W$ on the hyperelliptic Riemann surface $C$ defined
by \eqn\yeso{y^2=\prod_{i=1}^{2g+2}(x-e_i)} can indeed be expressed
in terms of the genus zero twist field correlation function
\eqn\eso{\big\langle\E(e_1)\E(e_2)\dots \E(e_{2g+2})\big\rangle.}
However, to do so requires an important detail that we have not yet
described.

There are at first sight several strange features in relating $Z_C$
to the twist field correlation function:

(1) The dimension of the twist field $\E$ seems to be wrong. The
Riemann surface $C$ degenerates as $e_i\to e_j$ for any $i$ and $j$.
As the complex structure of $C$ is invariant under the exchange
$e_i\leftrightarrow e_j$, a natural local parameter near the
degeneration locus on the moduli space is actually $w=(e_i-e_j)^2$.
(For example, this statement can be made precise using eqn. \clasfo\
below.) One expects $Z_C$ to behave for $e_i\to e_j$ as $w^{-k}$,
since the ground state energy is $-k$. However, as $\E$ has
dimension $3k/2$, the behavior of the twist field correlation
function as $e_i\to e_j$ is actually $(e_i-e_j)^{-3k}\sim
w^{-3k/2}$.

(2) This brings in relief the fact that if $k$ is odd, $\E$ actually
has half-integral dimension  and so is fermionic. For odd $k$, the
twist field correlation function \eso\  actually changes sign under
exchange of $e_i$ and $e_j$, while $Z_C$ is invariant under this
exchange.

(3) The twist field correlation function is most naturally
understood as a $3k/2$-differential in each variable.  We have
written the formulas \tofog\ and \helpfun\ in such a way as to
emphasize this.  It may not be immediately clear what this means for
the partition function $Z_C$.

The answer to all of these questions has to do with the proper
treatment of the conformal anomaly in the holomorphic context.  We
postpone a full explanation to section \role\ and for now we merely
write down the precise relation between the genus 1 partition
function and the twist field correlation function:
\eqn\yetro{Z_C(e_1,e_2,\dots,e_4)=2^{8k}\biggl(\prod_{1\leq i<j\leq
4}(e_i-e_j)^k\biggr) \big\langle\E(e_1)\E(e_2)\dots
\E(e_{4})\big\rangle.} The factor of $2^{8k}$ just reflects the way
we have normalized $\E$. The factor of $\prod_{i<j}(e_i-e_j)^k$
resolves problems (1) and (2) above.  In section \role, we will
explain the meaning of that factor and its relation to problem (3)
(as well as the generalization to higher genus). For now, we merely
mention that in eqn. \yetro, the twist field correlation function
should be understood as a function, without the factor of
$\prod_i(de_i)^{3k/2}$.

Now we will use the general relation \yetro\ along with our
previous formulas for the twist field correlation functions to
analyze the genus 1 partition function for $k=1,2$. We make an
$SL(2,\C)$ transformation to set $e_4=\infty$ and to fix
\eqn\romito{e_1+e_2+e_3=0.} The genus 1 partition function at
$k=1$ is then, using \tofog, \eqn\yopo{Z_{C}(k=1)={2^8\over
\prod_{i<j}(e_i-e_j)^2}\left((e_1-e_3)^3(e_2-e_3)^3+{\rm
cyclic}\right)+24.} We want to express this in terms of
$q=\exp(2\pi i\tau)$, the usual parameter on the genus 1 moduli
space, and to show explicitly that $Z_C$ has the expected behavior
near $q=0$. The classical formulas are
\eqn\clasfo{\eqalign{e_1-e_2&=\theta_3^4(0,\tau)\cr
                     e_3-e_2&=\theta_1^4(0,\tau)\cr
                     e_1-e_3&=\theta_2^4(0,\tau),\cr}}
where \eqn\lasfo{\eqalign{\theta_1(0,\tau)&=\sum_{n\in
\Z}q^{(n+1/2)^2/2}\cr
\theta_2(0,\tau)&=\sum_{n\in\Z}(-1)^nq^{n^2/2}\cr
\theta_3(0,\tau)&=\sum_{n\in\Z}q^{n^2/2}.\cr}} Expanding \yopo\ near
$q=0$, one can readily verify the expected behavior
$Z_C(k=1)=q^{-1}+{\cal O}(q)$.  Since $Z_C$ is manifestly
modular-invariant, this implies that $Z_C$ must equal $J$.  This can
also be verified directly by comparing \yopo\ to the classical
formula for the $j$-function
\eqn\memor{j(q)=2^5{\left((e_1-e_2)^2+(e_2-e_3)^2+(e_3-e_1)^2\right)^3
\over \prod_{i<j}(e_i-e_j)^2}} and recalling that $J=j-744$.

The case $k=2$ can be treated similarly.  Our previous formula
\helpfun\ for the twist field four-point function implies, via
\yetro, a formula for $Z_C(k=2)$.   If we take $e_4\to\infty$ and
$e_1+e_2+e_3=0$, this formula becomes
\eqn\milfik{\eqalign{Z_C(k=2)=&2^{16}\left({(e_1-e_3)^2(e_2-e_3)^2\over
(e_1-e_2)^4}+{\rm cyclic}\right)\cr &+3\cdot
2^{12}\left({(e_1-e_3)(e_2-e_3)\over (e_1-e_2)^2}+{\rm
cyclic}\right)+(12k+1).\cr}}   One can expand this near $q=0$ and
verify the expected behavior $Z_C(k=2)=q^{-2}+1+{\cal O}(q)$.
Together with modular invariance, this implies that
$Z_C(k=2)=J^2-393767$, as one can also verify directly using
\memor.

\def\R{{\cal R}}\def\M{{\cal M}}\def\L{{\cal L}}
\subsec{Role Of The Conformal Anomaly}\subseclab\role

Naively speaking, the partition ``function'' of a conformal field
theory on a Riemann surface $C$ depends only on the complex or
conformal structure of $C$.

Actually, because of the conformal anomaly, things are more
subtle.  The usual way to describe the situation is to endow $C$
with a metric  $h$, and to think of the partition function as a
function $Z(h)$ that transforms in a certain prescribed way
(depending on the central charge) under conformal rescaling $h\to
e^\phi h$ of the metric.

This description is natural in differential geometry.  However, it
has a few limitations.  One is that it really only gives an
adequate framework if the left and right central charges $c_L$ and
$c_R$ are equal.  A second limitation, more crucial for our
purposes, is that it is not well-adapted to holomorphic
factorization.  In a holomorphic CFT, everything will be much
simpler if we can incorporate the conformal anomaly in purely
holomorphic terms.

An alternative approach is to think of the partition function not
as a function but as a section of a suitable line bundle $\R$ over
$\M$, the moduli space of complex Riemann surfaces.  This approach
is developed in detail in \ref\segal{G. Segal, ``The Definition Of
Conformal Field Theory,'' in U. Tillmann, ed. {\it Topology
Geometry, and Quantum Field Theory} (Cambridge University Press,
2004).}.  Holomorphic factorization is incorporated naturally in
this approach by saying that in a holomorphic CFT, $\R$ is a
holomorphic line bundle over $\M$.

More specifically, $\R\cong \L^{c/2}$, where $c$ is the central
charge and $\L$ is a fundamental line bundle over $\M$ that is
known as the determinant line bundle.  $\L$ is defined as follows.
Let $p$ be a point in $\M$ corresponding to a Riemann surface $C$.
Then $\L_p$, the fiber of $\L$ at $p$, is the top exterior power
of the vector space $H^0(C,K_C)$ of holomorphic differentials on
$C$.  So if $C$ has genus $g$ and $\omega_1,\dots,\omega_g$ are
holomorphic differentials on $C$, then the expression
$\omega_1\wedge \omega_2\wedge\dots\wedge\omega_g$ defines a
vector in $\L_p$.  In our problem, $c=24k$, and the partition
function is a section of $\L^{12k}$.

On a hyperelliptic Riemann surface
\eqn\felfox{y^2=\prod_{i=1}^{2g+2}(x-e_i),} we can make this
completely explicit.  A basis of the space of holomorphic
differentials is given by \eqn\elfox{{dx\over y},~{x\,dx\over
y},\,{x^2\,dx\over y},\dots,{x^{g-1}\,dx\over y}.} A section of
$\L^{12k}$ over the space of hyperelliptic equations is hence an
expression of the form \eqn\ubelfox{\Theta=\left({dx\over
y}\wedge{x\,dx\over y}\wedge{x^2\,dx\over
y}\wedge\dots\wedge{x^{g-1}\,dx\over
y}\right)^{12k}F(e_1,\dots,e_{2g+2}).} To get a section of
$\L^{12k}$ over the moduli space of hyperelliptic Riemann
surfaces, $\Theta$ must be invariant under the action of
$SL(2,\C)$ on the space of hyperelliptic equations.  This action
takes the form \eqn\ubelfox{\eqalign{e_i&\to {ae_i+b\over
ce_i+d}\cr x&\to {ax+b\over cx+d}\cr y&\to {y\over
(cx+d)^{g+1}\prod_{i=1}^{2g+2}(ce_i+d)^{1/2}},\cr}} and the
condition for $\Theta$ to be invariant is that $F$ transforms by
\eqn\nubelfox{F(e_1,\dots,e_{2g+2})\to F(e_1,\dots,e_{2g+2})
\prod_{i=1}^{2g+2}(ce_i+d)^{-6kg}.}

The twist field correlation function, on the other hand, is of the
form
\eqn\zubelfox{\big\langle\E(e_1)\E(e_2)\dots\E(e_{2g+2})\big\rangle
=\prod_{i=1}^{2g+2}(de_i)^{3k/2} G(e_1,\dots,e_{2g+2})} for some
``function'' $G$. $SL(2,\C)$ invariance of the correlation
function means that $G$ transforms by
\eqn\yesso{G(e_1,\dots,e_{2g+2})\to
G(e_1,\dots,e_{2g+2})\prod_{i=1}^{2g+2}(ce_i+d)^{3k}.} Comparing
\nubelfox\ and \yesso, we see that the two transformation laws are
consistent if the relation between $F$ and $G$ is
\eqn\ubelfox{F(e_1,\dots,e_{2g+2})=A_g
\,G(e_1,\dots,e_{2g+2})\,\prod_{1\leq i<j\leq 2g+2}(e_i-e_j)^{3k}}
with some constant $A_g$.  Moreover, this is the most general
transformation between $F$ and $G$ that is holomorphic, invariant
under permutations of the $e_i$, and an isomorphism as long as the
$e_i$ are all distinct. So it must be correct with some choice of
the constant. (The relation between $F$ and $G$  is actually part
of a general theory that is briefly indicated below.)

This is not the whole story.  The partition function has another
important property in genus 1: it can be regarded as an ordinary
function, rather than a section of a line bundle.  Indeed, the genus
1 partition function can be defined as a trace, $Z(q)=\Tr\,q^{L_0}$,
and in this form it manifestly is equal to an ordinary function of
$q$.  Yet in the above presentation, the genus 1 partition function
is a section of $\L^{12k}$.  What reconciles the two points of view
is that the line bundle $\L^{12}$ is trivial in genus 1.  If we
describe a genus 1 Riemann surface $C$ by the hyperelliptic equation
$y^2=\prod_{i=1}^4(x-e_i)$, then as explained in \segal, $\L^{12}$
is trivialized by the section \eqn\yefort{s=\left({dx\over
y}\right)^{12}\prod_{1\leq i< j\leq 4}(e_i-e_j)^2.} The point of
this formula is that $s$ is holomorphic, $SL(2,\C)$-invariant,
invariant under permutations of the $e_i$, and nonzero as long as
the $e_i$ are all distinct (so that $C$ is smooth). A power $\L^n$
of the line bundle $\L$ can be trivialized in this fashion precisely
if $n$ is an integer multiple of 12 (and this gives one way to
understand the fact that holomorphic factorization is possible only
if $c$ is an integer multiple of 24).  The genus 1 partition
function $Z$ understood as an ordinary function is obtained from the
section $\Theta$ of $\L^{12k}$ by dividing by $s^k$.  Thus
\eqn\helmigo{Z={F \over \prod_{1\leq i<j \leq 4}(e_i-e_j)^{2k}}.}
Combining this with \ubelfox, we get the relation \yetro\ between
the genus 1 partition function, understood as a trace, and the
correlation function of twist fields: \eqn\elmingo{Z=A_1
\biggl(\prod_{1\leq i<j\leq 4}(e_i-e_j)\biggr)\big\langle
\E(e_1)\E(e_2)\E(e_3)\E(e_4)\big\rangle. } The numerical value of
the constant $A_1$ depends on how the twist field $\E$ has been
normalized.

What about $g>1$? What is the best formula depends on how one
prefers to treat the conformal anomaly, and there seems to be in
the physics literature no standard recipe for doing so in the
holomorphic context.  To make contact with the genus 2 formulas of
Tuite \tuite, we may proceed as follows.  A genus 2 Riemann
surface is always a hyperelliptic curve,
$y^2=\prod_{i=1}^6(x-e_i)$.  In genus 2, we cannot trivialize the
line bundle\foot{This fact is consistent with holomorphic
factorization at $c=24k$, since the genus 2 partition function
cannot be written as a trace and so need not be definable as an
ordinary function.} $\L^{12}$, but we can trivialize $\L^{10}$ by
an obvious analog of \yefort: \eqn\grosto{\tilde s=\left({dx\over
y}\wedge {x\,dx\over y}\right)^{10} \prod_{1\leq i< j\leq
6}(e_i-e_j)^2.} Hence, instead of regarding the genus 2 partition
function as a section $\Theta$ of $\L^{12k}$, we can regard it  as
a section $Z=\Theta/\tilde s^k$ of $\L^{2k}$. If we do this, then
the relation between $Z$ and the twist field correlation function
is the obvious analog of \elmingo: \eqn\zelmingo{Z=A_2
\biggl(\prod_{1\leq i<j\leq 6}(e_i-e_j)\biggr)\big\langle
\E(e_1)\E(e_2)\dots\E(e_6)\big\rangle. }  It can be shown that, in
genus 2, a section of $\L^r$ is the same as a Siegel modular
function of weight $r$.  So the formula \zelmingo\ is the one we
should use if we wish to regard the genus 2 partition function at
$c=24k$ as a Siegel modular function of weight $2k$, as was done
in \tuite. This probably gives the most intuitively appealing
formulas in genus 2.  For genus greater than 2, we leave the
choice of the most useful formalism to the reader.

\nref\deligne{P. Deligne, ``Le D\'{e}terminant de La
Cohomologie,'' in {\it Current Trends In Arithmetical Algebraic
Geometry}, Contemp. Math. {\bf 67} (Amer. Math. Soc., Providence,
1987).} We conclude by briefly sketching the theoretical context
for the formula \ubelfox.  In doing so, we will not limit
ourselves to the case of a hyperelliptic Riemann surface, but will
consider an arbitrary base curve $C_0$ and a two-fold cover $C\to
C_0$ branched at points $e_1,\dots,e_s$.  Let $K_i$ be the
cotangent bundle to $C_0$ at the point $e_i$.  Let ${\cal
L}_{C_0}$ and ${\cal L}_C$ be the determinant lines of $C_0$ and
$C$ respectively.  Then as has been shown by P. Deligne, using
results in \deligne,
  there is a natural isomorphism  ${\cal
L}_C^8\cong {\cal L}_{C_0}^{16} \otimes_iK_{C_i}$.  If $C_0$ is of
genus zero, then ${\cal L}_{C_0}$ is naturally trivial, and the
$3k/2$ power of this isomorphism is the map in \ubelfox.   (The
isomorphism has a natural square root, relevant if $k$ is odd.)

\appendix{A}{More Extremal Partition Functions}F
In section \extremal, we have already described the extremal
superconformal partition functions $F_{\k}$ for $\k\leq 4$. The
purpose of this appendix is to briefly describe these functions,
and the corresponding Ramond partition functions $H_{\k}$, for
$5\leq \k\leq 10$.

For $\k=5$, we have \eqn\tolmo{\eqalign{F_5=&\,\,K^5 - 1380 K^3 -
10239K^2 + 324871K + 2579929\cr =&\,\,q^{-5/2} + q^{-1} + q^{-1/2}
+ 1 + 924492 q^{1/2} + 185710868 q +
 11953414737 q^{3/2} \cr &\,\,+ 416513865728 q^2 + 9727092442216 q^{5/2}+\dots .\cr}}
The corresponding Ramond partition function is \eqn\obolmo{H_5=
143 + 371027968 q + 832984743936 q^2 + 340658459983872 q^3 +
 56418463131631616 q^4+\dots.}
The coefficients in these series cannot be expressed as linear
combinations of dimensions of monster representations with
reasonably small coefficients.  But strangely, at least one
difference can be so represented; the difference between the
coefficient of $q$ in $F_5$ and half the corresponding coefficient
in $H_5$ is $196884=1+ 196883$. (The coefficients in $H_5$, except
the first one, are even because of supersymmetry, so it is natural
to divide by 2.) Also, the first coefficient in $H_5$, namely 143,
is the dimension of the smallest nontrivial representation of a
sporadic finite group known as the Suzuki group. Unfortunately, the
other coefficients are too large, and the representations of the
Suzuki group too small, for it to be easy to get convincing evidence
of whether the Suzuki group is a symmetry of supergravity at $\k=5$.

For $\k=6$, the partition functions are
 \eqn\olmo{\eqalign{F_6& =
K^6 - 1656 K^4 - 12287K^3 + 618373K^2 +
 6487237K - 12026567\cr &=q^{-3} + q^{-3/2} + q^{-1} + q^{-1/2} + 1 + 3724378 q^{1/2} +
 1298410586 q\cr&\,\, + 127852130050 q^{3/2} + 6378693040128 q^2 +
 204402839559265 q^{5/2}+\dots \cr H_6&=1 + 2589372416 q + 12754796707840 q^2 \cr &
 \,\,+ 9529193701720064 q^3 +
 2622934801904828416 q^4+\dots.\cr}}

  The fact that the leading coefficient in the Ramond sector is 1
  suggests that the model may be invariant under a very large
  discrete symmetry group.  The monster $\MM$ does not work, since
  the first non-trivial NS coefficient 3724378 cannot be expressed
  nicely in terms of dimensions of monster representations.  It
  turns out that another large sporadic group, the baby monster
  $\Bbb{B}$, is a better candidate.  The dimensions
  $r_1,\dots,r_{12}$ of the first 12 irreducible representations
  of $\Bbb{B}$ are listed in Table 3.  In terms of those
  dimensions, we find that the first two non-trivial NS
  coefficients in the partition functions can be expressed as
  follows
  \eqn\regoto{\eqalign{     3724378&= 7r_1+4r_2+3r_3+3r_4\cr
                              1298410586&=14r_1+16r_2+7r_3+8r_4+4r_6+3r_7+r_8+2r_9.\cr}}
  These formulas are not unique, as the $r_i$ obey linear relations with
  small coefficients, for example $r_1+r_3+r_5+r_8=r_2+r_4+r_9$.
  If we let $h_1$ denote the first non-trivial coefficient in
  $H_6$, then it turns out that $h_1/2=
  f_1-f_{1/2}=1298410586-3724378.$  So by subtracting the formulas in \regoto,
  $h_1/2$ can also be expressed as
  a positive linear combination of the $r_i$ with fairly small
  coefficients.    If is also true that $h_2/2=f_2-f_1+f_{1/2}$,
  and so is consistent with $\Bbb{B}$ symmetry if $f_2$ can be suitably expressed in terms of the
  $r_i$.

  These results are suggestive of baby monster
  symmetry at $k^*=6$, though they are perhaps a little less striking than
  the evidence presented in section \extremal\ for monster symmetry at
  $k^*=4$, since the required coefficients are larger and the
  dimensions involved are smaller.  Also, it is harder to continue
  this analysis beyond the first few coefficients, as many
  representations of $\Bbb{B}$ come in.

\bigskip
\centerline{\vbox{\offinterlineskip
\def\tablerule{\noalign{\hrule}}
\halign to 4.5truein{\tabskip=1em plus 2em#\hfil&\vrule height12pt
depth5pt#&#\hfil&\vrule height12pt depth5pt#&#\hfil\tabskip=0pt\cr
 \tablerule
$r_1$&& 1 \cr $r_2$ && 4371\cr
 $r_3$ && 96255
\cr  $r_4$ && 1139374 \cr $r_5$ && 9458750 \cr
$r_6$ && 9550635 \cr $r_7$ && 63532485 \cr $r_8$ &&
347643114 \cr $r_9$ && 356054375 \cr $r_{10}$ &&
1407126890 \cr  $r_{11}$ && 3214743741 \cr $r_{12}$ &&
4221380670 \cr}}}\bigskip \centerline{
\vbox{\hsize=5.1truein\baselineskip=12pt \noindent Table 3.
Presented here from \atlas\ are the dimensions $r_i$ of the $i^{th}$
irreducible representation of the baby monster group $\Bbb{B}$, for
$i=1,\dots, 12$. }}\bigskip

 We present the remaining cases with little comment. For $\k=7$, we have
 \eqn\ugol{\eqalign{F_7 & = K^7 - 1932K^5 - 14335K^4 + 988051K^3 +
 11525041K2 - 75824563K - 840705550\cr &=q^{-7/2} + q^{-2} + q^{-3/2} + q^{-1} + q^{-1/2} + 2 +
 13404883 q^{1/2}\cr & \,\,+ 7740446996 q + 1126452195714 q^{3/2} +
 78170641348884 q^2+\dots.\cr
 H_7&=            262 + 15394525184 q + 156260255891456 q^2 + 203203950584774656 q^3
   +\dots           .\cr}}

 For $\k=8$, the analogous formulas are
 \eqn\zugol{\eqalign{F_8&=K^8 - 2208K^6 - 16383K^5 + 1433905K^4 +
 17693341K^3 - 213343055K^2\cr & \,\, - 3164679732K - 2780845557\cr
 &=\,\,q^{-4} + q^{-5/2} + q^{-2} + q^{-3/2} + q^{-1} + 2q^{-1/2} + 3 +
 44146598 q^{1/2} \cr &\,\,+ 40700662036 q + 8516908978515 q^{3/2}+\dots\cr
 H_8&=-213 + 80651894784 q + 1605169778655232 q^2 + 3496922597386551296
 q^3+\dots.\cr}}
This is the first case in which the leading coefficient in $H$ is
negative, so that it is necessary to add an integer to $F$ and
$H$.

For $\k=9$, we find \eqn\bugol{\eqalign{F_9& = K^9 - 2484K^7 -
18431K^6 + 1955935K^5 +
 24992137K^4 - 445606619K^3 \cr &\,\,- 7443672266K^2 +
 1774761946K + 223898812203\cr  &=q^{-9/2} + q^{-3} + q^{-5/2} + q^{-2} + q^{-3/2} + 2q^{-1}
  + 3q^{-1/2} + 3 + 135149374 q^{1/2}\cr & + 193216791918 q +
 56847816152503 q^{3/2}+\dots\cr
H_9&=453 + 381161021440 q + 14282018665201664 q^2 +
 50519288391656243200 q^3+\dots
  }}

  Finally, for $\k=10$, we find
  \eqn\zugol{\eqalign{F_{10}&=K^{10} - 2760K^8 - 20479K^7 + 2554141K^6 +
 33421429K^5 - 793639831K^4\cr & - 14145708496K^3 +
 30831695165K^2 + 1166011724825K + 2482063616019\cr
&=q^{-5} + q^{-7/2} + q^{-3} + q^{-5/2} + q^{-2} + 2q^{-3/2} +
3q^{-1} + 3q^{-1/2} + 3\cr &\,\,\, + 389274233 q^{1/2} +
842231630010 q +
 341925580784341 q^{3/2}+\dots \cr
 H_{10}&=-261 + 1652836102144 q + 112692628289650688 q^2 +
 630520566901614002176 q^3+\dots.\cr
 }}
This gives a second example in which the leading coefficient of
$H$ is negative, so that it is necessary to add an integer.
 \listrefs
\end